\documentclass[aip, reprint,twocolumn,superscriptaddress,notitlepage,showpacs,showkeys]{revtex4-1}
\usepackage{graphicx,amsmath,amsfonts,amssymb,upgreek,txfonts,color}
\usepackage[colorlinks,linkcolor=blue,citecolor=blue,urlcolor=blue,breaklinks=true]{hyperref}
\usepackage[utf8x]{inputenc}

\begin{document}

\title{Ultra-precision quantum sensing and measurement based on nonlinear hybrid optomechanical systems containing ultracold atoms or atomic Bose-Einstein condensate 
}

\author{Ali Motazedifard} 
\email{motazedifard.ali@gmail.com}
\address{Quantum Optics Group, Department of Physics, University of Isfahan, Hezar-Jerib, 81746-73441, Isfahan, Iran}
\address{Department of Physics, University of Isfahan, Hezar-Jerib, 81746-73441, Isfahan, Iran}
\address{Quantum Optics and Communication group, Quantum sensing and metrology group, Iranian Center for Quantum Technologies (ICQTs), Tehran, Iran}

\author{A. Dalafi} 
\email{a\_dalafi@sbu.ac.ir}
\address{Laser and Plasma Research Institute, Shahid Beheshti University, Tehran 19839-69411, Iran}

\author{M. H. Naderi} 
\email{mhnaderi@sci.ui.ac.ir}
\address{Quantum Optics Group, Department of Physics, University of Isfahan, Hezar-Jerib, 81746-73441, Isfahan, Iran}
\address{Department of Physics, University of Isfahan, Hezar-Jerib, 81746-73441, Isfahan, Iran}

\date{\today}
\begin{abstract}
In this review, we study how a hybrid optomechanical system (OMS), in which a quantum micro- or nano-mechanical oscillator (MO) is coupled to the electromagnetic (EM) radiation pressure, consisting of an ensemble of ultracold atoms or an atomic Bose-Einstein condensate (BEC), can be used as an ultra precision quantum sensor for measuring very weak signals. As is well-known in any precise quantum measurement the competition between the shot noise (SN) and the backaction noise of measurement executes a limitation on the measurement precision which is the so-called standard quantum limit (SQL). In the case where the intensity of the signal is even lower than the SQL, one needs to perform an ultra precision quantum sensing to beat the SQL. For this purpose, we review three important methods for surpassing the SQL in a hybrid OMS: (i) the backaction evading measurement of a quantum nondemolition (QND) variable of the system, (ii) the coherent quantum backaction noise cancellation (CQNC), and (iii) the so-called parametric sensing, the simultaneous signal amplification and added noise suppression below the SQL. Furthermore, we have shown in this article for the first time how the classical fluctuation of the driving laser phase, the so-called laser phase noise (LPN), affects the power spectrum of the output optical field in a standard OMS and induces an additional impression noise which makes the total system noise increase above the SQL. Also, for the first time in this review it has been shown that in the standard OMSs, it is impossible to amplify signal while suppressing the noise below the SQL simultaneously.

\end{abstract}


\maketitle

\tableofcontents

\section{Introduction}\label{secIntroduction}

During the past decade, the field of quantum \textit{optomechanics} \cite{aspelmeyerOMS,aspelmeyerOMSBOOk,chenOMS,meystreOMS,milburnBook,milburnOMS,isartReviewLevitated}, in which the mechanical (phononic) mode of a macroscopic quantum mechanical oscillator (MO) is coupled to an optical (or microwave) mode via the radiation pressure, as an interesting field in quantum \textit{science} and \textit{technologies} has been developed using the-state-of-the-art technologies both in theory and experiment for a variety of purposes, including testing the fundamentals of physics, exploring the quantum effects on the macro scale as well as controllable ultra-sensitive measurements such as force-sensing.

The optomechanical systems (OMSs) have been applied to a wide variety of research fields including Bell test for macroscopic mechanical entanglement \cite{optomechanicalBelltest1}, optomechanical teleportation \cite{teleportation}, detection of acoustic blackbody radiation \cite{blackbodyOMS2020PurdySingh}, quantum sensing \cite{xsensing1,cQNCPRL,cQNCPRX,aliNJP,complexCQNC,cQNCNatureexp,masssensing,conditionalbackactionevading,aliDCEforcesenning,fani2020BA,sillanpaasensing1,sillanpaasensing2,sillanpaa2020ForceFree,seokForce2020,kippenbergIntermodulationNoise,magnetOMS2014,thermometry2017,mehryMagneticsensing2020,xiongForcesensor2020,atomicforcemicroscopy2021,belloSqueezing2020,khaliliForce2020,xSensing2020,groblacherCQNC2020,GiovanniCQNC2020} such as position, force, magnetic, and temperature sensing; and quantum illumination or quantum radar \cite{quantumillumination,barznjeRadar}, cooling of the MO \cite{groundstatecooling,sidebandcooling,lasercooling}, generation of entanglement \cite{palomaki2,paternostro,genesentangelment,dalafiQOC,barzanjehentanglement,foroudcrystalentanglement,roomtemperatureEntanglement}, synchronization of MOs \cite{mari1,mianZhang,bagheri,foroudsynch}, generation of quadrature squeezing and amplification \cite{clerkdissipativeoptomechanics,pontinmodulation,optomechanicswithtwophonondriving,clerkfeedback,harris,bowen,twofoldsqueezing,foroudState,aliDCEsqueezing,sillanpaa1,sillanpaa2,sillanpaa3,sillanpaa4,sillanpaa5,wollman}, the dynamical Casimir effect (DCE)-based nonclassical radiation sources \cite{aliDCE1,aliDCE2,aliDCE3,noriDCEPRL,noriDCE2,noriDCE3}, quantum simulation of the curved space-time \cite{foroudCurvedspacetime} and the optomechanically induced transparency (OMIT) \cite{oMIT1,oMIT2,oMIT3,vitaliOMIT,marquardtOMIT,xiongOMIT} which is analogous to the familiar phenomenon of the Electromagnetically induced transparency (EIT) \cite{eIT}, realization of microwave nonreciprocity, unidirectional transport, isolator and circulator \cite{barzanjehnonreciprocity}, topological phonon transport \cite{paninterTopology2020}, detecting the acoustic blackbody radiation \cite{purdy2020acousticOMS}, and Green's function approach in optomechanics \cite{aliGreen,greenScarlatella1,greenScarlatella2}.

Ideally, quantum mechanics imposes a fundamental limit on measurement precision, called the Heisenberg limit (HL) \cite{heisenbergCite}, which can be only achieved for closed quantum systems which are completely noiseless. Nevertheless, for real situations where the quantum system is open, i.e., interacts with an environment and is exposed to the environmental noises, the HL is not achievable. In these real situations, environmental decoherence imposes a more severe limitation on precision instead of the HL, which is called the standard quantum limit (SQL). In the present article where we have focused on the OMSs as open quantum systems that are exposed to the environmental noises, the precision limit of measurement is determined with the SQL.

In the OMSs, the competition between the SN \cite{radiationNoise1} and the radiation pressure backaction noise \cite{radiationNoise2}, which have opposite dependences on the input power, determines the standard quantum limit (SQL) \cite{braginskyBook}. 
The SN limits high-precision interferometry at high frequencies \cite{lIGO1992}, while backaction noise becomes relevant only at large enough powers and will be limiting in the low-frequency regime for the next-generation gravitational-wave detectors \cite{radiationNoise3,radiationNoise4}. In fact, increasing the input power decreases the SN, on one hand, while causes an increase of the backaction noise on the other hand. Therefore, the greatest challenge in an ultra-precision quantum measurement \cite{Sillanpaahiddencorrelation2018,sillanpaanoiselessmeasurement2017} for surpassing the SQL is to find methods to suppress the backaction noise.

There are various proposals for quantum noise reduction and beating the SQL in force measurements, such as frequency-dependent squeezing of the input field \cite{shapirosensing}, variational measurements \cite{kimblesensing,khalilisensing}, the use of Kerr medium in a cavity \cite{bondurantsensing}, a dual MO setup \cite{pinardSensing}, the optical spring effect \cite{chenSensing}, two-tone measurements \cite{zimmermannSensing,clerkfeedback,meystreSensing,braginsky1980} as well as their experimental realization~\cite{hageSensing,grayEXPsensing,sheardSensingEXP,heidmannSensingEXP,pontin}, and quantum-nondemolition (QND) measurements \cite{meystreSensing,teufel}. 
In addition to these proposals which are based on noise reduction, the CQNC proposals are based on the noise cancellation via quantum interference. It should be noted that although in these methods the backaction noise of measurement is reduced or even canceled but the signal response is never amplified. 
In more recent proposals it has been shown that it is possible to suppress the added noise of measurement while amplifying the input signal simultaneously in a bare \cite{optomechanicswithtwophonondriving} or nonlinear hybrid \cite{aliDCEforcesenning} OMS through the parametric modulation of the phononic modes.

We should emphasize that one of the strategies to enhance the quantum effects in order to improve the quantum sensing is adding suitable \textit{nonlinearities} into the OMSs or \textit{hybridizing} the cavity which provide us more controllability on the systems in the context of the \textit{quantum control} protocols, and that is why many recent researches have been allocated to high-precision measurements based on feasible nonlinear hybridized systems.
Recently, hybrid OMSs containing Bose-Einstein condensates (BECs) \cite{meystreBEC2010,brennNatureBECexp,brennNatureBECOMS,ritterBECexp} in which the fluctuation of the collective excitation of the BEC, i.e., Bogoliubov mode, behaves like an effective mechanical mode \cite{meystreBEC2010} and the nonlinear atom-atom interaction simulates an atomic amplifier \cite{dalafi1,dalafi2,dalafi3}, have more controllability and can increase the quantum effects at macroscopic level \cite{dalafi1,dalafi2,dalafi3,dalafi4,dalafi5,dalafi7,dalafi8,bhattacherjeeNMS}. Besides, such hybrid systems are suitable for reduction of quantum noise \cite{bhattacherjeenoisereduction} or may act as a quantum amplifier/squeezer \cite{aliDCEsqueezing}. Moreover, by considering the quadratic optomechanical coupling in such hybrid systems, one can generate robust entanglement and strong mechanical squeezing beyond the SQL \cite{dalafiQOC}. Here, it is worth reminding that during the recent decades many researches have been done experimentally and theoretically in atomic-BEC systems in the context of quantum measurement \cite{onofrioBEC1996RMP,onofrioBEC2002PRAmeasurement,onofrioBEC2002PRAGross,onofrioDecoherenceAtom1998}. Very recently, it has been shown experimentally that the BEC has the potential to perform high-precision quantum Gravimetry \cite{gravimetryBEC2020PRL} and 
high-performance quantum memory \cite{memoryBEC2020Exp}.

Also, in recent years, hybrid OMSs equipped with an atomic gas have attracted considerable attention. It has been shown that the additional atomic ensemble can improve optomechanical cooling \cite{rabl2012,genes2009,hammerer2010,geraci2015,camerer2011} and also provide the possibility of ground state cooling outside the resolved sideband regime \cite{meystre2014,jockel2014}. Moreover, the coupling of the MO to an
atomic ensemble can generate a mechanical squeezed state \cite{nori2008}, or robust EPR-type entanglement between collective spin variables of the atomic medium and the MO \cite{zoller2009,tombesi2007}.

In this review article, we are going to focus on three important methods in ultra-precision quantum measurements based on hybrid OMSs for surpassing the SQL: (i) backaction noise \textit{evasion}, (ii) \textit{coherent} quantum noise \textit{cancellation} (CQNC), and (iii) simultaneous noise-suppression and signal-amplification, the so-called  \textit{parametric} sensing (PS).
The \textit{first} one \cite{fani2020BA}, i.e., the backaction evading measurement is a special and more realistic type of the ideal quantum non-demolition (QND) measurements. In this method, a so-called QND variable is defined out of the quantum dynamical variables of the system which is not affected by the backaction noise. Here, we specifically review \cite{fani2020BA} how one can perform a backaction evading measurement on the collective mode of the BEC in a hybrid OMS in which the BEC mode is coupled to the radiation pressure of the cavity. Similar to the first one, in the second method the signal is never amplified. 

The \textit{second} one \cite{aliNJP} is a scheme based on the CQNC in a hybrid OMS containing an ultra-cold atomic ensemble which plays the role of a negative mass oscillator (NMO) and leads to the backaction noise cancellation due to the destructive quantum interference (DQI). Although in this method the backaction noise can be perfectly vanished, but the signal is never amplified.
Finally, in the \textit{third} one \cite{aliDCEforcesenning} as a golden method, we investigate the situation where the noise is strongly suppressed below the SQL while the signal-response can be substantially amplified. For this purpose, we consider a BEC-based optomechanical cavity where the \textit{s}-wave scattering frequency of the BEC atoms as well as the spring coefficient of the MO are \textit{parametrically} time-modulated. Under special conditions, the mechanical response of the system to the input signal is enhanced substantially which leads to the signal amplification while the added noise of measurement can be suppressed much below the SQL. Furthermore, because of its large mechanical gain, this modulated nonlinear hybrid system, which can be identified as a quantum linear amplifier-sensor, is much better in comparison to the modulated-bare one.

Finally, it should be reminded that the presented methods in this article is general and can be applied to any quantized system as well as generic electro-optomechanical systems. Furthermore, the exact results and approach can be applied to any quantum optical system whose effective Hamiltonian is exactly the same as the OMSs with linear coupling between the field and phononic modes such as levitated systems.

\subsection{Outline}
In sec.~(\ref{secOMS}), we present a short introduction to cavity optomechanics and its dynamics in the linearized regime. Also, in this section we introduce the SN, backaction, and laser phase noise (LPN) which are appeared in the output phase spectrum of the cavity optical mode. Besides it is shown how the SQL is determined by the competition between the SN and backaction noise.

In sec.~(\ref{secBEC}), we show that an atomic BEC in the dispersive regime of atom-field interaction inside the optical cavity is effectively analogous to the OMS where the Bogoliubov mode in the BEC behaves effectively as a MO and the atom-atom collision is responsible for the atomic parametric amplifier analogous to the optical parametric amplifier (OPA).

We present the sensing-mechanism of CQNC, backaction noise evasion and parametric measurement in an atom-assisted OMS in Sections~(\ref{secCQNC}), (\ref{secBAnoise}), and (\ref{secParametricsensing}), respectively.
Finally, we summarize our conclusion remarks and our perspectives in Sec.~(\ref{secOutlook}).

\section{quantum optomechanics \label{secOMS}}

\subsection{dynamics of the standard OMS}

\begin{figure}
	\includegraphics[width=7cm]{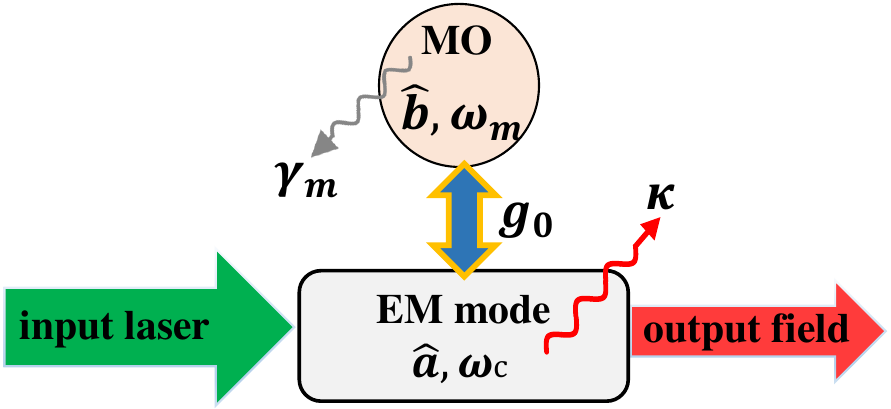}
	\caption{(Color online) Schematic of a general OMS in which a mechanical-phononic mode $ \hat b $ is coupled to a classically driven quantized electromagnetic mode (EM) $ \hat a $ via the radiation pressure with strength $ g_0 $. The natural frequencies and dissipation rates of the EM mode and mechanics are respectively $ \omega_c,\kappa $ and $ \omega_m, \gamma_m $.}
	\label{fig1}
\end{figure}

Fig.~(\ref{fig1}) shows schematically an OMS in which the mechanical mode $ \hat b $ with natural frequency $ \omega_m $ is coupled to the EM mode $ \hat a $ with frequency $ \omega_c $ via the radiation pressure with single-photon optomechanical coupling strength $ g_0 $. Both the EM and mechanical modes are considered as open quantum systems with dissipation rates of $ \kappa$ and $ \gamma_m $, respectively. The EM (cavity) mode is classically driven at rate $ E_L=\sqrt{\kappa P_L/\hbar \omega_L} $ where $ \omega_L $ and $ P_L $ are, respectively, the laser frequency and input laser power.

Note that in the presented model, the \textit{single}-mode approximations for the EM and mechanical modes are valid provided that the cavity free spectral range (FSR) is much larger than the mechanical frequency \cite{singlecavitymoderegime} and the detection bandwidth is chosen such that it includes only a single, isolated, mechanical resonance and mode-mode coupling is negligible\cite{singlemechanicalmoderegime}. Furthermore, in an OMS the cavity frequency depends on the MO displacement, $q$, as ${\omega _c}(q) =\omega_c+ (c/L ){\cos ^{ - 1}}\left( {\left| {{r_c}} \right|\cos (4\pi q/{\lambda _c})} \right)$ \cite{jayich} where $r_c$ and $\lambda_c$ are, respectively, the reflectivity of the MO and the wavelength of the EM mode, and the position of the MO is considered as its distance from the anti-node of the cavity field. It has been shown \cite{soltanolkotabi1} that this dependence leads to a nonlinear coupling (phonon number-dependent optomechanical coupling) between the radiation pressure field and the MO through multi-phonon excitations of the vibrational sidebands. However, by considering the first excitation of the vibrational sideband in the limit of very small values of the Lamb-Dicke parameter, $\eta  = (4\pi /{\lambda _c})\sqrt {\pi \hbar /(m{\omega _m})}$ with m being the motional mass of the MO, and for low values of the membrane reflectivity, the phonon number dependence of the optomechanical coupling can be neglected \cite{soltanolkotabi2}.

Under the single-mode approximation, the total Hamiltonian of the OMS in a frame rotating at driving laser frequency $ \omega_L $ can written as \cite{aspelmeyerOMS,aspelmeyerOMSBOOk}
\begin{eqnarray}\label{hamiltonianOMSNL1}
\hat H_{\rm OMS } &=&\hbar \Delta_c \hat a^\dag \hat a+\hbar\omega_{m}\hat b^\dag \hat b -\hbar g_0 \hat a^\dag \hat a(\hat b+\hat b^\dag)\nonumber\\
&& + i\hbar E_L(\hat a^\dag -\hat a ),
\end{eqnarray}
where $ \Delta_c= \omega_c-\omega_L$ is the cavity detuning. Here, we have considered only the linear coupling between the mechanics and radiation pressure with coefficient $ g_0= x_{\rm zpf} \partial\omega_c/\partial q $ where $ x_{\rm zpf}= \sqrt{\hbar / (2m \omega_m)} $ being the zero-point-position quantum vacuum fluctuation. It should be reminded although it may seem that our description of OMSs is limited to those with the moving-end mirror or  memberane-in-the-middle geometries, but it can be generalized to any other kinds of OMSs like those consisting of levitating dielectrics \cite{isartReviewLevitated} and having an effective Hamiltonian like Eq.(\ref{hamiltonianOMSNL1}).

The dynamics of the system is fully characterized by the fluctuation-dissipation processes affecting both the optical and the mechanical modes. We describe the effect of the fluctuations of the vacuum radiation input and the Brownian noise associated with the coupling of the MO to its thermal environment within the input-output formalism of quantum optics \cite{gardinerBook,wallsBook}. For the given Hamiltonian (\ref{hamiltonianOMSNL1}) this results in the nonlinear quantum Langevin equations (QLEs) 
\begin{subequations} \label{qLEsOMS}
	\begin{eqnarray}
	&& \dot {\hat a} =  - i{\Delta _c}\hat a + i{g_0}\hat a (\hat b + {{\hat b}^\dag }) + {E_L} - \frac{\kappa }{2}\hat a + \sqrt \kappa  {{\hat a}_{in}}, \\
	&& \dot {\hat b }=  - i{\omega _m}\hat b + i{g_0}{{\hat a}^\dag }\hat a  - \frac{{{\gamma _m}}}{2}\hat b + \sqrt {{\gamma _m}} {{\hat b}_{in}},
	\end{eqnarray}
\end{subequations} 
in which the cavity-field quantum vacuum fluctuation $\hat a_{in}$ and the motional quantum fluctuation $\hat b_{in}$ satisfy the commutation relations $\left[ {{{\hat a}_{in}}(t),\hat a_{in}^\dag ( t')} \right] =\left[ {{{\hat b}_{in}}(t),\hat b_{in}^\dag ( t')} \right]= \delta (t - t')$. In the limit of high mechanical quality factor ${Q_m} = {\omega _m}/{\gamma _m} \gg 1$ and when $\hbar {\omega _m} \ll {k_B}T$ where $k_B$ is the Boltzmann constant and $T$ is the temperature of the mechanical bath, $\hat b_{in}$ satisfies the nonvanishing Markovian correlation functions \cite{tombesi} $\langle {\hat b_{in}^\dag (t){{\hat b}_{in}}(t')}\rangle  = n_{th}\delta (t - t')$, $\langle {{{\hat b}_{in}}(t)\hat b_{in}^\dag (t')}\rangle  = (n_{th} + 1)\delta (t - t')$ where $ n_{th} = [\exp (\hbar {\omega _m}/{k_B}T) - 1]^{-1}$ is the mean thermal excitation number of the MO. Furthermore, the nonvanishing correlation function of the input vacuum noise is given by \cite{gardinerBook} $\langle {{{\hat a}_{in}}(t)\hat a_{in}^\dag (t')}\rangle  = \delta (t - t')$.

If the cavity is intensely driven so that the intracavity field is strong which is realized for high-finesse cavities and enough driving power, the QLEs \ref{qLEsOMS}(a) and \ref{qLEsOMS}(b) can be solved analytically by adopting a linearization scheme in which the operators are expressed as the sum of their classical mean values and small fluctuations, i.e., $\hat a = \alpha + \delta \hat a$ and  $\hat b = \beta + \delta \hat b$ with the assumption $\langle \delta {{{\hat{r}}}^{\dag }}\delta \hat{r} \rangle / \langle {{{\hat{r}}}^{\dag }}\hat{r}\rangle \ll 1$ $(r=a, b)$. The classical amplitudes $\alpha  = \langle \hat a \rangle$ and $\beta  = \langle \hat b \rangle$ are governed by equations $\dot \alpha  =  - ({\kappa  \mathord{\left/{\vphantom {\kappa  2}} \right.\kern-\nulldelimiterspace} 2} + i{\Delta})\alpha  + {E_L}$ and $\dot \beta  =  - ({{{\gamma _m}} \mathord{\left/ {\vphantom {{{\gamma _m}} 2}} \right. \kern-\nulldelimiterspace} 2} + i{\omega _m})\beta  + i{g_0}|\alpha {|^2}$ where $\Delta = {\Delta _c} -2 {g_0} {\rm Re} \beta$. The dynamics of the quantum fluctuations can be described by the \textit{linearized} QLEs~\cite{aspelmeyerOMS,aspelmeyerOMSBOOk,milburnBook,milburnOMS,meystreOMS,chenOMS}	
\begin{eqnarray} 
	&& \delta {\dot {\hat a}} \! = \!  - (i{{\Delta }} + \frac{\kappa }{2})\delta {\hat a} + ig(\delta \hat b + \delta{{\hat b}^\dag })\! +\! \sqrt \kappa  {{\hat a}_{in}}, \label{qLEsLinearOMS1a}\\
	&&  \delta \dot {\hat b} \! = \! - (i{\omega _m} \! + \! \frac{{{\gamma _m}}}{2}) \delta \hat b + i g(\delta \hat a + \delta{{\hat a}^\dag })\!  +\! \sqrt {{\gamma _m}} {{\hat b}_{in}}, \label{qLEsLinearOMS1b}
\end{eqnarray}
where $g = {g_0}{\alpha _{ss}} $ is the coherent intracavity-field-enhanced optomechanical coupling strength with $\alpha_{ss} = {E_L}/(\kappa /2 + i{{\Delta }})$ being the steady-state value of $\alpha$ which is always possible to take as a real number by an
appropriate redefinition of phases. Note that one can usually approximate ${\Delta} \approx {\Delta _c}$ for small values of the MO displacement which means that $|\Delta _c| \gg {g_0} \rm Re \beta $.

Using the Fourier transform, $ G(\omega)= \int_{-\infty}^{\infty}d\tau G(\tau)  e^{i\omega \tau} $, the QLEs can be written as follows
\begin{eqnarray} 
&& -i \omega \delta \hat a(\omega)=  - (i{{\Delta }} + \frac{\kappa }{2})\delta {\hat a}(\omega) + ig(\delta \hat b(\omega) + \delta{{\hat b}^\dag(\omega) })\! \nonumber \\
&& \qquad \qquad \qquad + \sqrt \kappa  {{\hat a}_{in}}(\omega),  \label{qLEs_wOMS1a} \\
&&- i \omega  \delta \hat b(\omega) \! = \! - (i{\omega _m} \! + \! \frac{{{\gamma _m}}}{2}) \delta \hat b(\omega) + i g(\delta \hat a(\omega) + \delta{{\hat a}^\dag(\omega) })\! \nonumber  \\
&& \qquad \qquad \qquad  +\! \sqrt {{\gamma _m}} {{\hat b}_{in}}(\omega). \label{qLEs_wOMS1b}
\end{eqnarray}
Note that in the Fourier space $ \hat f^\dag (\omega)= (\hat f(-\omega))^\dag $. This set of linear operator equations can be analytically solved. To obtain the output operators one should use the input-output relation $ \hat a_{\rm out}= \sqrt{\kappa} \delta \hat a - \hat a_{in} $ \cite{gardinerBook,wallsBook}.

\subsection{Shot noise, backaction noise and laser phase noise in optomechanical systems}\label{subSQL}

One of the most important aspects of the OMSs is their capability for precise quantum measurements. Ideally, at zero temperature and in the absence of any kind of classical noises the ultimate limit of the measurement precision in an OMS, the so-called standard quantum limit (SQL), is determined by the competition between the backaction noise and the imprecision shot noise. However, in realistic situations, the thermal fluctuations due to a nonzero environmental temperature as well as the presence of some classical noises like the laser phase noise (LPN) which is the classical fluctuation in the phase of the external laser driving the cavity, affect the measurement precision. In this subsection, we review the SQL in a standard OMS and, here for the fist time, we investigate how the LPN induces an additional impression noise which make the total system noise increase above the SQL.

In order to model the LPN \cite{dalafi6}, we assume that the phase of the driving laser has random fluctuations which is described by the classical stochastic variable $\phi(t)$ whose time derivative $\dot{\phi}(t)\equiv\delta\zeta(t)$ satisfies the following stochastic differential equation \cite{kennedyLPN1,abdi2011LPN2}
\begin{eqnarray}\label{zetaStch}
	\delta\ddot{\zeta}(t)+\tilde{\gamma}\delta\dot{\zeta}(t)+\omega_{N}^{2}\delta\zeta(t)=\omega_{N}^{2}\sqrt{2\Gamma_{L}}\delta\epsilon(t),
\end{eqnarray} 
where $ \omega_{N} $ and $ \tilde{\gamma} $ are, respectively, the central frequency and the bandwidth of the zero-mean noise $ \delta\zeta $ while $ \Gamma_{L} $ is the laser linewidth. $ \delta\epsilon(t) $ is a classical white noise with correlation $ \langle\delta\epsilon(t)\delta\epsilon(t^{\prime})\rangle_{\rm cl}=\delta(t-t^{\prime}) $ which is the input noise for the linear stochastic differential equation (\ref{zetaStch}). Based on the theory of classical stochastic processes \cite{papoulisbookstochastic}, the power spectrum of the output noise $ \delta\zeta(t) $ is determined by the equation 
\begin{eqnarray}
&& S_{\delta\zeta}(\omega)=|h(\omega)|^{2} S_{\delta\epsilon}(\omega) , 
\end{eqnarray}
where $ S_{\delta\epsilon}(\omega)=1 $ is the power spectrum of the input white noise and
\begin{eqnarray}
&&	h(\omega)=\frac{\omega_{N}^{2}\sqrt{2\Gamma_{L}}}{(\omega^{2}-\omega_{N}^{2})+i\tilde{\gamma}\omega},
\end{eqnarray}
is the linear response function of Eq.~(\ref{zetaStch}). Therefore the power spectrum of the output noise $ \delta\zeta(t) $ is obtained as
\begin{eqnarray}
&&	S_{\delta\zeta}(\omega)=\frac{2\Gamma_{L}\omega_{N}^{4}}{(\omega^{2}-\omega_{N}^{2})^{2}+\tilde{\gamma}^{2}\omega^{2}}.
\end{eqnarray}

If we would like to measure a classical weak signal $f(t)$ which has been coupled to the position quadrature $\hat q=(\hat b+\hat b^\dag)/\sqrt{2}$ of the MO, the Hamiltonian (\ref{hamiltonianOMSNL1}) in the presence of the LPN in the frame rotating at the frequency $\omega_L+\dot{\phi}$ can be rewritten as
\begin{eqnarray}\label{HOMSLPS}
	\hat H&=&\hbar(\Delta_c-\delta\zeta)\hat a^\dagger\hat a+\frac{1}{2}\hbar\omega_m(\hat q^2+\hat p^2)\nonumber\\
	&&-\hbar g_0\hat q\hat a^\dagger\hat a+i\hbar E_L(\hat a^\dag-\hat a)+\hbar f(t)\hat q,
\end{eqnarray}
where the single photon optomechanical coupling has been defined as $g_0=\frac{\omega_c}{L}\sqrt{\frac{\hbar}{m\omega_m}}$. It should be emphasized again that this opromechanical coupling can be generalized to any other kinds of OMSs or any system with effective optomechanical Hamiltonian like Eq.(\ref{hamiltonianOMSNL1}), e.g, levitated OMSs \cite{isartReviewLevitated}.

In the presence of the LPN, the QLE of the phase quadrature $\hat P_a \equiv \delta\hat Y=(\delta\hat a-\delta\hat a^\dag)/\sqrt{2}i$ of the optical field is coupled to the classical stochastic equation (\ref{zetaStch}) because of the presence of $\delta\zeta$ in the first term of the Hamiltonian (\ref{HOMSLPS}). Similarly, the amplitude quadrature of the optical field is defined as $\delta\hat X=(\delta\hat a+\delta\hat a^\dag)/\sqrt{2}$. On the other hand, by defining the conjugate variable of $\delta\zeta(t)$ as another classical stochastic variable $\delta\theta(t)\equiv\delta\dot\zeta(t)/\omega_{N}$, the second order differential equation (\ref{zetaStch}) can be rewritten as a set of two first order differential equations which are coupled to the QLEs of the OMS quadratures. In this way, the equations of the system dynamics can be written in the following compact form
\begin{eqnarray} \label{u1}
	\frac{d}{dt} \hat{ \boldsymbol{u}}(t)= \boldsymbol{A} \hat{ \boldsymbol{u}}(t) + \hat{ \boldsymbol{u}}_{in}(t), 
\end{eqnarray}
where the vector of the system fluctuations has been defined as
\begin{equation}
	\hat{ \boldsymbol{u}}(t)=\Big(\delta\hat X(t), \delta\hat Y(t), \delta\hat q(t), \delta\hat p(t), \delta\zeta(t), \delta\theta(t)\Big)^{\rm T},
\end{equation}
and the vector of noises are given by
\begin{eqnarray}
	\hat{ \boldsymbol{u}}_{in}(t)&=&\Big(\sqrt{\kappa} \delta\hat X_{in}(t),\sqrt{\kappa} \delta\hat Y_{in}(t),0,\nonumber\\
	&& \sqrt{2\gamma_m} \delta\hat P_{in}(t)+f(t) , 0, \omega_{N}\sqrt{2\Gamma_N}\delta\epsilon(t) \Big)^{\rm T}.
\end{eqnarray}
Here, there are two important points that should be reminded before proceeding further. Firstly, one can also investigate the continuous variables quantum mechanics, which has been studied in the Heisenberg picture in the present review, through the approach of the master equation in the Schrödinger picture\cite{serafiniBook}.  
Secondly, the stochastic variables $ ( \delta\zeta(t), \delta\theta(t)) $ are classical and cannot be considered as a quantum part of the two-mode OMS, and thus, the system is a bipartite quantum system. Nevertheless, since the equation of motion corresponding to these classical variables has been coupled to the QLEs of the OMS, they have been incorporated into the compact form of Eq.(\ref{u1}). In fact, it is a trick for considering the classical stochastic noises together with the quantum noises in same level.
It should also be noted that, we have here considered the interaction of the MO with its reservoir in the \textit{absence} of the rotating-wave approximation (RWA) \cite{milburnBook,scullybook,fordQnoise1988} in spite of Eq.(\ref{qLEsLinearOMS1b}) where the interaction has been considered in the \textit{presence} of the RWA \cite{milburnBook,scullybook,fordQnoise1988}. Actually, in the high quality factor regime of an MO together with the small mechanical bandwidth regime $ B$ is very smaller than the mechanical frequency ($ B \ll \omega_m $), the interaction of the MO with its environment can be modeled such that the quantum noise only affects the mechanical momentum\cite{milburnBook,scullybook}. Besides, the drift matrix $ \boldsymbol{A} $ is as follows
\begin{eqnarray} \label{chi0OMS}
	&&\!\!\!\!\!\!\!\!\!\!\!\!  \boldsymbol{A}\!=\! \left( \begin{matrix}
		{-\frac{\kappa}{2}} & {\Delta} & 0 & 0 & 0 & 0  \\
		{-\Delta} & {-\frac{\kappa}{2}} & {\sqrt{2} g } & {0} & {\sqrt{2}\alpha} & {0} \\
		{0} & {0} & {0} & {\omega_m} & {0} & {0} \\
		{\sqrt{2} g} & {0}& {-\omega_m} & {-\gamma_m} & {0} & {0}  \\
		{0} & {0} & {0} & {0} & {0} & {\omega_N} \\
		{0} & {0} & {0} & {0} & {-\omega_{N}} & {\tilde{\gamma}} \\
	\end{matrix} \right).
\end{eqnarray}

The input signal $f(t)$ can be detected experimentally by measuring the spectrum of the optical output phase $\delta\hat Y_{out}=\delta\hat Y_{in}-\sqrt{\kappa}\delta\hat Y$ which can be obtained by solving the QLEs in the frequency domain. Under the resonance condition of $\Delta=0$, $\delta\hat Y_{out}(\omega)$ is obtained as follows \cite{milburnBook}
\begin{eqnarray}\label{YoutwOM}
&& \delta\hat Y_{out}(\omega)=-\Bigg(\frac{\kappa/2+i\omega}{\kappa/2-i\omega}\Bigg)\delta\hat Y_{in}(\omega)-2\gamma_m C_{\rm eff}(\omega)\chi_m(\omega) \delta\hat X_{in}(\omega)\nonumber\\
&&\qquad -\frac{h(\omega)}{g_0}\sqrt{2\gamma_m C_{\rm eff}(\omega)}\delta\epsilon(\omega)+2\gamma_m\chi_m(\omega)\sqrt{C_{\rm  eff}(\omega)}\delta\hat P_{in}(\omega)\nonumber\\
&&\qquad -\chi_m(\omega)\sqrt{2\gamma_m C_{\rm eff}(\omega)}f(\omega).
\end{eqnarray}
where  $C_{\rm  eff}(\omega)=\frac{4g^2}{\kappa\gamma_m}(1-2i\omega/\kappa)^{-2}$ is the effective cooperativity where in the steady-state limit $ \omega \ll \kappa $ becomes $C_{\rm  eff}\simeq 4g^2/\kappa \gamma_m $. Furthermore, $\chi_m(\omega)=\omega_m/ (\omega_m^2-\omega^2-i\omega\gamma_m)$ is the mechanical response of the system to the input noise. The first two terms in Eq.~(\ref{YoutwOM}) are, respectively, the imprecision shot noise and the radiation pressure backaction noise while the third and fourth terms are, respectively, the LPN and the thermal noise. The last term corresponds to the input signal. It should be noted that the LPN has been appeared as another imprecision noise which has the classical nature.
Since we would like to measure the input signal $f(t)$, we define the detected output field as
\begin{eqnarray}
&& \delta\hat f_{\rm det}(\omega)=\frac{\delta\hat Y_{out}(\omega)}{\chi_m(\omega)\sqrt{2\gamma_m C_{\rm eff}(\omega)}}
\end{eqnarray}
which is just the rescaled optical output phase quadrature. The power spectrum of the detected output field which is defined as
\begin{eqnarray}
&& S_{\! \rm det}(\omega)=\frac{1}{4\pi}\int_{-\infty}^{+\infty}d\omega'\langle\delta\hat f_{\rm det}(\omega)\delta\hat f_{\rm det}(\omega')+\delta\hat f_{\rm det}(\omega')\delta\hat f_{\rm det}(\omega)\rangle , \nonumber \\
\end{eqnarray}
can be obtained as follows
\begin{eqnarray}
S_{\rm det}(\omega)&=&\frac{1}{4\gamma_m|\chi_m(\omega)|^2|C_{\rm eff}(\omega)|}+\gamma_m|C_{\rm eff}(\omega)|\nonumber\\
&& +\frac{|h(\omega)|^2}{g_0^2|\chi_m(\omega)|^2}+S_{\rm ff}(\omega) + \gamma_m(2\bar n_{th}+1).
\end{eqnarray}
Here, the first two terms are, respectively, the power spectra of the optical SN and radiation pressure backaction noise while the second two terms corresponds, respectively, to the LPN and the power spectrum of the input signal. The last term is the thermal noise power spectra associated to the MO with nonvanishing correlation function  $\left\langle {\delta \hat P_{in}(\omega ) \delta \hat P_{in}( - \omega ')} \right\rangle  = ({{\bar n}_{th}} + \frac{1}{2})\delta (\omega  - \omega ') $ where in the \textit{classical} limit of $ k_{\rm B} T \gg \hbar \omega_m $, $ \bar n_{th}+1/2 \simeq k_{\rm B} T /\hbar \omega_m $. Note that, usually the MOs even at high frequencies and mK-temperature still satisfy the mentioned approximation.

In the absence of the LPN and at zero temperature, there are still two quantum noises, i.e., the imprecision optical SN and the radiation pressure backaction noise, which always exist due to the quantum nature of the system. The power spectrum corresponding to these quantum noises has a
minimum value at a specified value of the effective cooperativity where the derivative of the power spectrum versus $|C_{\rm eff}(\omega)|$ is zero. This minimized power spectrum which occurs at
\begin{equation}
|C_{\rm eff}^{\rm opt}(\omega)|=\frac{1}{2\gamma_m|\chi_m(\omega)|},
\end{equation}
and has been known as the standard quantum limit (SQL) \cite{milburnBook}, quantifies the best measurement precision that can be achieved for a given mechanical susceptibility and is obtained as
\begin{equation}
S_{\rm SQL}(\omega)=\frac{1}{|\chi_m(\omega)|}.
\end{equation}
It means that the total noise power spectrum defined as $S_N(\omega)=S_{\rm det}(\omega)-S_{\rm ff}(\omega)$ cannot be smaller than $S_{\rm SQL}(\omega)$. Therefore, the minimum of $S_N(\omega)$ which occurs at $|C_{\rm eff}^{\rm opt}(\omega)|$ is obtained as
\begin{equation}
S_N^{opt}(\omega)= 2\gamma_m \frac{{k_{\rm B}} T}{\hbar \omega_m}+ \frac{1}{|\chi_m(\omega)|}+\frac{|h(\omega)|^2}{g_0^2\chi_m(\omega)|^2}.
\end{equation}
In this way, $S_N^{opt}(\omega)$ determines the minimum of the signal power spectrum which is detectable by the optomechanical sensor. In other words, unless $S_{\rm ff}(\omega)\ge S_N^{opt}(\omega)$ the signal $f(\omega)$ is not detectable.

\begin{figure}
	\includegraphics[width=8.5cm]{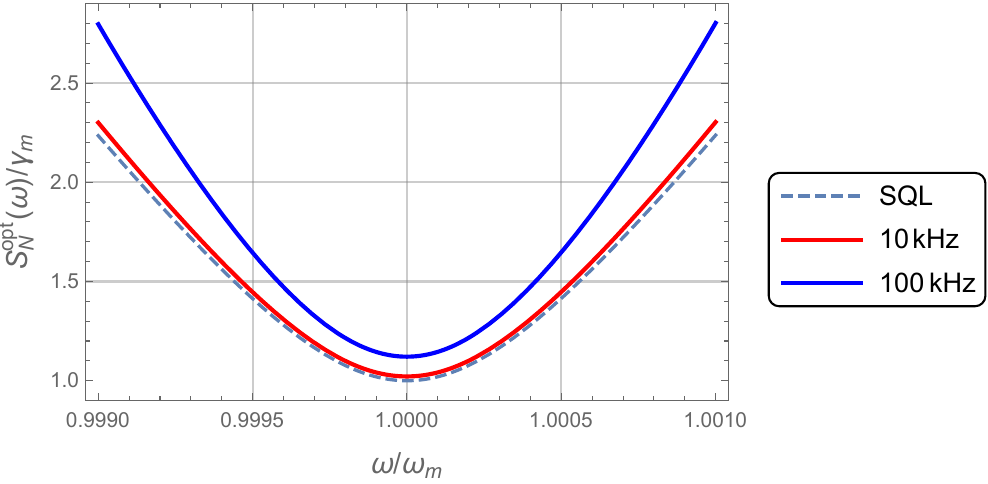}
	\caption{(Color online) The optimized total noise power spectrum $S_N^{opt}(\omega)/\gamma_m$ at temperature $ T\simeq 0.1 \mu \rm K $, versus the normalized frequency $\omega/\omega_m$ for two values of the laser line width 10kHz (red solid curve) and 100kHz (blue solid curve). The dashed curve shows the SQL.}
	\label{fig2}
\end{figure}

In Fig.~(\ref{fig2}), the SQL power spectrum (dotted curve) as well as the optimized total noise power spectrum $S_N^{opt}(\omega)/\gamma_m$ have been plotted versus the normalized frequency $\omega/\omega_m$ for two values of the laser linewidths $\Gamma_L=10 \rm kHz$ (red curve) and $\Gamma_L=100 \rm kHz$ (blue curve) under the on-resonance condition of $\Delta=0$ based on the experimental data \cite{ritterBECexp} where the cavity has a length of $L=178 \rm \mu m$ with a damping rate of  $\kappa=2\pi\times1.3 \rm MHz$ which is pumped by a laser with wavelength $\lambda=780 \rm nm$. The movable end mirror with mass $m=10 \rm ng$ and damping rate $\gamma_m=2\pi\times100 \rm Hz$ oscillates with frequency $\omega_m=2\pi\times100 \rm kHz$. Furthermore, it has been assumed that the central frequency and the bandwidth of the LPN are respectively, $\omega_N=2\pi\times140 \rm kHz$ and and $\tilde \gamma=\omega_N/2$.

As is seen from Fig.~(\ref{fig2}), in a standard OMS whose MO has been cooled down to a temperature of order $0.1\mu\rm K$, the optimized total noise $S_N^{opt}(\omega)$ gets near to the SQL at frequencies very near to the mechanical resonance frequency while away from the mechanical resonance frequency the total noise increases, especially for the larger values of the laser line width. It is also evident from the plot that the total noise corresponding to a laser linewidth of $\Gamma_L=10\rm kHz$ is very near to the SQL while for $\Gamma_L\ge 100\rm kHz$ the destructive nature of the LPN is manifested very clearly.

\section{Dispersive atomic-BEC as an analog optomechanical system  \label{secBEC}}
In this section we are proceeding to show that the interaction Hamiltonian of an atomic-BEC, which is trapped inside an optical cavity (see Fig.~(\ref{fig3})), with the cavity mode in the dispersive regime of atom-field interaction is analogous to an OMS with strong parametric and cross-Kerr nonlinearities.

\subsection{second quantized Hamiltonian of the atomic-BEC trapped in a quantum optical lattice}

\begin{figure}
	\includegraphics[width=7cm]{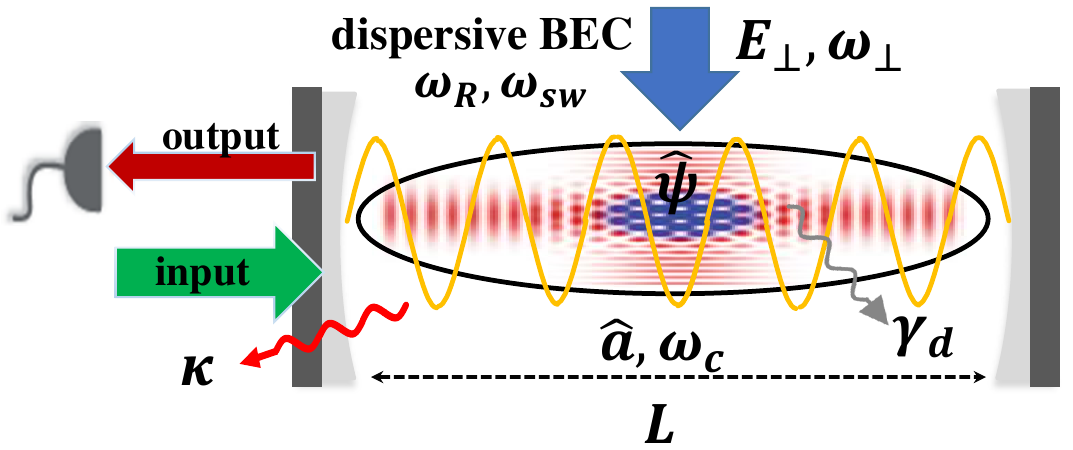}
	\caption{(Color online) Schematic of a cigar-shape dispersive atomic-BEC trapped inside the optical cavity which is analogous to an OMS. }
	\label{fig3}
\end{figure}

As shown in fig.~(\ref{fig3}), consider $ N $ ultracold two-level atoms, like Rb-atoms, with mass $m_a$ which have been confined in a cylindrically symmetric optical trap inside the optical lattice of a single-mode, high-finesse Fabry-Perot cavity with length $L$  with a transverse trapping frequency $ \omega_\bot $ and negligible longitudinal confinement frequency $ \omega_\parallel $ along the axis of the cavity (x axis). The cavity is driven at rate $E_L$  through one of its mirrors by a laser with frequency $\omega_p$. Under theses conditions, one can describe the dynamics within an effective one-dimensional model by quantizing the atomic motional degree of freedom along the $ x $ axis only.

In the dispersive regime of  atom-field interaction where the laser pump is far-detuned from the atomic resonance ($ \Delta_a=\omega_a - \omega_L \gg \Gamma_a $ where $ \Gamma_a$ is the atomic linewidth), the excited electronic state of the atoms can be adiabatically eliminated and spontaneous emission can be neglected \cite{ritsch}. In the frame rotating at the pump frequency, the total Hamiltonian of the system can be written as
\begin{eqnarray} \label{h1_BEC}
\hat  H=\hbar\Delta_c\hat a^{\dagger}\hat a+i\hbar E_L(\hat a^{\dagger}-\hat a)+\hat H_{BEC},
\end{eqnarray}
where $\hat a$ is the annihilation operator of the optical field, $\Delta_c=\omega_{c}-\omega_{L}$, and $\hat H_{BEC}$ is the Hamiltonian of the atomic BEC which is given by the following equation in the framework of second quantization formalism
\begin{eqnarray} \label{H1_BEC}
&&\!\!\!\!\!\!\!\!\!\!\!\!\!\!\! \hat  H_{BEC}=\int_{-L/2}^{L/2}dx \hat \psi^\dag(x) \Big[\frac{-\hbar^2}{2m_a} \frac{d^2}{dx^2}+V_{ext}^{\lVert}(x) \nonumber\\ 
&& + \hbar U_0 \cos^2(k_0 x) \hat a^\dag \hat a+\frac{1}{2}U_{s} \hat \psi^{\dag}(x)  \hat \psi^{\dag}(x)\Big] \hat \psi(x).
\end{eqnarray}
Here, $L$ is the cavity length, $ \hat \psi(x) $ is the annihilation operator of the atomic field, $ U_0=-g_a^2/\Delta_a $ is the optical lattice barrier height per photon which represents the atomic backaction on the field in the dispersive regime, $ g_a $ is the vacuum Rabi frequency or atom-field coupling constant, $ U_s=4\pi \hbar^2 a_s/m_a $, and $ a_s $ is the two-body \textit{s}-wave scattering length \cite{ritsch,domokos}. 

In Eq.~(\ref{H1_BEC}), the longitudinal trapping potential $ V_{ext}^{\lVert}(x)= m_a \omega_{\lVert}^2 x^2/2 $ can be approximately ignored because it is very weak in comparison with other terms. In an effective one-dimensional model this approximation is valid as long as $ \omega_\bot \gg \omega_\lVert $ so that the periodic potential of the optical lattice in the longitudinal direction is only slightly modified by $ V_{ext}^{\lVert}(x) $. Additionally, the energy arising from the atom-atom interaction has to be smaller than the energy splitting of the transverse vibrational states $ \hbar \omega_\bot $ which implies that the linear density of the condensate is smaller than $ 1/2a_s $ \cite{morsch,cigar}.

In the weakly interacting regime, i.e., $ U_0 \langle \hat a^\dag \hat a  \rangle \le 10\omega_R $ where $ \omega_R=\hbar k_0^2/2m_a $ is the recoil frequency of the condensate atoms, one can restrict the atomic field operator $\hat \psi(x) $ to the first two symmetric momentum side modes with momenta $ \pm 2\hbar k_0 $ which are excited by the atom-light interaction \cite{domokos2}.  In this manner, because of the parity conservation and considering the Bogoliubov approximation \cite{nagy}, the atomic field operator can be expanded as the following single-mode quantum field \cite{dalafi1}
\begin{eqnarray} \label{si1}
&& \hat \psi(x)= \sqrt{N/L}+ \phi_2(x) \hat d ,
\end{eqnarray}
where $ \phi_2(x)= \sqrt{2/L}\cos(2k_0x) $ and the Bogoliubov mode $ \hat d $ corresponds to the quantum fluctuations of the atomic field around the classical condensate mode $ \sqrt{N/L} $. In this expansion we have not only neglected terms proportional to $ \cos(2mk_0x) $ with $ m \ge 2 $ but also the terms with nonzero quasimomenta \cite{dalafi1}. Based on the numerical results shown in Ref.\cite{cigar}, even the transition probability to the state with $ m=2 $ is very low. On the other hand, as has been shown in Refs. \cite{meystreBEC, dalafi_qpt}, the scattering to these extra modes due to the atom-atom interaction, or interaction with  $ V_{ext}^{\lVert}(x) $, can be simulated as a damping process and may be incorporated into the noise that affects the matter field. It means that the extra modes of the BEC as well as the fluctuations of longitudinal trapping potential can effectively behave as a kind of atomic reservoir for the Bogoliubov mode of the BEC which injects noise to that mode and also lead to dissipation (for more details, see Ref \cite{dalafi4}). 
Moreover, similar to the standard OMS, one can also take into account the classical laser phase and amplitude noises as well as the trap potentials noises as have been already investigated in the atomic systems \cite{lPN1,lPN2}.

In the case that the system does not have parity symmetry, for example, when the BEC is trapped inside a ring cavity, one should also consider $ \rm sine $ modes, which in the present model, have been set aside \cite{meystreBEC2,meystreBEC3}. By substituting the atomic field operator, Eq. (\ref{si1}), into the Hamiltonian of Eq. (\ref{H1_BEC}), we arrive at the following form for the Hamiltonian of the atomic BEC subsystem
\begin{subequations}
\begin{eqnarray}   
&&  \hat H_{BEC}\! = \! \!  \hbar\delta_0 \hat a^\dag \hat a \!+\!  \hbar \omega_d \hat d^\dag \hat d \!+ \! \hbar G_0\hat a^\dag \hat a (\hat d  + \hat d^\dag) \! +  \! \hat H_{sw} \!\! + \hat H_{CK} , \label{hBEC}\quad \\
&& \hat H_{sw}= \hbar\frac{\omega_{sw}}{4} (\hat d^2 + \hat d^{\dag 2}) , \label{h_sw} \\
&& \hat H_{CK}=  \hbar g_{CK} \hat a^\dag \hat a \hat d^\dag \hat d  , \label{h_ck}
\end{eqnarray}	
\end{subequations}
where $ \delta_0=NU_0/2 $, $ \omega_d=4\omega_R + \omega_{sw} $ is the effective frequency of the Bogoliubov mode in the atomic BEC, $ G_0=\sqrt{2N}U_0/4 $ is the strength of an optomechanical-like coupling between the Bogoliubov mode of the BEC and the intracavity field, $ \omega_{sw}= 8\pi\hbar N a_s/(m_a L w^2) $ is the \textit{s}-wave scattering frequency of atom-atom interaction (with $ w $ being the waist radius of the optical mode), and  $ g_{CK}=U_0/2 $ is the cross-Kerr (CK) coefficient.

The first term in Eq.~(\ref{hBEC}) leads to a shift of the cavity detuning, i.e., $ \Delta_c \to \Delta_0=\Delta_c + NU_0/2 $ which can be interpreted as an effective Stark-shifted detuning. The second term describes the energy of the Bogoliubov mode $ \hat d $. The third term is an optomechanical-like interaction which corresponds to the linear radiation pressure coupling of the Bogoliubov mode and the optical field. In this manner, the Bogoliubov mode plays the role of another MO. The fourth term is the atom-atom interaction energy  which plays the role of the \textit{atomic parametric amplifier} and is responsible for the generation of atomic squeezed state. The last term denotes the CK nonlinear coupling between the intracavity field and the Bogoliubov mode. As has been shown\cite{dalafi4,dalafi7}, in the case of $ g_{CK}/ G_0\ll 1 $ the effect of this term is negligible and can be ignored while it has the observable effects on the entanglement if the driving laser is strong enough \cite{dalafi7}. In the following sections we will ignore the CK nonlinearity against the interatomic interaction because in the force sensor model of the hybrid OMS the driving laser is not necessary to be strong.

\subsection{the analogy between an OMS and dispersive BEC}
As was shown in the previous subsection, the Hamiltonian of the trapped dispersive BEC inside the optical cavity in the regime of $ g_{CK}/ G_0\ll 1 $ can be rewritten as
\begin{equation} \label{hBECfinalAnalogy}
\hat H_{BEC} = \hbar\delta_0 \hat a^\dag \hat a + \hat H_d,
\end{equation}
in which $\hat H_d$ has been defined as
\begin{equation}\label{Hd}
\hat H_d = \hbar \omega_d \hat d^\dag \hat d +  \hbar G_0\hat a^\dag \hat a (\hat d  + \hat d^\dag)  +   \hbar\frac{\omega_{sw}}{4} (\hat d^2 + \hat d^{\dag 2}).
\end{equation}
Let us compare the above Hamiltonian of the BEC to the standard optomechanical Hamiltonian (\ref{hamiltonianOMSNL1}). It is evident that the interaction of the radiation pressure of the optomechanical cavity with the mechanical mode of the moving mirror is exactly the same as the interaction of the optical lattice mode with the Bogoliubov phononic mode of the atomic BEC. In this manner, the atomic BEC inside the cavity is analogous to the OMS having a nonlinear opto-atomic interaction (the second term in Eq.(\ref{Hd})). Nevertheless, the atomic system has an extra nonlinear term corresponding to the atom-atom interaction (the last term in Eq.(\ref{Hd})) which provide more controllability compared to the OMS. The atom-atom interaction is originally nonlinear as is seen from the last term of Eq.(\ref{H1_BEC}). However, in the Bogoliubov approximation where the lowest BEC mode (the first term in Eq.(\ref{si1})) is considered as a c-number, the interatomic interaction takes the form of the last term in Eq.(\ref{Hd}). It should be reminded that the \textit{s}-wave scattering frequency of interatomic interaction is experimentally controllable through the frequency of the BEC transverse trap \cite{morsch}.

We will show that this nonlinearity helps us to engineer the hybrid OMSs containing a BEC with more controllability to achieve high precision quantum sensors. That is why many people are interested in hybridizing the OMS with atomic BEC which can be realized experimentally\cite{jaskulaBECSCE,recati}.

\section{Backaction evading measurement using hybrid OMS \label{secBAnoise}}
The QND measurement which is an idealized version of the backaction evasion measurement was first introduced by Braginsky \textit{et al.} \cite{braginsky1980} for the purpose of the gravitational wave detection and then it was applied for other cases such as spin measurement \cite{takahashi1999}, atomic  magnetometery \cite{shah2010}, single photon detection \cite{nogues1999}, and measurements based on optomechanical cavities \cite{heidmann1997,jacobs1994}.  A backaction evading measurement of a cantilever has been also proposed in a standard optomechanical cavity by a modulated input field which also leads to the conditional squeezing of the mechanical oscillator \cite{clerkfeedback}. This scheme which is called the \textit{two-tone} backaction evading measurement has been realized experimentally \cite{kippenbergBA2019,forcedetection2}. In the present section we are going to review a backaction evading measurement based on a hybrid OMS consisting of a trapped BEC \cite{fani2020BA}.

\begin{figure}
	\centering
	\includegraphics[width=7.5cm]{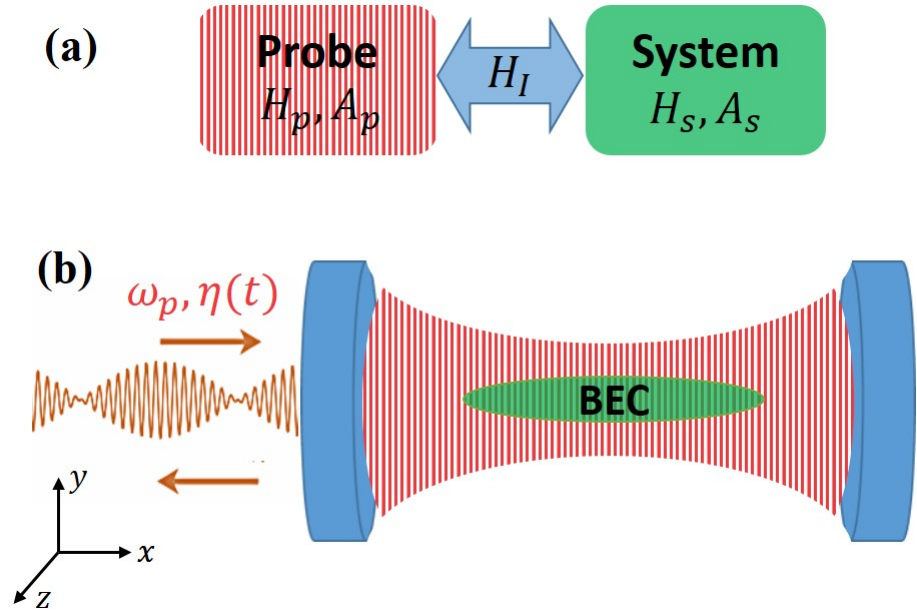}
	\caption{(Color online) (a) Schematic diagram for an indirect measurement, (b) Schematic of a one-dimensional Bose Einstein condensate trapped inside the optical lattice of an optical cavity  which is pumped coherently by a laser with a modulated amplitude. Here, the Bogoliubov mode of the BEC (the quantum system indicated by the green color) is measured indirectly by the optical mode of the cavity which plays the role of the quantum probe system (indicated by the vertical red stripes). Reprinted with permission from M. Fani and A. Dalafi, J. Opt. Soc. Am. B 37, 1263 (2020).}
	\label{fig4}
\end{figure}

Firstly, let us remind that in a QND measurement, in order to evade the backaction noise, one should measure a so-called QND variable, $ \hat A_s $, of a quantum system with Hamiltonian $\hat H_s$ which should satisfy the condition, $ [\hat A_s(t),\hat A_s(t')] = 0 $ in the ideal situation. The most well-known requirement on the continuous QND variables is that they are conserved during the free evolution, i.e., they satisfy the equation $ i \hbar \frac{\partial\hat A_s}{\partial t} + [\hat A_s , \hat H_s] = 0 $ \cite{braginsky1980,cQNCPRX}. In addition, the QND measurement is an indirect measurement so that the system variable is measured indirectly by an observable $\hat A_p$  of another quantum system, the so-called probe system, with Hamiltonian $\hat H_p$ which has been coupled to the system through an interaction Hamiltonian, $ \hat H_I $ as has been demonstrated in Fig.\ref{fig4}(a).

Secondly, in an ideal QND measurement the interaction Hamiltonian between the system and the probe, i.e., $ \hat H_I$, has to fulfill the following requirements \cite{scullybook}: (i) $\hat H_I$ should be directly dependent on $\hat A_s$, (ii) $[\hat A_s , \hat H_I] = 0$, and (iii) $ [\hat A_p , \hat H_I]\neq 0 $. The second condition ensures that the coupling of the system to the probe dose not affect the dynamics of the system variable (backaction evasion criterion). The third condition requires that the interaction Hamiltonian affects the dynamics of $\hat A_p$ so that the information corresponding to $\hat A_s$ is transformed to $\hat A_p$ which is measured directly.

In the following, we show how the two-tone backaction evading method \cite{clerkfeedback} can be generalized to an effective optomechanical system consisting of a BEC in order to do a QND measurement of the collective excitation of the BEC (the Bogoliubov mode). For this purpose, we consider an atomic BEC trapped in an effectively one-dimensional trap inside an optical cavity which is pumped by an external laser with a \textit{modulated} amplitude. Here, the collective mode of the BEC is considered as the quantum system to be measured while the radiation pressure of the cavity field is used as the probe system (Fig.~(\ref{fig4})). In this way, we introduce a QND observable for the BEC and determine the circumstances in which the QND conditions are met. Then, it is shown that the the collective excitation of the BEC can be measured at the output of the cavity with a precision under the SQL without disturbing the BEC dynamics. 

\subsection{definition of the QND variables for BEC}
Here, we consider a cavity consisting of a cigar-shaped BEC like the one investigated in section (\ref{secBEC}) described by the Hamiltonian (\ref{h1_BEC}) with the difference that here we assume the cavity is pumped coherently by a classical laser light with the time-dependent amplitude $\eta(t)$ \cite{fani2020BA}. In this way, the total Hamiltonian of the system is given by
\begin{eqnarray} \label{h1_BEC QND}
\hat  H=-\hbar\Delta\hat a^{\dagger}\hat a+i\hbar\eta(t)(\hat a-\hat a^{\dagger})+\hat H_{d},
\end{eqnarray}
where $\Delta=\omega_L-\omega_c-\delta_0$ is the effective detuning and $\hat H_d$ is given by Eq.(\ref{Hd}). The Hamiltonian $\hat H_d$ can be diagonalized by the following Bogoliubov transformations \cite{nagy2013}
\begin{subequations}
	\begin{eqnarray}
	&& \hat b = \frac{1}{2 \chi} [(\chi^2 + 1) \hat d + (\chi^2 - 1) \hat d^{\dagger} ],\\
	&& \hat b^{\dagger} = \frac{1}{2 \chi} [(\chi^2 - 1) \hat d + (\chi^2 + 1) \hat d^{\dagger} ],
	\end{eqnarray}
\end{subequations}
where $\chi = (\Omega_+ / \Omega_-)^{1/4} $ and $ \Omega_\pm = \omega_d \pm \frac{1}{2} \omega_{sw} $. In this way, the total Hamiltonian of the system in terms of the Bogoliubov mode, $b$, is given by 
\begin{eqnarray}\label{Hb}
\hat H &=&- \hbar \Delta_c \hat a^{\dagger} \hat a + \hbar \omega_m \hat b^{\dagger} \hat b + \hbar G \hat a^{\dagger} \hat a (\hat b + \hat b^{\dagger}) \nonumber \\
&+& i \hbar \eta (t) (\hat a  - \hat a^{\dagger}  ) + \hat H_{diss},
\end{eqnarray}
where $ \omega_m = \sqrt{\Omega_- \Omega_+} $ is the effective Bogoliubov frequency, and $ G = G_0 /\chi $ is the coupling strength. It should be noticed that the Hamiltonian of Eq.~(\ref{Hb}) is quite similar to a standard driven optomechanical system, but with the time-dependent pump amplitude.

In the linearized regime, the optical and atomic fields can be written as the sum of their classical mean values and quantum fluctuations as $\hat a=\alpha(t)+\delta\hat a$ and $\hat b=\beta(t)+\delta\hat b$, respectively. Now, by defining the optical quadratures $\hat X = \frac{1}{\sqrt{2}} (\delta \hat a + \delta \hat a^{\dagger})$ and $\hat Y = \frac{-i}{\sqrt{2}} (\delta \hat a - \delta \hat a^{\dagger})$, and the generalized Bogoliubov quadratures which rotate at arbitrary frequency $\Omega$,
\begin{subequations}
	\begin{eqnarray}
	\hat Q_{\Omega} = \frac{1}{\sqrt{2}} (\delta \hat b e^{i \Omega t} + \delta \hat b^{\dagger} e^{-i \Omega t}), \label{QOmega}\\
	\hat P_{\Omega} = \frac{-i}{\sqrt{2}} (\delta \hat b e^{i \Omega t} - \delta \hat b^{\dagger} e^{-i \Omega t}),\label{POmega}
	\end{eqnarray}
\end{subequations}
the equations of motion of the system quadratures are obtained as
\begin{subequations}\label{eqmX}
	\begin{eqnarray}\label{eqmXa}
	\dot{\hat X} &=&- \Delta' \hat Y - \frac{\kappa}{2} \hat X + \sqrt{\kappa} \hat X_{in} \nonumber\\ 
	&+& 2 G {\mathop{\rm Im}\nolimits} \alpha (t) (\hat Q_{\Omega} \cos \Omega t + \hat P_{\Omega} \sin \Omega t) 	,\\ \label{eqmXb}
	\dot{\hat Y}&=& \Delta' \hat X -  \frac{\kappa}{2} \hat Y + \sqrt{\kappa} \hat \hat Y_{in} \nonumber \\
	&-& 2 G {\mathop{\rm Re}\nolimits} \alpha (t) (\hat Q_{\Omega} \cos \Omega t + \hat P_{\Omega} \sin \Omega t), \\\label{eqmXc}
	\dot{\hat Q}_{\Omega} &=& (\omega_m - \Omega)\hat P_{\Omega} -  \frac{\gamma}{2} \hat Q_{\Omega} + \sqrt{\gamma} \hat Q_{in} \nonumber \\
	&+& 2 G \sin \Omega t ( {\mathop{\rm Re}\nolimits} \alpha \hat X + {\mathop{\rm Im}\nolimits} \alpha \hat Y)	,\\ \label{eqmXd}
	\dot{\hat P}_{\Omega}&=& - (\omega_m - \Omega)\hat Q_{\Omega} - \frac{\gamma}{2} \hat P_{\Omega} + \sqrt{\gamma} \hat P_{in} \nonumber \\
	& -& 2 G \cos \Omega t ( {\mathop{\rm Re}\nolimits} \alpha \hat X + {\mathop{\rm Im}\nolimits} \alpha \hat Y) .
	\end{eqnarray}
\end{subequations}
where the mean-fields $\alpha(t)$ and $\beta(t)$ are time-dependent due to the time dependence of the pump amplitude and satisfy the following equations of motion
\begin{subequations}\label{mean}
	\begin{eqnarray}\label{mean1}
	&&\dot{\alpha} = (i\Delta' - \kappa/2)\alpha+ \eta (t),\\ \label{mean2}
	&&\dot{\beta} = (-i \omega_m - \gamma/2) \beta - i G |\alpha|^2,
	\end{eqnarray}
\end{subequations}
in which $ \Delta' = \Delta - G (\beta + \beta^*)$ is the new effective detuning.

The equations of motion (\ref{eqmXa})-(\ref{eqmXd}) are quite general for any time-dependence of the pump amplitude and any $ \Omega$, but since we are interested in the QND measurement on the BEC, we have to choose the parameters $ \Omega $ and $ \eta (t) $, so that the QND conditions are satisfied. Before that, let us remind that the effective Hamiltonian from which the linearized equations (\ref{eqmXa})-(\ref{eqmXd}) are obtained can be written as follows
\begin{subequations}
	\begin{eqnarray}
	&& \hat H_{eff}= \hat H_s +\hat H_p + \hat H_ I +\hat H_{diss} ;\label{Heff}\\ 
	&&\hat H_s =\frac{1}{2} \hbar \omega_m (\hat Q_{\Omega}^2 + \hat P_{\Omega}^2),\label{Hs} \\ 
	&&\hat H_p = -\frac{1}{2} \hbar \Delta'_c (\hat X^2 + \hat Y^2),\\ \label{Hp}
	&&\hat H_I = 2 \hbar G (\hat Q_{\Omega} \cos\Omega t +  \hat P_{\Omega} \sin \Omega t) (\hat X {\mathop{\rm Re}\nolimits} [\alpha] + \hat Y {\mathop{\rm Im}\nolimits} [\alpha] ). \label{HI}\quad\quad
	\end{eqnarray}
\end{subequations}
Here, the Hamiltonian of the system to be measured, i.e., $\hat H_s$, is that of the Bogoliubov mode of the BEC, $\hat H_p$ is the Hamiltonian of the optical mode of the cavity which plays the role of the probe system and $\hat H_I$ is the interaction Hamiltonian between them. In addition, it is obvious that the Heisenberg equations for $\hat Q_{\Omega}(t)$ and $\hat P_{\Omega}(t)$ which lead to the equations (\ref{eqmXc}) and (\ref{eqmXd}) should be considered as $\dot {\hat A}=\frac{1}{i\hbar}[\hat A,\hat H_{eff}]+\frac{\partial}{\partial t}\hat A$ for $\hat A=\hat Q_{\Omega}, \hat P_{\Omega}$ because of their explicit time dependence.

Now, based on the definitions of (\ref{QOmega}) and (\ref{POmega}) together with the system Hamiltonian of (\ref{Hs}), the only way to have the continuous QND variables, i.e., to fulfill the condition, $ i \hbar \frac{\partial \hat A_s}{\partial t} + [\hat A_s , \hat H_s] = 0 $, is that $ \Omega = \omega_m $. In this way, the QND variables of the BEC can be defined as $ \hat Q \equiv \hat Q_{\omega_m} $ and $ \hat P \equiv \hat P_{\omega_m} $. In fact, these quadratures rotate at the effective Bogoliubov frequency $ \omega_m $. Thus in the phase diagram description, the error box is stationary with respect to theses quadratures, so measuring one of them does not lead to any back action due to the free evolution \cite{wallsBook}. From now on, we choose $ \hat Q $ as $ \hat A_s $, i.e., the QND operator of the system which is going to be measured. By this choice, the first condition for $\hat  H_I $, i.e., the condition (i), is obviously fulfilled by the interaction Hamiltonian of (\ref{HI}) because it depends on $\hat Q$. Furthermore, without loss of generality, we can assume $ {\mathop{\rm Im}\nolimits} [\alpha] = 0 $, which is not very restricting and can be achieved by adjusting the pump phase. Therefore, if we choose $ \hat Y $ as the probe operator $ \hat A_P$, then the third QND condition, i.e., the condition (iii), is also satisfied by the interaction Hamiltonian (\ref{HI}).

Nevertheless, by the above-mentioned choices the second QND condition, i.e., the condition (ii), is not still satisfied because $ [\hat Q , \hat H_I] = 2 i\hbar G  \hat X \alpha(t) \sin \omega_m t\neq 0 $. However, if the Fourier transform of $ \alpha (t) \sin \omega_m t $ only contains terms with frequencies greater than or equal to $2\omega_m$, and also if the spectral density of the Bogoliubov mode is so narrow that it does not respond to the frequencies much larger than $ \omega_m $, i.e., if $\gamma\ll\omega_{m}$, then the effect of the above mentioned commutator in the system dynamics will be negligible. Although in this case the measurement is not an ideal QND one, it can be considered as a backaction evading measurement which can surpass the SQL as will be shown in the next sections. One of the simplest forms of $ \alpha(t) $ which satisfies this condition is
\begin{equation}\label{alpha}
\alpha(t) = \alpha_{max} \cos \omega_m t,
\end{equation}
where we will consider it as the optical mean field and will also specify the explicit form of the time modulation of the pump amplitude which leads to this mean-field amplitude in the cavity (see equation. (\ref{etat})). By substituting Eq.~(\ref{alpha}) into the set of Eqs.~(\ref{eqmXa})-(\ref{eqmXd}) the Heisenberg-Langevin equations take the following form
\begin{subequations}\label{eqmfinal}
	\begin{eqnarray}
	\dot{\hat X}&=&- \Delta'_c \hat Y - \frac{\kappa}{2} \hat X + \sqrt{\kappa} \hat X_{in},\label{eqmfinala} \\
	\dot{\hat Y}&=& \Delta'_c \hat X -  \frac{\kappa}{2} \hat Y - G \alpha_{max} (1 +\cos (2 \omega_m t) ) \hat Q \nonumber \\
	&-& G \alpha_{max} \sin (2\omega_m t)  \hat P  + \sqrt{\kappa} \hat Y_{in},\\ \label{eqmfinalb}
	\dot{\hat Q}&=&  - \frac{\gamma}{2} \hat Q +  G \alpha_{max} \sin (2 \omega_m t) \hat X  + \sqrt{\gamma} \hat Q_{in},\label{eqmfinalc}\\
	\dot{\hat P}&=& - \frac{\gamma}{2} \hat P -  G  \alpha_{max} (1+\cos (2 \omega_m t)) \hat X +  \sqrt{\gamma} \hat P_{in}.\label{eqmfinald}
	\end{eqnarray}
\end{subequations}
As is seen from the set of equations of (\ref{eqmfinala})-(\ref{eqmfinald}), for $ \Delta'_c=0 $ the backaction noise of measurement which is injected to the quadrature $\hat Q$ through the second term in the right-hand side of Eq.~(\ref{eqmfinalc}) is minimized because in this case, the extra noises corresponding to $\hat Y$ are no longer transferred to $\hat Q$ due to decoupling of $\hat X$ from $\hat Y$.

Using the mean-field Eq.~(\ref{mean1}) and considering the above-mentioned conditions, the pump amplitude $ \eta (t) $ corresponding to the intra-cavity field amplitude (\ref{alpha}) can be specified as
\begin{equation}\label{etat}
\eta(t)= \eta_{max} \cos (\omega_m t + \phi),
\end{equation}   
where the amplitude and the phase are, respectively, given by $ \eta_{max} = \alpha _{max}\sqrt {\frac{{{\kappa ^2}}}{4} + \omega _m^2} $ and $\phi  = \arctan(2\omega_{m}/\kappa)$. The expression (\ref{etat}) shows that the pump amplitude should be modulated at the frequency of the Bogoliubov mode which means the cavity should be driven by two lasers tuned at both of the first sidebands of the cavity, i.e., $ \omega_p \pm \omega_m $, with the same amplitude $ \eta_{max}/2 $ and with the phase difference $ \phi $, where $\omega_p$ is determined by the resonance condition $ \Delta'_c=0 $. In addition, putting $\alpha(t)$ of Eq.~(\ref{alpha}) and the Fourier series of $\beta (t)$, i. e., $\beta(t) = \sum_{n} \beta_n e^{i n \omega_m t}$ in Eq.~(\ref{mean2}), we find that the only nonzero Fourier components of $\beta(t)$ are
\begin{eqnarray}
&&\beta_0 = -\frac{i G \alpha_{max}^2}{2(i \omega_m + \gamma/2)} \approx - G \alpha_{max}^2 / (2 \omega_m)  ,\\
&&\beta_2 = -\frac{i G \alpha_{max}^2}{4(3 i \omega_m + \gamma/2)} \approx - G \alpha_{max}^2 / (12 \omega_m) ,\\
&&\beta_{-2} = -\frac{i G \alpha_{max}^2}{4(-i \omega_m + \gamma/2)} \approx  G \alpha_{max}^2 / (4 \omega_m ),
\end{eqnarray}
where the approximated expressions are valid for $ \gamma \ll \omega_m $, which is compatible with the experimental data. In order to calculate the effects of the back action and the added noise due to the coupling of the BEC to the cavity field, in the next section, we calculate the spectra of the Bogoliubov mode. 

\subsection{added noise in the QND measurement of the BEC}
In order to obtain the added noise in the QND measurement of the $\hat Q$ operator of the BEC, the stationary power spectrum of the $\hat Q$ which is defined as \cite{malz2016}
\begin{subequations}
	\begin{eqnarray}
	&& S_Q(\omega) = \lim_{T \rightarrow \infty} \int_{-T/2}^{T/2} dt S_Q(\omega,t),\label{SsQA} \\
	&& S_Q(\omega,t) = \frac{1}{2} \int_{-\infty}^{\infty} d\tau e^{i \omega \tau} \left\langle \hat Q(t) \hat Q(t+\tau) +  \hat Q(t+\tau) \hat Q(t) \right\rangle ,\label{SsQb} \nonumber\\
	\end{eqnarray}
\end{subequations}
should be calculated. For this purpose, the QLEs (\ref{eqmfinala})-(\ref{eqmfinald}) should be solved in the frequency space. Since we are interested in the linear response of the Bogoliubov mode, the BEC quadratures $\hat Q(\omega)$ and $\hat P(\omega)$ should be calculated to the first order of $G\alpha_{max}$.For this purpose, it is enough to obtain $ \hat X(\omega) $ to the zeroth order of $ G \alpha_{max} $ as \cite{fani2020BA}
\begin{equation} \label{X}
\hat X(\omega) = \chi_c (\omega) \sqrt{\kappa} \hat X_{in} (\omega),
\end{equation}
where $ \chi_c (\omega) = (\frac{\kappa}{2} - i \omega)^{-1} $ is the optical susceptibility. Based on the definitions (\ref{SsQA})-(\ref{SsQb}), the stationary power spectrum of the quadrature $Q$ is obtained as
\begin{equation}\label{SQ}
S_Q (\omega) = \gamma |\chi_m (\omega)|^2 \left[  \frac{1}{2} +  
\bar{n}^{th}_b +  n_{bad} (\omega) \right]  ,
\end{equation}
where $ \chi_m(\omega) = (\frac{\gamma}{2} - i \omega)^{-1} $ is the Bogoliubov mode susceptibility and $n_{bad}(\omega) $ is defined as
\begin{equation}\label{nbadomega}
n_{bad} (\omega) =  \frac{\kappa}{8\gamma} (G \alpha_{max})^2 [|\chi_c (\omega + 2\omega_m)|^2 + |\chi_c (\omega - 2\omega_m)|^2].
\end{equation}
This expression shows the number of quanta in the frequency domain which is added to the spectrum of $ \hat Q $ due to the interaction with the cavity field. Indeed, $ n_{bad}$ corresponds to the backaction noise injected to the $ \hat Q$ spectrum arising from the coupling of the Bogoliubov mode of the BEC to the probe system. The subscript $bad$ refers to the bad-cavity limit where $ \kappa \gg\omega_m $. Notice that in the limit of $ \kappa \ll\omega_m $, i.e., in the good-cavity limit, $n_{bad}$ at resonance is approximated as $ n_{bad} (0) \approx \frac{\kappa (G \alpha_{max})^2}{16 \gamma \omega_m ^2},$ which goes to zero. It shows that the interaction-induced noise can be neglected in this limit and therefore we can perform a backaction evading measurement on $\hat Q$. 

On the other hand, the spectrum of the conjugate operator $ \hat P $ is given by \cite{fani2020BA} 
\begin{equation}
S_P (\omega) = \gamma |\chi_m (\omega)|^2 \left[  \frac{1}{2} +  
\bar{n}^{th}_b +  n_{bad} (\omega) +  n_{BA} (\omega) \right]   ,
\end{equation}
where 
\begin{equation}\label{nBAw}
n_{BA} (\omega) = \frac{\kappa}{2\gamma} (G \alpha_{max})^2 |\chi_c (\omega)|^2 ,
\end{equation}
is the backaction noise added to the observable $ \hat P $. In the good cavity limit, $n_{BA}$ at resonance is approximated as $ n_{BA} (0) \approx \frac{2}{ \kappa \gamma} (G \alpha_{max})^2 $ which is much greater than $n_{bad}(0)$ as is expected from the uncertainty relation.  

As has been mentioned before, the probe operator is the phase quadrature of the cavity field, i.e., $\hat Y$. It means that to measure the QND variable, i.e., $\hat  Q $, one has to do a homodyne measurement on the phase quadrature of the output field. Thus in order to show the possibility of backaction evading measurement of $ \hat Q $ and obtain the necessary conditions, in the following we calculate the $ S_{Y_{out}}(\omega) $ and show that the important advantage of the hybrid system consisting of a BEC in comparison to the bare optomechanical systems is that the effective frequency $ \omega_m $ of the Bogoliubov mode can be controlled by $ \omega_{sw}$ such that $ n_{bad} $ can be decreased by increasing $ \omega_{sw} $.

The phase quadrature of the output field is given by $ \hat Y_{out} = \hat  Y_{in} - \sqrt{\kappa} \hat Y $ \cite{clerkquantumnosie1} which can be calculated by solving the QLEs (\ref{eqmfinala})-(\ref{eqmfinald}) in the Fourier space. Consequently, the output spectrum of the phase quadrature of the cavity can be written as follows \cite{fani2020BA}
\begin{equation}\label{syout}
S_{Y_{out}} (\omega) = |{\cal G}(\omega)|^2 A(\omega) \Bigg[\frac{1}{2} +  \bar{n}^{th}_b +  n_{add}(\omega)\Bigg],
\end{equation}
where, we have defined
\begin{eqnarray}
&&{\cal G}(\omega) =  \sqrt{\kappa} G \alpha_{max} \chi_c(\omega),\\ \label{Aomega}
&&A(\omega) = \gamma |\chi_m (\omega)|^2 \nonumber \\
&& \,\,\,\,\,\,\,\,\,\,\,\,\,\,\,\,\,+ \frac{\gamma}{2} [|\chi_m(\omega + 2 \omega_m)|^2 + |\chi_m(\omega - 2 \omega_m)|^2]  .
\end{eqnarray}
The function, $ {\cal G}(\omega) $ is called the gain coefficient \cite{clerkfeedback} because the output spectrum can be written as $ S_{Y_{out}} (\omega) = |{\cal G} (\omega)|^2 \{S_Q(\omega) + S_{1}(\omega)\} $ where
\begin{equation}
S_{1}(\omega) = A(\omega) \left[ \frac{1}{2} + \bar{n}_b^{th} + n_{add} \right]  - \gamma |\chi_m(\omega)|^2 \left[ \frac{1}{2} + \bar{n}_b^{th} + n_{bad} \right].\nonumber
\end{equation} 
As is seen, the spectrum of the phase quadrature of the output field relates to the $ \hat Q $ spectrum by the coefficient ${\cal G}(\omega)$. Furthermore, the added noise in Eq.(\ref{syout}) is given by 
\begin{eqnarray}\label{naddomega}
n_{add}(\omega) &=& \frac{1}{2 |{\cal G}(\omega)|^2 A (\omega)} + n_{bad}(\omega) \nonumber \\
&+& \frac{\gamma n_{BA}(\omega)}{4 A(\omega)} [|\chi_m(\omega + 2 \omega_m)|^2 + |\chi_m(\omega - 2 \omega_m)|^2] \nonumber\\ 
&+& \frac{\kappa}{2 \alpha^2_{max} A(\omega)} \left\lbrace \bar{\beta}_0^2 |\chi_c(\omega)|^2 \right. \nonumber \\
&+& \left.  \bar{\beta}_2 \bar{\beta}_{-2} [|\chi_c(\omega + 2 \omega_m)|^2 + |\chi_c(\omega - 2 \omega_m)|^2] \right\rbrace ,
\end{eqnarray}
which represents  the increase in the number of the optical field quanta due to the measurement. The SQL is determined by $ n_{add} (\omega) = 1/2 $, so that if $ n_{add} (\omega) <1/2 $ it is said that the SQL has been beaten \cite{caves1980RMPHYS,clerk2004} and the measurement is an ultra-precision measurement. To find the conditions for beating the SQL we write the on-resonance added noise in the regime of $\gamma \ll \omega_m , \kappa $ as 

\begin{equation}\label{Nadd0}
n_{add}(0) \approx \frac{1}{16 n_{BA}(0)} +\frac{1}{8} \Big(\frac{\kappa^2}{4\omega_m^2 + \kappa^2/4}\Big) n_{BA}(0).
\end{equation}

Now, we are going to find the optimum value of $\eta_{max}$ which minimizes the added noise. Based on Eq.(\ref{nBAw}) $n_{BA}$ in Eq.(\ref{Nadd0}) depends on $\alpha_{max}$ which itself depends on $\eta_{max}$. So, we can find the optimum value for the pump amplitude which minimizes the added noise as
\begin{equation}\label{etaopt}
\eta_{max}^{opt} = \left[  \frac{\gamma (\omega_m^2 + \kappa^2/4)\sqrt{4\omega_m^2 + \kappa^2/4}}{2 \sqrt{2} G^2} \right]^{1/2}.
\end{equation}
The minimum value of the added noise at resonance which occurs at $\eta_{max}=\eta_{max}^{opt}$ is given by
\begin{equation}\label{n-add-min}
n_{add}^{min}(0) \approx \frac{\sqrt{2}}{4} \frac{ \kappa}{\sqrt{\kappa^2 + 16 \omega_m^2}}.
\end{equation}

As is seen from Eq.(\ref{n-add-min}), the minimum value of the on-resonance added noise is always smaller than $1/2$ for the optimized value of the pump amplitude, i.e., the SQL is beaten anyway. Nevertheless, $n_{add}^{min}(0)$ can be decreased much below the SQL by increasing the effective frequency of the Bogoliubov mode. Since $\omega_{m}=\sqrt{(4\omega_{R}+\frac{1}{2}\omega_{sw})(4\omega_{R}+\frac{3}{2}\omega_{sw})}$, one can decrease $n_{add}^{min}(0)$ by increasing $\omega_{sw}$ which itself can be controlled by the transverse frequency of the optical trap \cite{morsch} or alternatively, via a Feshbach resonance by the application of an appropriate magnetic field \cite{marte2002,donley2001}. The interesting point is that even in the bad cavity limit the added noise can be less than the SQL. However, the larger the ratio of $ \omega_m / \kappa $, the smaller the added noise and therefore the more precise the measurement is.

\begin{figure}
	\centering
	\includegraphics[width=7cm]{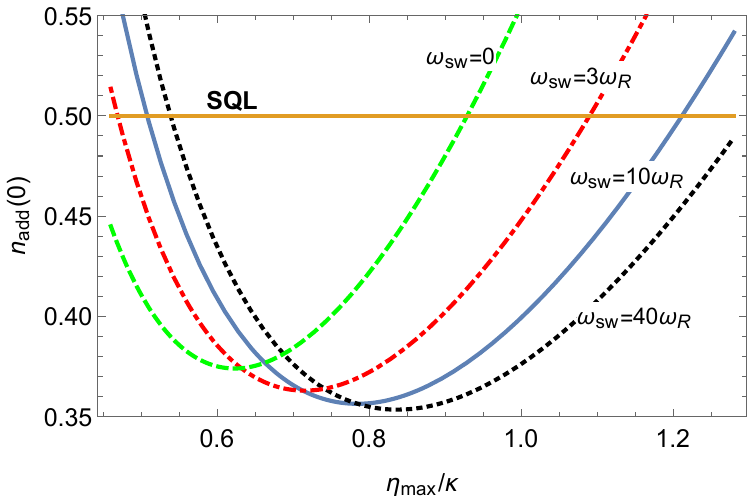}
	\caption{(Color online) The on-resonance added noise versus $ \eta_{max}/\kappa$ for four different values of $\omega_{sw} $. Reprinted with permission from M. Fani and A. Dalafi, J. Opt. Soc. Am. B 37, 1263 (2020)}
	\label{fig5}	
\end{figure}

To see more details, in Fig.~(\ref{fig5}) the on-resonance added noise $ n_{add}(0) $ has been plotted versus $\eta_{max}/\kappa $ for four different values of the $ \omega_{sw} $. The results has been obtained based on the experimentally feasible data \cite{ritterBECexp}. For this purpose, we consider a BEC of $^{87}$Rb consisting of $ N= 5 \times 10^4 $ atoms prepared in the $5\text{S}_{1/2}$ ground state. The atoms have been trapped in a cavity of length $L=178\mu$m with a bare frequency $ \omega_c =  2.41494 \times 10^{15} \text{Hz} $ corresponding to a wavelength of $ \lambda = 780 \text{nm} $ which couples to the atomic D$_2$ transition corresponding to the atomic transition frequency $ \omega_a = 2.41419 \times 10^{15} \text{Hz} $ and coupling strength $ g_a = 2 \pi \times 14.1 \text{MHz}$. The recoil frequency of the atoms in the optical lattice of the cavity is  $ \omega_R = \frac{\hbar k_c^2}{2 m_a} = 2\pi \times 3.77 \text{kHz}$. Due to stability considerations, here we only consider repulsive BEC with positive s-wave scattering frequency. In addition, the cavity decay and Bogoliubov damping rates are $ \kappa = 2\pi \times 13 \text{MHz} $ and  $ \gamma = 0.001 \kappa $, respectively.  

As is seen from Fig.~(\ref{fig5}) and as is expected from Eq.~(\ref{etaopt}), for each value of $ \omega_{sw}$ there is an optimum value for the pump amplitude which minimizes the added noise. By increasing the \textit{s}-wave scattering frequency, the effective frequency of the Bogoliubov mode increases which leads to the increase of the optimum value of $\eta_{max}$ based on Eq.~(\ref{etaopt}) and the decrease of the minimum of the added noise $n_{add}(0)$. In this way, for lager values of $\omega_{sw}$ the minimum of the on-resonance added noise gets lower.

\begin{figure}
	\centering
	\includegraphics[width=7cm]{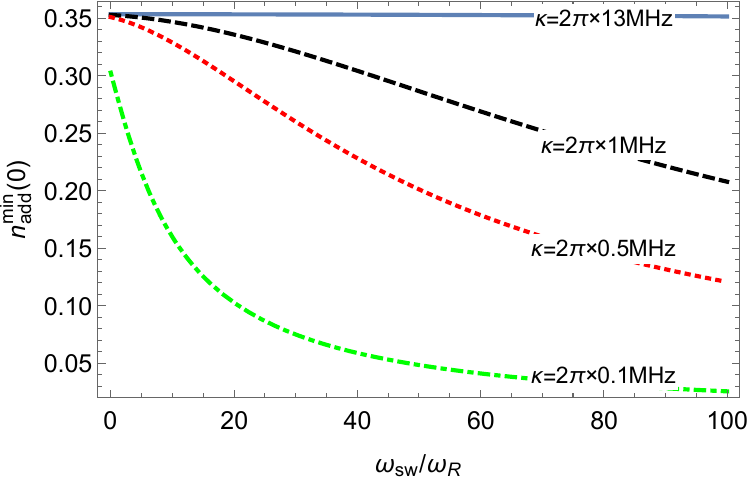}
	\caption{(Color online) The minimum value of the on-resonance added noise versus the \textit{s}-wave scattering frequency for three different values of cavity decay rates. Reprinted with permission from M. Fani and A. Dalafi, J. Opt. Soc. Am. B 37, 1263 (2020)}
	\label{fig6}
\end{figure}

On the other hand, in order to show how much the added noise of measurement in the present hybrid system can be decreased below the SQL, in Fig.~(\ref{fig6}) we have plotted $n_{add}^{min}(0) $, i.e., Eq.~(\ref{n-add-min}), versus the \textit{s}-wave scattering frequency for four different values of the cavity decay rates. It is obviously seen that the minimum of the added noise decreases with the increase of $\omega_{sw}$ for any value of $\kappa$. Nevertheless, the important point is that for the lower cavity decay rates the decrease of $n_{add}^{min}(0) $ is more severe versus $\omega_{sw}$. For example, if the cavity decay rate is of the order of $0.1 \text{MHz}$ the minimum of the added noise can be decreased as low as $0.01$ for sufficiently large values of the \textit{s}-wave scattering frequency. It is because of the fact that the increase of $\omega_{sw}$ makes the effective frequency of the Bogoliubov mode increase and consequently causes the system to go to the good cavity regime.

It should be emphasized that it is one of the most important advantages of the hybrid optomechanical systems consisting of BEC in comparison to the bare ones, that one can control the effective frequency of the atomic mode while the mechanical oscillators in the bare optomechanical systems have fixed natural frequencies which cannot be changed after fabrication. Therefore, the controllability of such hybrid systems gives us the possibility to change the system regime from the bad cavity limit to the good cavity limit through manipulation of the \textit{s}-wave scattering frequency.

Furthermore, as will be discussed in Sec.~\ref{secParametricsensing} the signal-to-nose ratio (SNR) of such hybrid system is proportional to $ 1/\sqrt{(\bar n_{b}^{th}+n_{add}+1/2)} $. Therefore, in order to increase the SNR, one needs to decrease both the temperature of the system and the added noise of measurement. At very low temperatures where $\bar n_{b}^{th}\approx 0$ the only way to increase the SNR is the reduction of the added noise; especially for $n_{add}<1/2$ where SQL is surpassed the SNR is maximized. Since the effective temperature of the Bogoliubov mode of the BEC is generally much lower than that of the moving mirror of a bare optomechanical system, the present hybrid system can have a larger SNR in comparison to the bare one.

It should also be noticed that there is a one-to-one correspondence between the amount of splitting between the normal modes of the transmitted field of the cavity and the \textit{s}-wave scattering frequency of the atomic collisions \cite{dalafi5}. In fact, by measuring the frequency splitting of the two peaks of the phase noise power spectrum which is experimentally feasible by the homodyne measurement of the light reflected by the cavity, one can estimate the value of the \textit{s}-wave scattering frequency of the atoms. In this way, the effective frequency of the Bogoliubov mode can be measured in terms of the transverse frequency of the optical trap of the BEC which can be calibrated based on the amount of splitting between the normal modes of the transmitted field of the cavity. So, by knowing the effective frequency of the Bogoliubov mode, one can obtain the optimum value of the pump amplitude as well as the minimum value of the on-resonance added noise necessary for a back action evading measurement through Eqs.~(\ref{etaopt}) and (\ref{n-add-min}), respectively.

\section{Coherent quantum noise cancellation (CQNC): destructive quantum noise interference   \label{secCQNC} }
In the previous section, we showed that using the QND method, one can evade the backaction noise and surpass the SQL in a quantim measurement based on a hybrid OMS. In this section, we are going to introduce a quantum noise interference method to cancel the backaction in measurements using OMSs.

Among several different schemes to enhance the force sensing or quantum measurement \cite{cQNCYan2020,segalQND2020,woollyForce2020,hammerer2020force,kippenberg2020Force,taylorBAforce2020,cQNCpolzikPRD,masselBA2020,fedorov2019sensing,jingForce2020,foglianoForcesensing2019,cripeBAaudio2019,cQNCthermonoise2019Ramp,cQNCMehmood2018,schliesserForce2019,kippenbergBA2019,zhangForce2017,khaliliSQL2017,gyroscopeDavuluri2017,teufelXsensing2008,sillanpaaBA2016,nunnenkampForce2014,vitali2001Force,pontinForce2014,malzBA2016,clerkBA2016,genes2016Force,davuluri2016Forcerotation,agarwalFreeForceOMS2017,guccioneRotationForce2016}, the CQNC of backaction noise is based on quantum interference and has been introduced for the first time in Refs.\cite{cQNCPRL,cQNCPRX}.
The idea is based on introducing an \textit{anti}-noise path in the dynamics of the system via the addition of an ancillary oscillator which manifests an equal and opposite response to the light field, i.e, an effective negative mass oscillator (NMO). It is worthwhile to remind that to fulfill the CQNC condition, the interaction between the ancillary mode and the cavity mode must be the same as the interaction between the mechanics and radiation pressure in the linearized regime. 
In the context of atomic spin measurements an analogous idea for coherent backaction cancellation was proposed independently \cite{polzikBAspin2009,polzik2015Trajectrory}, and has been applied for magnetometry below the SQL \cite{polzikmagnetormetry2010}, demonstrating that Einstein-Podolski-Rosen (EPR)-like entanglement of atoms generated by a measurement enhances the sensitivity to pulsed magnetic fields.

The original proposal of CQNC \cite{cQNCPRL} focused on the use of an ancillary cavity that is red-detuned from the optomechanical cavity. A QND coupling of the electromagnetic fields within the two cavities yields the necessary anti-noise path, so that the backaction noise is coherently canceled. Ref.~\cite{maximilian} considered in more detail the all-optical realization of the CQNC proposal put forwarded in Refs.~\cite{cQNCPRL,cQNCPRX}, and found that the requirements for its experimental implementation appear to be very challenging, especially for the experimentally relevant case of low mechanical frequencies and high-quality MO such as gravitational wave detectors. Other setups, which provide effective negative masses of ancillary systems for CQNC, have been suggested based on employing BEC \cite{meystreCQNBEC2013}, or the combination of a two-tone drive technique and positive-negative mass oscillators \cite{woolley2013BA}.
Furthermore, a theoretical scheme for CQNC based on a dual cavity atom-based OMS has been proposed \cite{meystreCQNC2015}. In this scheme, a MO used for force sensing is coupled to an ultracold atomic ensemble trapped in a separate optical cavity which behaves effectively as an effective NMO. The two cavities are coupled via an optical fiber. This system is a modification of the setup suggested for hybrid cooling and electromagnetically induced transparency (EIT) \cite{meystrecoolingCQN2014} and the interaction between the optomechanical cavity and the atomic ensemble leads to the CQNC. The atomic ensemble acts as a more flexible NMO, for which the impedance-matching condition of a decay rate identical to the mechanical damping rate is easier to satisfy with respect to the full-optical implementation \cite{maximilian}.

\begin{figure}
	\begin{center}
		\includegraphics[width=8.7cm]{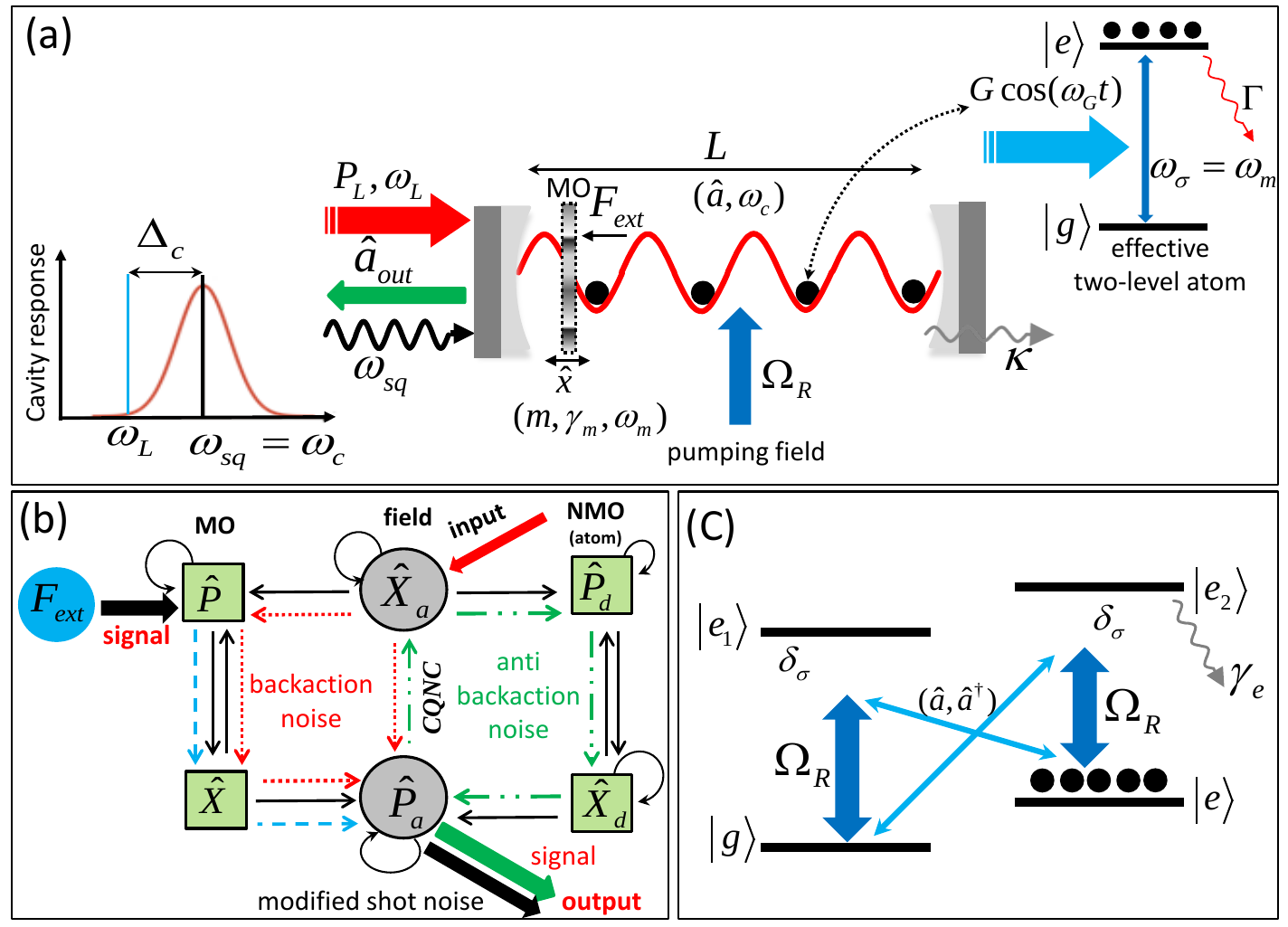}
	\end{center}
	\caption{(Color online) (a) Schematic description of the single-mode nonlinear hybrid OMS assisted with squeezed vacuum injection, which had been introduced for CQNC-based force-sensing. A mechanical oscillator with frequency $\omega_m$ is placed within a single-mode Fabry-P\'{e}rot cavity containing an atomic ensemble that can be controlled by a classical pumping field with Rabi frequency $\Omega_R$ with effective transition rate $\omega_\sigma = \omega_m$. An external force $F_{ext}$ is exerted on the mechanical oscillator acting as a sensor. The cavity is driven by a classical laser field with power $P_L$ and frequency $\omega_L$, and also a squeezed light field, resonant with the cavity mode, $\omega_{sq}=\omega_c$ , is injected into the cavity.
		(b) Flow chart representation of the backaction noise cancellation caused by the anti-noise path associated with the interaction of the cavity mode with the atomic ensemble acting as a NMO.
		(c) Atomic scheme leading to the effective Faraday interaction, with a double $\Lambda$ atomic system coupled to the intracavity mode $\hat a$ (thin blue line) and driven by a classical control field (thick blue line) of frequency $\omega_G=\omega_c$ resonant with the cavity mode. Reprinted with permission from A. Motazedifard, F. Bemani, M. H. Naderi, R. Roknizadeh, and D. Vitali, New J. Phys. 18, 073040 (2016). Licensed under a Creative Commons 3.0 License. }
	\label{fig7}
\end{figure}

In the following, we review the proposed\cite{aliNJP} hybrid system by considering only a \emph{single} optomechanical cavity and a \textit{single} cavity mode, coupled also to an atomic ensemble, which is also injected by \textit{squeezed}\cite{aliNJP} vacuum instead of thermal noise (see Fig.\ref{fig7}(a)).
The atomic ensemble is coupled to the radiation pressure and the coupling strength of the atom-field interaction is \textit{coherently} modulated. We show that the interaction between the optomechanical cavity and the atomic ensemble leads to an effective NMO that can provide CQNC conditions able to eliminate the mechanical backaction noise. In fact, destructive quantum interference between the collective atomic noise and the backaction noise of the MO realizes an `anti-noise' path, so that the backaction noise can be canceled (Fig.\ref{fig7}(b)). We show that CQNC conditions are realized when the optomechanical coupling strength and the mechanical frequency are equal to the coupling strength of the atom-field interaction and to the effective atomic transition rate, respectively\cite{aliNJP}. Furthermore, the dissipation rate of the MO needs to be matched to the decoherence rate of the atomic ensemble.
In addition, the injection of appropriately squeezed vacuum light leads to control and improve the noise reduction for force detection. It is well known that the injection of a squeezed state in the unused port of a Michelson interferometer can improve interferometric measurements \cite{caves1980,caves1981,reynaud1990,pace,mcKenzie,braginskyBook,chenOMS}, as recently demonstrated in the case of gravitational wave interferometers~\cite{lIGO1992}.
The improvement of the performance of measurement via squeezing injection has also been demonstrated in other interferometers, such as the Mach-Zehnder \cite{mach-Zehnder}, Sagnac \cite{sagnac}, and polarization interferometers \cite{polarization}. Squeezing-enhanced measurement have been realized also within optomechanical setups: an experimental demonstration of squeezed-light enhanced mechanical transduction sensitivity in microcavity optomechanics has been reported in \cite{transduction}. Moreover, by utilizing optical phase tracking and quantum smoothing techniques, improvement in the detection of optomechanical motion and force measurements with phase-squeezed state injection has also been verified experimentally \cite{phase-squeezed state}. Finally the improvement in position detection by the injection of squeezed light has been recently demonstrated also in the microwave domain ~\cite{clark}. Also, it has been recently theoretically shown that even the intracavity squeezing generated by parametric down conversion can enhance quantum-limited optomechanical position detection through de-amplification~\cite{intracavity squeezing}. More recently, by investigating the response of the MO in an optomechanical cavity driven by a squeezed vacuum it has been shown that the system can be used as a high sensitive nonclassical light sensor \cite{lotfipor}.
In this section, we show that in the presence of CQNC if the cavity mode is injected with squeezed light with an appropriate phase, backaction noise cancellation provided by CQNC is much more effective because squeezing allows to suppress the SN contribution at a much smaller input power, and one has a significant reduction of the force noise spectrum even with moderate values of squeezing and input laser power.

As shown in Fig.~\ref{fig7}(a), we consider an OMS system consisting of a single Fabry-P\'{e}rot cavity in which a MO, serving as a test mass for force sensing, is directly linearly coupled to the radiation pressure of an optical cavity field. Furthermore, the cavity contains an ensemble of effective two-level atoms that is coupled to the intracavity mode.
As is shown in Fig.~\ref{fig7}(c), the two-level atomic ensemble with coherently time-modulated coupling constant considered in this scheme is achievable by considering a double $\Lambda$-type atomic ensemble driven by the intracavity light field and by a classical control field.

We consider an standard optomechanical setup with a single cavity mode\cite{singlecavitymoderegime}  driven by a classical laser field with frequency $\omega_L$, input power $P_L$, interacting with a single\cite{singlemechanicalmoderegime} mechanical mode treated as a quantum mechanical harmonic oscillator with effective mass $m$, frequency $\omega_m$, and canonical coordinates $\hat x$ and $\hat p$, with commutation relation $[\hat x, \hat p]=i\hbar $. 
Moreover, the cavity is injected by a squeezed vacuum field with central frequency $\omega_{sq}$ which is assumed to be resonant with the cavity mode $\omega_{sq}=\omega _c$. In this manner, the total Hamiltonian describing the system is given by
\begin{eqnarray}
&& \hat H = {\hat H_{c}} + {\hat H_m} + {\hat H_{om}} + {\hat H_{d}} + {\hat H_{at}} + {\hat H_{F}},
\label{E1}
\end{eqnarray}
where $\hat H_{c}$ describes the cavity field, $\hat H_m$ represents the MO in the absence of the external force $F_{\rm ext}$, $\hat H_{om}$ denotes the optomechanical coupling, $\hat H_{d}$ accounts for the driving field, $\hat H_{at}$ contains the atomic dynamics, and $\hat H_F$  denotes the contribution of the external force. The first four terms in the Hamiltonian of Eq.~(\ref{E1}) are given by
\begin{subequations}\label{E2}
	\begin{eqnarray}
	&&{{\hat H}_c} = \hbar {\omega _c}{{\hat a}^\dag }\hat a ,\\
	&&{{\hat H}_m} = \hbar {\omega _m}{{\hat b}^\dag }\hat b = \frac{{{{\hat p}^2}}}{{2m}} + \frac{1}{2}m\omega _m^2{{\hat x}^2},\\
	&&{{\hat H}_{om}} =  \hbar {g_0}{{\hat a}^\dag }\hat a(\hat b + {{\hat b}^\dag }),\\
	&&{{\hat H}_d} = i\hbar {E_L}({{\hat a}^\dag }{e^{ - i{\omega _L}t}} - \hat a{e^{i{\omega _L}t}}),
	\end{eqnarray}
\end{subequations}
where $\hat a$ and $\hat b$ are the annihilation operators of the cavity field and the MO, respectively, whose only nonzero commutators are $[ \hat a,\hat a^{\dag}] =[ \hat b,\hat b^{\dag}]=1$. Furthermore, $\hat x = { x_{\rm zpf}}(\hat b + {{\hat b}^\dag })$ and $\hat p = i{ p_{\rm zpf}}({{\hat b}^\dag } - \hat b)$ with ${p_{\rm zpf}} = \hbar /2{ x_{\rm zpf}}$. ${g_0}$ is the single-photon optomechanical strength, while $E_L = \sqrt{P_L \kappa_{\rm in}/\hbar \omega_L}$, with $\kappa_{\rm in}$ being the coupling rate of the input port of the cavity.

For the atomic subsystem, we consider an ensemble of $N$ ultracold four-level atoms interacting non-resonantly with the intracavity field and with a classical control field with Rabi frequency $\Omega_R$ and frequency $\omega
_G$ (see Fig.~\ref{fig7}(c)). Considering the far off-resonant interaction, the two excited states $\left| {{e_1}} \right\rangle$ and $\left| {{e_2}} \right\rangle$ will be only very weakly populated. In this limit, these off-resonant excited states can be adiabatically eliminated so that the light-atom interaction reduces the coupled double - $\Lambda$ system to an effective two-level system, with upper level $\left| {{e}} \right\rangle$ and lower level  $\left| {{g}} \right\rangle$, (Fig.~\ref{fig7}(c)), driven by the so-called Faraday or quantum non-demolition interaction \cite{faradayInteraction}. Apart from the light-matter interface, the Faraday interaction has important applications also in continuous non-demolition measurement of atomic spin ensembles \cite{atomic2}, quantum-state control/tomography \cite{tomography} and magnetometry \cite{magnetometry}. In the system under consideration, we also assume that a static external magnetic field tunes the Zeeman splitting between the states $\left| {{e}} \right\rangle$ and $\left| {{g}} \right\rangle$ into resonance with the frequency $\omega_m$ of the MO.

By introducing the collective spin operators $ \hat S_+ $ and $ \hat S_z $ which obey the commutation relations $\left[ {{{\hat S }_ + },{{\hat S }_ - }} \right] = 2{{\hat S }_z}$ and $\left[ {{{\hat S }_ \mp },{{\hat S }_z}} \right] =  \pm {{\hat S }_z}$, the effective Hamiltonian of the atomic ensemble can be written as \cite{aliNJP}
\begin{eqnarray}
&& {{\hat H}_{at}} = \hbar {\omega _m }{{\hat S }_z} + \hbar G_0\cos ({\omega _{G}}t)(\hat a + {{\hat a}^\dag })({{\hat S }_ + } + {{\hat S }_ - }),
\label{E4}
\end{eqnarray}
where $G_0 = {E_0}({\Omega _R}/{\delta _\sigma})$ is the atom-field coupling, with $E_0$ and $\delta_{\sigma}$ denoting the cavity-mode Rabi frequency and the detuning of the control beam from the excited atomic states, respectively. Now, we assume that the atoms are initially pumped in the hyperfine level of higher energy, $\left| {{e}} \right\rangle$, which results in an inverted ensemble that can be approximated for large $N$ by a harmonic oscillator of negative effective mass. This fact can be seen formally using the Holstein-Primakoff mapping of angular momentum operators onto bosonic operators~\cite{primakoff1940}. In our case, we have a total spin equal to $N/2$ and one can introduce an effective atomic bosonic annihilation operator $\hat{d}$ such that $\hat S_z= N/2-\hat{d}^{\dag} \hat{d}$, $\hat S_+ = \sqrt{N}\left[1-\hat{d}^{\dag} \hat{d}/N\right]^{1/2} \hat{d}$, $\hat S_- = \sqrt{N}\hat{d}^{\dag}\left[1-\hat{d}^{\dag} \hat{d}/N\right]^{1/2} $, so that the commutation rules are preserved. As long as the ensemble remains close to its fully inverted state, we can take $\hat{d}^{\dag} \hat{d}/N \ll 1$ and approximate $\hat S_- \simeq \sqrt{N}\hat{d}^{\dag}$, $\hat S_+ \simeq \sqrt{N}\hat{d}$. Therefore, under the bosonization approximation, we can rewrite Eq.~(\ref{E4}) as
\begin{eqnarray}
&& {{\hat H}_{at}} =  - \hbar {\omega _m}\hat{d}^{\dag} \hat{d} + \hbar G\cos ({\omega _{^G}}t)(\hat a + {{\hat a}^\dag })({{\hat{d}+\hat{d}^{\dag} }}),
\label{E5}
\end{eqnarray}
which shows that the atomic ensemble can be effectively treated as a NMO, coupled with the collective coupling $G=G_0\sqrt{N}$ with the cavity mode.
Moving to the frame rotating at laser frequency $\omega_L$, where $\hat a \to \hat a e^{-i\omega_L t}$, choosing the resonance condition $\omega_G=\omega_L$, and applying the RWA in order to neglect the fast rotating terms, i.e., the terms proportional to $e^{\pm i(\omega_G+\omega_L)t}$, one gets
\begin{eqnarray}
&& {{\hat H}_{at}}=  - \hbar {\omega _m}\hat{d}^{\dag} \hat{d} + \hbar \frac{G}{2}(\hat a + {{\hat a}^\dag })({{\hat{d}+\hat{d}^{\dag} }}).
\label{E6}
\end{eqnarray}
Therefore, the total Hamiltonian of the system in the frame rotating at laser frequency $\omega_L$ is time-independent and can be written as
\begin{eqnarray}
&&\hat H = \hbar {\Delta_{c0}}{{\hat a}^\dag }\hat a + \hbar {\omega _m}{{\hat b}^\dag }\hat b - \hbar {\omega _m}\hat{d}^{\dag} \hat{d} + \hbar {g_0}{{\hat a}^\dag }\hat a(\hat b + {{\hat b}^\dag }) \nonumber \\
&&\qquad + \hbar \frac{G}{2}(\hat a + {{\hat a}^\dag })({{\hat{d}+\hat{d}^{\dag} }}) + i\hbar {E_L}({{\hat a}^\dag } - \hat a) ,
\label{E7}
\end{eqnarray}
where $\Delta_{c0}=\omega_c-\omega_L$. The quantum dynamics of the system is determined by the QLEs obtained by adding damping and noise terms \cite{vitalinoisemembrane} to the Heisenberg equations associated with the  above Hamiltonian \cite{aliNJP},
\begin{subequations}\label{E8}
	\begin{eqnarray}
	&&\!\!\!\!\!\!\!\!  \dot {\hat x} = \hat{p}/m,\\
	&&\!\!\!\!\!\!\!\!  \dot {\hat p} =  - m{\omega _m^2}\hat x -2 p_{\rm zpf} g_0{{\hat a}^\dag }\hat a - \gamma _m \hat p +\eta +\tilde{F}_{\rm ext} ,\\
	&& \!\!\!\!\!\!\!\! \dot {\hat a} = \! - i{\Delta_{c0}}\hat a \! - i g_0 \hat a \frac{\hat{x}}{ x_{\rm zpf}}  - i\frac{G}{2}({{\hat{d}+\hat{d}^{\dag} }}) \! + \! E_L \! - \! \frac{\kappa }{2}\hat a \!+ \! \sqrt \kappa  {{\hat a}^{in}} ,\\
	&&\!\!\!\!\!\!\!\!  \dot{\hat d} = i{\omega _m }\hat d - i\frac{G}{2}(\hat a + {{\hat a}^\dag }) - \frac{\Gamma }{2}\hat d + \sqrt \Gamma  \hat d^{in} ,
	\end{eqnarray}
\end{subequations}
where $\gamma_m$, $\Gamma$ and $\kappa$ are, respectively, the mechanical damping rate, the collective atomic dephasing rate, and the cavity photon decay rate. We have also considered an external classical force $\tilde{F}_{\rm ext}$ which has to be detected by the MO. The system is also affected by three noise operators: the thermal noise acting on the MO, $\eta(t) $, the optical input vacuum noise, ${{\hat a}^{in}}$, and the bosonic operator describing the optical vacuum fluctuations affecting the atomic transition, $\hat d^{in}$ \cite{gardinerBook}. These noises are uncorrelated, and their only nonvanishing correlation functions are $\langle {{\hat a}^{in}}(t) {{\hat a}^{in}}(t)^{\dag}\rangle = \langle {{\hat d}^{in}}(t) {{\hat d}^{in}}(t)^{\dag}\rangle = \delta (t - t')$ \cite{gardinerBook}.  Here, we have assumed that the external classical force has no quantum noise.

In the regime of large mechanical quality factor, $ Q_m=\omega_m/\gamma_m \gg 1 $, the Brownian mechanical thermal noise operator, $\eta(t)$, obeys the following symmetrized correlation function \cite{vitalinoisemembrane}
\begin{eqnarray}
&& \! \left\langle {\eta (t) \eta (t') + \eta (t') \eta (t) } \right\rangle /2  \simeq \!  \hbar  m \gamma_m \omega_m (2{{\bar n}_m} +\!  1)\delta (t - t') ,
\label{E16}
\end{eqnarray}
where ${{\bar n}_m} = [\exp (\hbar {\omega _m}/{k_B}T) - 1]^{ - 1}$ is the mean thermal phonon number with $T$ being the temperature of the thermal bath of the MO. 
We define the optical and atomic quadrature operators ${{\hat X}_a} = ({{\hat a}^\dag } + \hat a)/\sqrt 2$, $\hat P_a = i({{\hat a}^\dag } - \hat a)/\sqrt 2$, ${{\hat X}_d } = ({\hat d } + {\hat d }^{\dagger})/\sqrt 2$, $ \hat P_d = i({\hat d }^{\dagger}- {\hat d })/\sqrt 2$ and their corresponding noise operators $\hat X_a^{in} = (\hat a^{in,\dag}  + {{\hat a}^{in}})/\sqrt 2$, $\hat P_a^{in} = i(\hat a^{in,\dag}  - {{\hat a}^{in}})/\sqrt 2$, $\hat X_d^{in} = (\hat d^{in,\dag} + \hat d^{in})/\sqrt 2$ and $\hat P_d^{in} = i(\hat d^{in,\dag } - \hat d^{in})/\sqrt 2$. Moreover we adopt dimensionless MO position and momentum operators $\hat {X} = \hat{x}/ \sqrt{2}x_{\rm zpf}$ and $\hat {P} = \hat{p}/ \sqrt{2}p_{\rm zpf}$, so that $[\hat {X},\hat {P}]=i$.
We then consider the usual regime where the cavity field and the atoms are strongly driven and the weak coupling optomechanical limit, so that we can linearize the dynamics of the quantum fluctuations around the semiclassical steady state. After straightforward calculations, the linearized quantum Langevin equations for the quadratures' fluctuations are obtained as
\begin{subequations} 
	\begin{eqnarray}
	&& \delta \dot{\hat X }= {\omega _m}\delta \hat P , \label{fluctuationa}\\
	&& \delta {{\dot {\hat X}}_d } =  - {\omega _m }\delta {{\hat P}_d } - \frac{\Gamma }{2}\delta {{\hat X}_d } + \sqrt \Gamma  \hat X_d ^{in} , \\
	&& \delta {{\dot {\hat X}}_a} = {\Delta _c}\delta {{\hat P}_a} - \frac{\kappa }{2}\delta {{\hat X}_a} + \sqrt \kappa  \hat X_a^{in} ,  \\
	&& \delta \dot {\hat P} =\!  - {\omega _m}\delta \hat X \! - \! {\gamma _m}\delta \hat P \! - g\delta {{\hat X}_a} \! + \! \sqrt {{\gamma _m}} (\hat f \! +\! {F_{\rm ext}}) , \\
	&& \delta {{\dot {\hat P}_a}} =\! - \! {\Delta _c}\delta {{\hat X}_a} \!  - \! g\delta \hat X \! - \! G\delta {{\hat X}_d } \! - \! \frac{\kappa }{2}\delta {{\hat P}_a} + \sqrt \kappa  \hat P_a^{in}\! ,\\
	&& \delta {{\dot{ \hat P}}_d } = {\omega _m }\delta {{\hat X}_d } - G\delta {{\hat X}_a} - \frac{\Gamma }{2}\delta {{\hat P}_d } + \sqrt \Gamma  \hat P_d ^{in} , \label{fluctuationf}
	\end{eqnarray}
\end{subequations}
where the effective linearized optomechanical coupling constant is $g =  2 g_0 \alpha_s$, ${\Delta _c}=\Delta_{c0}-g_0^2 |\alpha_s|^2/\omega_m$ is the effective cavity detuning, and $\alpha_s$ is the intracavity field amplitude, solution of the nonlinear algebraic equation $(\kappa/2+i\Delta_c)\alpha_s = E_L -i G^2 \omega_m {\rm Re}\alpha_s/(\Gamma^2/4 + \omega_m^2)$, which is always possible to take as a real number by an appropriate redefinition of phases. Finally, we have rescaled the thermal and external force by defining $f(t)= \eta(t)/\sqrt {\hbar m{\omega _m}{\gamma _m}}$ and $F_{\rm ext}=\tilde{F}_{\rm ext}(t)/\sqrt {\hbar m{\omega _m}{\gamma _m}}$.
These equations are analogous to those describing the CQNC scheme proposed in Ref.\cite{cQNCPRX} and then adapted to the case when the NMO is realized by a blue detuned cavity mode \cite{maximilian}, and by an inverted atomic ensemble in Ref.\cite{meystreCQNC2015}. Compared to the latter paper, the cavity mode tunnel splitting $2J$ is replaced by the effective cavity detuning $\Delta_c$.

As suggested by the successful example of the injection of squeezed light in the LIGO detector \cite{lIGO2013} and more recently in an electro-mechanical system \cite{clark}, we now show that the force detection sensitivity of the present scheme can be further improved and can surpass the SQL when the cavity is driven by a squeezed vacuum field, with a spectrum centered at the cavity resonance frequency $\omega_{sq}=\omega_c$.

The squeezed field driving is provided by the finite bandwidth output of an optical parametric oscillator (OPO), shined on the input of our cavity system, implying that the cavity mode is subject to a non-Markovian squeezed vacuum noise\cite{jahne2009}. In the white noise limit which is satisfied whenever the bandwidths of OPO are larger than the mechanical frequency and the cavity linewidth, the noise correlation functions can be written in Markovian form\cite{jahne2009} as $\left\langle {{{\hat a}_{in}}(t){{\hat a}_{in}}(t')} \right\rangle \! =\! M\delta (t - t')$ and $\left\langle {\hat a^{in,\dag} (t){{\hat a}^{in}}(t')} \right\rangle \! =\! N\delta (t - t')$. In the case of pure squeezing, $M = (1/2)\sinh (2r) \exp({  i\phi })$ and $N = {\sinh ^2}r$, with $r$ and $\phi$ being, respectively, the strength and the phase of squeezing, so that ${\left| M \right|^2} = N(N + 1)$.

In the following, we show how the backaction noise of measurement can be eliminated using the CQNC method.
\subsection{CQNC-based force sensing}

When an external force acts on the MO of an OMS, shifts its position and changes the effective length of the cavity which leads to a variation in the phase of the optical cavity output. As a consequence, the signal associated to the force can be extracted by measuring the optical output phase quadrature, $\hat P_a^{out}$, with heterodyne or homodyne detection. The expression for the output field can be obtained from the standard input-output relation \cite{gardinerBook,wallsBook}, i.e., $\hat P_a^{out} = \sqrt \kappa  \delta {{\hat P}_a} - \hat P_a^{in}$, and solving Eqs.~(\ref{fluctuationa})-(\ref{fluctuationf}) in the frequency domain as
\begin{eqnarray}
&&\hat P_a^{out} = \sqrt \kappa  {{\chi '}_a}\left\{ { - g{\chi _m}\sqrt {{\gamma _m}} \left( {\hat f + {\tilde F_{\rm ext}}} \right)} \right.\nonumber \\
&&\qquad \quad+\sqrt \kappa  \left[ {\left(1 - \frac{1}{{{{\chi '}_a}\kappa }}\right)\hat P_a^{in} - {\Delta _c}{\chi _a}\hat X_a^{in}} \right] \nonumber \\
&&\qquad\quad - G{\chi_d }\sqrt \Gamma  \left[ {\hat P_d ^{in} - \hat X_d^{in}\left(\frac{{\Gamma /2 + i\omega }}{{{\omega _m }}}\right)} \right] \nonumber \\
&&\qquad\quad +\left. \sqrt \kappa  {\chi _a}\left({g^2}{\chi _m} + {G^2}{\chi_{d }} \right) \hat X_a^{in} \right\} ,
\label{Pout}
\end{eqnarray}
where we have defined the susceptibilities of the cavity field, the MO, and of the atomic ensemble, respectively, as
\begin{eqnarray}
&&{\chi _a}(\omega ) = \frac{1}{{\kappa /2 + i\omega }} ,\nonumber \\
&&{\chi _m}(\omega ) = \frac{{{\omega _m}}}{{\left( {\omega _m^2 - {\omega ^2}} \right) + i\omega {\gamma _m}}} ,\nonumber \\
&&{\chi_d }(\omega ) = \frac{{ - {\omega _m }}}{{\left( {\omega _m ^2 - {\omega ^2} + {\Gamma ^2}/4} \right) + i\omega \Gamma }} ,
\label{susceptibilities}
\end{eqnarray}
and the modified cavity mode susceptibility as
\begin{eqnarray}
&& \frac{1}{{{{\chi '}_a}}} = \frac{1}{{{\chi _a}}} - {\chi _a}{\Delta _c}\left( {{g^2}{\chi _m} + {G^2}{\chi_d } - {\Delta _c}} \right) .
\label{effectivesusceptibility}
\end{eqnarray}
Since the output cavity phase is the experimental signal which, after calibration, is used for estimating the external force, it can be appropriately rescaled by defining the force noise operator as\cite{aliNJP,aspelmeyerOMS}
\begin{eqnarray}
&& {\hat F_{\rm ext}^{\rm est}} \equiv \frac{{ - 1}}{{g{{\chi '}_a}{\chi _m}\sqrt {\kappa {\gamma _m}} }}\hat P_a^{out} \equiv {F_{\rm ext}} + {\hat F_{N}},
\label{Force}
\end{eqnarray}
where the added force noise is defined as
\begin{eqnarray}
&&{{\hat F}_{N}} = \hat f - \sqrt {\frac{\kappa }{{{\gamma _m}}}} \frac{1}{{g{\chi _m}}}\left[ {\left(1 - \frac{1}{{{{\chi '}_a}\kappa }}\right)\hat P_a^{in} - {\Delta _c}{\chi _a}\hat X_a^{in}} \right] \nonumber \\
&&\qquad\qquad\qquad + \frac{{G{\chi_d }}}{{g{\chi _m}}}\sqrt {\frac{{{\Gamma }}}{{{\gamma _m}}}} \left[ {\hat P_d ^{in} - \frac{{\Gamma /2 + i\omega }}{{{\omega_m }}} \hat X_d^{in}} \right] \nonumber \\
&&\qquad\qquad\qquad - \frac{{{g^2}{\chi _m} + {G^2}{\chi_d }}}{{g{\chi _m}}}\sqrt {\frac{\kappa }{{{\gamma _m}}}} {\chi _a}\hat X_a^{in}.
\label{added force}
\end{eqnarray}
Eq.~(\ref{added force}) shows that in the present scheme for force detection there are four different contributions to the force noise spectrum. The first term, $\hat { f}= \hat \eta/\sqrt {\hbar m{\omega _m}{\gamma _m}}$, corresponds to the mechanical thermal noise of the MO with nonvanishing correlation function $\left\langle {\hat { f}(\omega )\hat {f}( - \omega ')} \right\rangle  = ({{\bar n}_m} + \frac{1}{2})\delta (\omega  - \omega ') $ where in the limit of $  { k_{\rm B}}T \gg \hbar {\omega _m} $ becomes $ ( k_{\rm B}T/\hbar {\omega _m}) \delta (\omega - \omega ') $. The second term corresponds to the shot noise associated with the output optical field, which is the one eventually modified by the squeezed input field. The third term is the contribution of the atomic noise due to its interaction with the cavity mode, while the last term describes the backaction noise due to the coupling of the intracavity radiation pressure with the MO and with the atomic ensemble which are coherently added to each other, and thus, can lead to quantum interference.

\subsection{CQNC conditions}
The CQNC will lead to the perfect backaction noise cancellation at all frequencies, and thus significantly lower noise in force detection. From the last term in Eq.~(\ref{added force}), it is evident that for $g=G$ and $\chi_m=-\chi_{d}$ the contributions of the backaction noises from the mechanics and from the atomic ensemble cancel each other for all frequencies. As shown in Fig.~\ref{fig7}(b), they can be thought of as `noise' and `anti-noise' path contributions to the signal force $F_{\rm ext}$. Therefore an effective NMO, in this case realized by the inverted atomic ensemble, is necessary for realizing $\chi_m=-\chi_{d}$. Therefore, CQNC is realized whenever:
\begin{itemize}
	\item[(i)]
	the coupling constant of the optical field with the MO and with the atomic ensemble are perfectly matched, $g=G$, which is achievable by adjusting the intensity of the fields driving the cavity and the atoms;
	\item[(ii)]
	the atomic dephasing rate between the two lower atomic levels $\Gamma$ must be perfectly matched with the mechanical dissipation rate $\gamma_m$ (we have assumed the atomic Zeeman splitting perfectly matched with the MO frequency $\omega_m$ from the beginning);
	\item[(iii)]
	the MO has a high mechanical quality factor, or equivalently, $\Gamma  \ll {\omega _m}$ so that the term ${\Gamma ^2}/4$ can be neglected in the denominator of $\chi_{d} $ (see Eq. (\ref{susceptibilities})).
\end{itemize}
Mechanical damping rates of high quality factor MO are quite small, not larger than 1 kHz. As already pointed out \cite{meystreCQNC2015}, the matching of the two decay rates is easier in the case of atoms because ground state dephasing rates can also be quite small \cite{heinze,dudin}. On the contrary, matching the dissipative rates in the case when the NMO is a second cavity mode, as in the fully optical model \cite{maximilian}, is more difficult because it requires having a cavity mode with an extremely small bandwidth which can be obtained only assuming large finesse and long cavities.

Note that under CQNC conditions the effective susceptibility of Eq.~(\ref{effectivesusceptibility}) becomes ${\chi '}_a^{CQNC} = (1/{\chi _a} + {\chi _a}\Delta _{c}^2)^{ - 1}$.
It is clear that under the CQNC conditions the last term in the noise force of Eq.~(\ref{added force}) is identically zero, and we can rewrite
\begin{eqnarray}
&&{{\hat F}_{N}} = \hat f - \sqrt {\frac{\kappa }{{{\gamma _m}}}} \frac{1}{{g{\chi _m}}}\left[ {\left(1 - \frac{1}{{{{\chi '}_a}\kappa }}\right)\hat P_a^{in} - {\Delta _c}{\chi _a}\hat X_a^{in}} \right] \nonumber\\
&&\qquad\quad- \left[ {\hat P_d^{in} - \frac{{\Gamma /2 + i\omega }}{{{\omega _m }}} \hat X_d^{in}} \right] .
\label{FadeCQNC}
\end{eqnarray}
In order to quantify the sensitivity of the force measurement, we consider the spectral density of added force noise which is defined by \cite{maximilian}
\begin{eqnarray}
{S_F}(\omega )\delta (\omega - \omega ') = \frac{1}{2}\left( {\left\langle {\hat F_{N}(\omega )\hat F_{N}(-\omega ')} \right\rangle  + c.c} \right).
\label{spectrum}
\end{eqnarray}
Under perfect CQNC conditions one gets the force noise spectrum in the presence of squeezed-vacuum injection which, in the experimentally relevant case $\kappa \gg \omega$, reads \cite{aliNJP}
\begin{eqnarray}
&&{S_F}(\omega ) = \frac{{{k_B}T}}{{\hbar {\omega _m}}} + \frac{1}{2}\left( {1 + \frac{{{\omega ^2} + \gamma _m^2/4}}{{\omega _m^2}}} \right) \nonumber\\
&&\qquad \quad \quad \! + \frac{\kappa }{{{g^2}{\gamma _m}}}\!\frac{1}{{{{\left| {{\chi _m}}\! \right|}^2}}}\! \left[ \frac{1}{2}\! \left(\! \frac{1}{2}\! +\! \frac{2 \Delta_c^2}{\kappa^2 }\right)^2 \! \! \!+ \! \Sigma (M,N,\Delta _c/\kappa) \right]\! ,
\label{final spectrum}
\end{eqnarray}
where
\begin{eqnarray}
&&\Sigma (M,N,\Delta _c/\kappa ) = N{{\left(\frac{1}{2} +\!  {{\frac{{{2 \Delta_c^2}}}{\kappa^2 }}}\right)}^2} \! + 2\frac{{{\Delta _c}}}{\kappa }{\mathop{\rm Im}\nolimits} M\left(4\frac{{\Delta _c^2}}{{{\kappa ^2}}} - 1\right) \nonumber \\
&&\qquad\qquad\qquad \qquad + {\mathop{\rm Re}\nolimits} M\left[ \frac{8 \Delta _c^2}{\kappa ^2} - \left(\frac{1}{2} + \frac{2 \Delta_c^2}{\kappa^2 }\right)^2 \right],
\label{squeezingterm}
\end{eqnarray}
is the contribution of the injected squeezing to the optomechanical SN.
Equation (\ref{final spectrum}) shows that when CQNC is realized, the noise spectrum consists of three contributions: the first term denotes the thermal Brownian noise of the MO, the second term describes the atomic noise, and the last one represents the optomechanical SN modified by squeezed-vacuum injection. We recall that with the chosen units, the noise spectral density is dimensionless and in order to convert it to ${\rm N^2} \rm H{\rm z^{ - 1}}$ units we have to multiply by the scale factor $\hbar m{\omega _m}{\gamma _m}$.
This noise spectrum has to be compared with the force noise spectrum of a standard optomechanical setup formed by a single-mode cavity coupled to a MO at zero detuning ($\Delta_c=0$) and the steady-state limit $ \omega/\kappa \ll 1 $,
\begin{eqnarray}
&& {S_F^{st}}(\omega ) = \frac{{{k_B}T}}{{\hbar {\omega _m}}} + \frac{1}{2}\left[ {\frac{1}{4}\frac{\kappa }{{{g^2}{\gamma _m}}}\frac{1}{{{{\left| {{\chi _m}} \right|}^2}}} + 4{g^2}\frac{1}{{\kappa {\gamma _m}}}} \right].
\label{sopt}
\end{eqnarray}
As can be seen, the first term is the contribution of the thermal white noise of the mechanical mode which is flat in frequency, while there are still two quantum noises, i.e., the imprecision optical SN proportional to $ 1/g^2 $ and the radiation pressure backaction noise proportional to $ g^2 $, which always exist due to the quantum nature of the system (for more details see Sec.\ref{subSQL}). In order to enhance the quantum effects, one should increase the enhanced optomechanical coupling $ g $ through the increase of the input power laser which leads to decrease of imprecision noise and increase of the backaction noise. 

\begin{figure}
	\includegraphics[width=8.6cm]{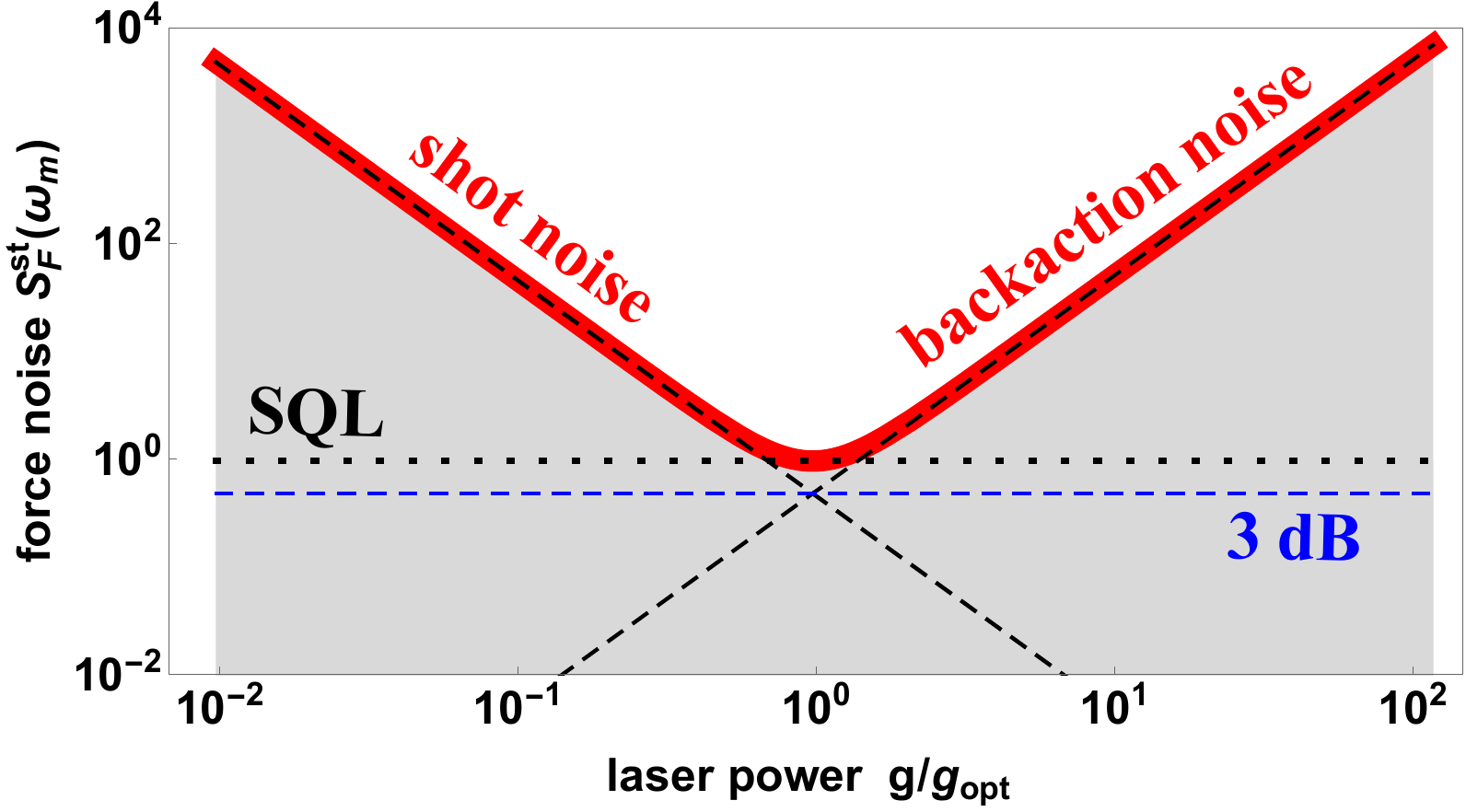}
	\caption{(Color online) Evaluated standard force noise spectrum in the absence of LPN and input signal (Eq.~(\ref{s_f_standard})) at zero-temperature, $ S_{F}^{\rm st}(g) $, versus the normalized optomechanical coupling or laser power $ g/g_{\rm opt} $. The black dotted curve is referred to the SQL while the blue dashed line is referred to 3dB-limit where the lower and higher branches in the force noise spectrum cross each other. The red solid line is the force noise spectrum at resonance mechanical frequency.  }
	\label{fig8}
\end{figure}

At zero temperature, this power spectrum has a minimum value at a specified value of the optomechanical coupling. As has been shown in Sec.\ref{subSQL}, the SQL for stationary force detection comes from the minimization of the noise spectrum at a given frequency over the driving power or over the squared optomechanical coupling $g^2(\omega) \equiv  g^2_{\rm opt}= \frac{\kappa}{4} \frac{1}{\vert \chi_m(\omega) \vert}$ where at the resonance frequency of the MO  $ \omega=\omega_m $ corresponds to the optimized cooperativity $ \mathcal{C}_{\rm opt}:=4g^2/\kappa \gamma_m = 1  $, and thus, it leads to
\begin{eqnarray}
&& {S_F^{st}}(\omega ) \ge {S_{\rm SQL}} = \frac{1}{{{\gamma _m}\left| {{\chi _m}(\omega )} \right|}},
\label{SQL}
\end{eqnarray}
which at the resonance frequency of the MO ($ \omega=\omega_m $) becomes unity. For more clarity, one can rewrite the standard optomechanical force noise spectrum of Eq.(\ref{sopt}) at zero temperature in terms of optimum optomechanical coupling $ g_{\rm opt} $ as
\begin{eqnarray} \label{s_f_standard}
&& S_F^{{\rm st}}(g)=\frac{1}{2} \Big[ (\frac{g}{g_{\rm op}})^2 +  (\frac{g_{\rm op}}{g})^2 \Big].
\end{eqnarray}
Fig.~(\ref{fig8}) shows qualitatively the force noise spectrum at the resonance frequency of the MO versus the normalized enhanced optomechanical coupling ($ g/g_{\rm opt} $). As is evident form Fig.~(\ref{fig8}) at lower powers, the imprecision shot noise dominates, while at higher powers the radiation pressure backaction noise is dominant. Note that the SQL ($ S_{\rm SQL}=1 $) is reached at the intermediate power corresponding to $ g=g_{\rm opt} $. 
The so-called 3dB-limit which corresponds to $50\%$ noise reduction below the zero-point level \cite{dalafiQOC}, occurs where the lower and higher branches of the force noise spectrum cross each other and leads to $ S_F^{\rm st}=1/2 $.

Let us return to the present CQNC scheme, in which the complete cancellation of the backaction noise term proportional to $g^2$ has the consequence that force detection is limited only by SN, and therefore, the optimal performance is achieved at very large power. In this limit, measurement is limited only by the additional SN-type term that is independent of the measurement strength ${{g^2}}$ corresponding to atomic noise (see Eq. (\ref{final spectrum})), and which is the price to pay for the realization of CQNC,
\begin{eqnarray}
&& {S_{\rm CQNC}} = \frac{1}{2}\left( {1 + \frac{{{\omega ^2} + \gamma _m^2/4}}{{\omega _m^2}}} \right),
\label{sCQNC}
\end{eqnarray}
(here we neglect thermal noise and other technical noise sources which are avoidable in principle). In the limit of sufficiently large driving powers when SN (and also thermal noise) is negligible, CQNC has the advantage of significantly increasing the bandwidth of quantum-limited detection of forces, well out of the mechanical resonance \cite{meystreCQNC2015}.
This analysis can be applied also for the present scheme employing a single cavity mode, and it is valid also in the presence of injected squeezing, which modifies and can further suppress the SN contribution. This is relevant because it implies that one can achieve the CQNC limit of Eq.~(\ref{sCQNC}), by making the SN contribution negligible, much easily, already at significantly lower driving powers. In this respect one profits from the ability of injected squeezing to achieve the minimum noise at lower power values.

Let us now see in more detail the effect of the injected squeezing by optimizing the parameters under perfect CQNC conditions. To be more specific, the injected squeezed light has to suppress as much as possible the SN contribution to the detected force spectrum, and therefore we have to minimize the function within the square brackets of Eq.~(\ref{final spectrum}), over the squeezing parameters $N$, $M$ and the detuning $\Delta_c$. Defining the normalized detuning $y=\Delta_c/\kappa$, one can rewrite this function as
\begin{eqnarray}
&& h(M,N,y) = \left(N + \frac{1}{2}\right){\left(\frac{1}{2} + 2{y^2}\right)^2} \nonumber \\
&&\qquad \qquad\qquad - |M|\left[a(y) \sin \phi + b(y) \cos\phi\right],
\end{eqnarray}
where $M=|M|e^{i\phi}$, and we have introduced the detuning-dependent functions
\begin{eqnarray}
a(y)&=&2y(1-4{y^2}) \nonumber \\
b(y)&=&\left(\frac{1}{2} + 2y^2\right)^2 -8y^2 .
\label{Fsimple}
\end{eqnarray}
$h(M,N,y)$ can be further rewritten as
\begin{eqnarray}
&& \!\!\!\!\!\!\!\!\! \!\!\!\!\!\!\!\!\! h(M,N,y) = \left(N + \frac{1}{2}\right){\left(\frac{1}{2} + 2{y^2}\right)^2} \nonumber \\
&&\qquad  -|M|\sqrt{a(y)^2+b(y)^2}\cos\left[\phi-\phi_{\rm opt}(y)\right], 
\end{eqnarray}
where $\tan \phi_{\rm opt}(y)=a(y)/b(y)$ and it is straightforward to verify that $\sqrt{a(y)^2+b(y)^2}=\left(1/2 + 2y^2\right)^2$. From this latter expression it is evident that, for a given detuning $y$, and whatever value of $N$ and $|M|$, the optimal value of the squeezing phase minimizing the shot noise contribution is just $\phi=\phi_{\rm opt}(y)$, for which one gets
\begin{eqnarray} \label{h_general}
&& h(|M|,N,y) = \left(N + \frac{1}{2}-|M|\right){\left(\frac{1}{2} + 2{y^2}\right)^2} .
\end{eqnarray}
The expression (\ref{h_general}) can be easily further minimized by observing that its minimal value is obtained by assuming pure squeezed light $|M|=\sqrt{N(N+1)}$ and also taking zero detuning $y=0$, i.e., driving the cavity mode at resonance, so that for a given value of the (pure) squeezing parameter $N$, one gets
\begin{equation}\label{h_particular}
h_{\rm min}(N) = \frac{1}{4} \left[N+1/2-\sqrt{N(N+1)}\right],
\end{equation}
which tends to zero quickly for large values of the squeezing parameter $N$, i.e., $h_{\rm min}(N \gg 1) \to 1/(32N)$. As a consequence, the SN contribution can be rewritten after optimization over the squeezing and detuning parameters as,
\begin{equation}\label{speopt}
{S_F}^{\rm shot, opt}(\omega )=\frac{\kappa }{4 g^2\gamma _m \left| \chi _m \right|^2} \left[N+1/2-\sqrt{N(N+1)}\right] .
\end{equation}
We notice that the optimal value of the detuning, $\Delta_c=0$ can be taken only in the present model with a single cavity mode and not in the dual-cavity model \cite{maximilian} where the parameter $\Delta_c$ is replaced by the coupling rate between the two cavities $2J$, which cannot be reduced to zero. This is an important advantage of the single cavity mode case considered here.
Equation (\ref{speopt}) shows that injected squeezing greatly facilitates achieving the ultimate limit provided by CQNC of Eq.~(\ref{sCQNC}) because in the optimal case and for large values of the squeezing parameter $N$, the SN term is suppressed by a factor $1/(4N)$ with respect to the case without injected squeezing (compare Eq.~(\ref{h_general}) in the case $M=N=y=0$ with Eq.~(\ref{h_particular}) in the case when $N \gg 1$). This is of great practical utility because it means that one needs a much smaller value of $g$, and therefore much less optical driving power in order to reach the same suppression of the SN contribution.

\begin{figure}
	\begin{center}
		\includegraphics[width=8cm]{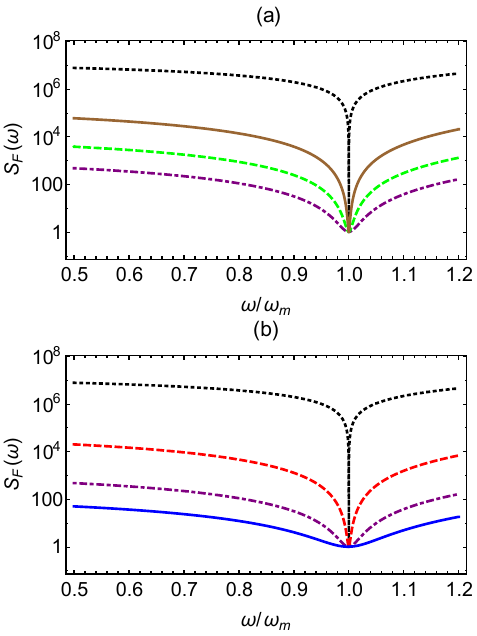}
	\end{center}
	\caption{(Color online) Force noise spectral density versus $\omega/\omega_m$ in the presence of perfect CQNC, with an optimized squeezed injected light with phase $\phi= \phi_{\rm opt}(0) = 0$ and $|M|=\sqrt{N(N+1)}$. (a) refers to the case with fixed squeezing parameter $N=10$ and different detunings: $\Delta_c/\kappa=0$ (dot-dashed purple line), $\Delta_c/\kappa=1/2$ (dashed green line), $\Delta_c/\kappa=1$ (full brown line). (b) refers to the optimal case $\Delta_c/\kappa=0$ and different values of the squeezing parameter, $N=0$ (dashed, red line), $N=10$ (dot-dashed, purple line), and $N=100$ (full, blue line). The dotted black line corresponds to the SQL. The other parameters are ${\omega _m}/2\pi  = 300$ KHz, ${\gamma _m}/2\pi  = 30 $ mHz, ${g_0}/2\pi  = 300$ Hz, ${\lambda _L} \simeq 780$ nm, $P_L=24 \mu $W and $\kappa/2\pi=1$ MHz. Reprinted with permission from A. Motazedifard, F. Bemani, M. H. Naderi, R. Roknizadeh, and D. Vitali, New J. Phys. 18, 073040 (2016). Licensed under a Creative Commons 3.0 License.}
	\label{fig9}
\end{figure}

Let us now illustrate how the combination of backaction cancellation by the atomic ensemble under CQNC and of the squeezing injected in the cavity may significantly improve force detection sensitivity. We consider an experimentally feasible scheme based on a membrane-in-the-middle setup \cite{thompson}, coupled to an ultracold atomic gas confined in the cavity and in a magnetic field, like the one demonstrated for light storage \cite{dudin}. 
A system of this kind has not been demonstrated yet, but the coupling of an atomic ensemble with a membrane has been already demonstrated \cite{camerer2011,jockel2014}. We assume typical mechanical parameter values for SiN membranes, ${\omega _m}/2\pi  = 300$ kHz, ${\gamma _m}/2\pi  = 30 $ mHz, ${g_0}/2\pi  = 300$ Hz, ${\lambda _L} \simeq 780$ nm, $P_L=24 \mu $W and $\kappa/2\pi=1$ MHz (see also the caption of Fig. 2). The ground state sub-levels of the ultracold atomic gas \cite{dudin} could be prepared in order to satisfy the CQNC condition, i.e., the Zeeman splitting tuned in order that the effective atomic transition rate coincides with $\omega_m$, the driving of the laser fields adjusted so that the two linearized couplings with the cavity mode, $G$ and $g$, coincide. Matching the dephasing rate $\Gamma$ with the damping rate $\gamma_m$ is less straightforward but one can decrease and partially tune the atomic dephasing rate using the magic-value magnetic field technique and applying dynamical decoupling pulse sequences, as demonstrated~\cite{dudin}.

In Fig.~(\ref{fig9}), the force noise spectral density ${S_F}(\omega )$ optimized over the squeezing parameters, i.e., $|M| = \sqrt{N(N+1)}$, $\phi =0$, is plotted versus frequency. In Fig.~\ref{fig9}(a) we fix the squeezing parameter $N=10$ and consider different values of the detuning: as shown above, force noise is minimum at the optimal case of resonant cavity driving $\Delta_c = 0$. This plot clearly shows the advantage of the present single-cavity scheme compared to the double-cavity setup \cite{meystreCQNC2015}, where the role of $\Delta_c$ is played by the mode splitting $2J$ associated with the optical coupling $J$ between the cavity that cannot be put to zero. In Fig.~\ref{fig9}(b) we fix the detuning at this optimal zero value, and we consider different values of the injected squeezing parameter $N$. At resonance ($\omega=\omega_m$), CQNC and injected squeezing does not improve with respect to the SQL spectrum (dotted black line), but force noise suppression is remarkable in a broad band around the resonance peak, and becomes more relevant for increasing injected squeezing $N$. Notice that injected squeezing allows a further reduction of the off-resonance ($\omega \neq \omega_m$) force noise with respect to what can be achieved with CQNC alone (see in Fig.~\ref{fig9}(b) the full blue line and the dot-dashed purple line compared to the dashed red line which refers to the case of no-injected squeezing, $N=0$.)

\begin{figure}
	\begin{center}
		\includegraphics[width=8.5cm]{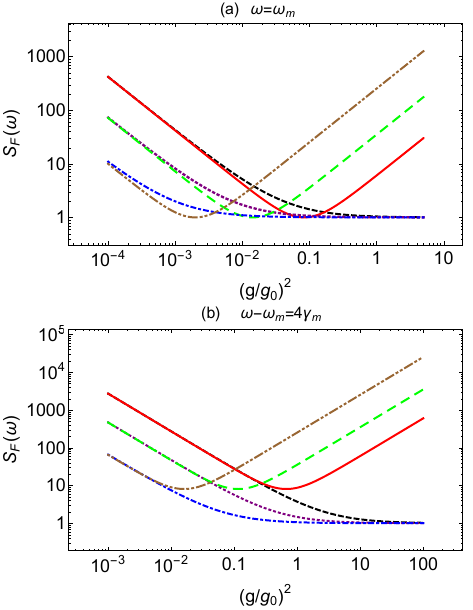}
	\end{center}
	\caption{(Color online) Force noise spectral density versus ${(g/{g_0})^2}$ (proportional to the input laser power) at the optimal value for the detuning $\Delta_c/\kappa=0$ and with an optimized squeezed injected light with phase $\phi= \phi_{\rm opt}(0) = 0$ and $|M|=\sqrt{N(N+1)}$. (a) refers to the value at the mechanical resonance, $\omega=\omega_m$, while (b) refers to the off-resonant case $\omega=\omega_m+4\gamma_m$. In both subfigures we compare the force noise spectrum with perfect CQNC and for a given (optimized) squeezing $N$ with the corresponding spectrum with the same injected squeezing but without atomic ensemble and CQNC. The full red line refers to standard optomechanical case $N=0$ without the atomic ensemble of Eq. (\ref{sopt}), while the dashed black line refers again to $N=0$ \emph{with} the atomic ensemble and perfect CQNC. The case with injected squeezing with $N=1$ corresponds to the long-dashed green line (without atoms and CQNC), and to the dotted purple line (with atoms and CQNC). Finally, the case with injected squeezing with $N=10$ corresponds to the double-dot dashed brown line (no atoms and CQNC), and to the dot-dashed blue line (with atoms and perfect CQNC). The other parameter values are as in Fig.~(\ref{fig9}). Reprinted with permission from A. Motazedifard, F. Bemani, M. H. Naderi, R. Roknizadeh, and D. Vitali, New J. Phys. 18, 073040 (2016). Licensed under a Creative Commons 3.0 License.}
	\label{fig10}
\end{figure}

In Fig.~(\ref{fig10}) the force noise spectral density is instead plotted versus $(g/g_0)^2$, which is proportional to the input laser power $P_L$, both at the mechanical resonance (Fig.~\ref{fig10}(a), $\omega=\omega_m$) and off-resonance (Fig.~\ref{fig10}(b), $\omega=\omega_m+4\gamma_m$). In both subfigures we compare the force noise spectrum with perfect CQNC and for a given optimized squeezing $N$, with the corresponding spectrum with the same injected squeezing but without atomic ensemble and CQNC, for three different values of $N$, $N=0, 1, 10$.
Back-action noise cancellation manifests itself with a significant noise suppression at large power, where minimum force noise is achieved. Without atoms and CQNC, force noise diverges at large power due to backaction, and one has the usual situation where minimum force noise is achieved at the SQL, at a given optimal power. In both cases, either with or without CQNC, injected squeezing with $\phi= 0$ and $|M|=\sqrt{N(N+1)}$ is not able to improve force sensitivity and to lower the noise at resonance (see Fig. \ref{fig10}(a)), i.e., the SQL value remains unchanged, but one has the advantage that for increasing $N$, the SQL is reached at decreasing values of input powers \cite{caves1980}. As already suggested by Fig.~(\ref{fig9}), instead one has a significant force noise suppression off-resonance and at large powers due to backaction cancellation (Fig. \ref{fig10}(b)).

As an outlook, in order to exploit the ponderomotive squeezing for stronger SN suppression within the CQNC method, one may use the optimized-rotated output phase quadrature of the cavity mode to enhance the quantum measurement.

\subsection{imperfect CQNC conditions}
backaction noise cancellation requires the perfect matching of atomic and mechanical parameters. As we discussed above, one can tune the effective atomic transition rate by tuning the magnetic field, and make it identical to the mechanical resonance frequency $\omega_m$. 
But, the perfect CQNC is an ideal case which is not satisfied in the real experiments. Let us here consider an imperfect CQNC \cite{aliNJP} to see what happens to the force noise spectrum. We still assume such a perfect frequency matching which, even though not completely trivial, can always be achieved due to the high tunability of Zeeman splitting. As we discussed in the previous subsection, one can also make the two couplings with the cavity mode $G$ and $g$ identical, by adjusting the cavity and atomic driving, and finally even the two decay rates, $\Gamma$ and $\gamma_m$. However, both coupling rates matching and decay rate matching are less straightforward, and therefore it is important to investigate the robustness of the CQNC scheme with respect to imperfect matching of these two latter parameters.

The total force noise spectrum in the presence of the mismatch in optimal case of zero detuning  $\Delta_c=0$ is obtained as \cite{aliNJP}
\begin{eqnarray}  \label{S_zerodetuning}
&& S(\omega)\vert_{\Delta_c=0}= \frac{k_{\rm B} T}{\hbar \omega_m}+\frac{\kappa}{4g^2 \gamma_m} \frac{1}{\vert \chi_m(\omega) \vert^2}\left(N+\frac{1}{2}-{\rm Re} M \right) \nonumber \\
&& \qquad \qquad + \frac{4g^2}{\kappa \gamma_m} \left(N+\frac{1}{2}+{\rm Re} M \right) \left\vert 1+\frac{G^2}{g^2}R(\omega) \right\vert^2 \nonumber  \\
&& \qquad \qquad + \frac{1}{2}\left(\frac{G}{g}\right)^2 \frac{\Gamma}{\gamma_m} \vert R(\omega)\vert^2 \left( 1+\frac{\omega^2 + \Gamma^2/4}{\omega_m^2} \right) \nonumber \\
&& \qquad \qquad  + {\rm Im} \left[  \frac{\left(2i{\rm Im} M-1\right)}{\gamma_m \chi_m^{\ast}(\omega)}   \left[1+\frac{G^2}{g^2}R(\omega)\right]\right] .
\end{eqnarray}
with 
\begin{eqnarray}
&& R(\omega)=\frac{\chi_d(\omega)}{\chi_m(\omega)}= - \left( 1+r(\omega) \right), \\
&& r(\omega) =\frac{i \omega (\gamma_m-\Gamma)}{(\omega_m^2-\omega^2)+i\omega \Gamma}.
\end{eqnarray}

To see the effects of mismatches, we have restricted our analysis to the parameter regime  corresponding to the optimal case under perfect CQNC conditions, i.e., the resonant case  $\Delta_c=0$, with an optimized pure squeezing, $\phi= \phi_{\rm opt}(0) = 0$ and $|M|=\sqrt{N(N+1)}$. We have also fixed the squeezing value, $N= 10$, and considered again the parameter values of the previous subsection, but now considering the possibility of nonzero mismatch of the couplings and/or of the decay rates. The calculations show \cite{aliNJP} that CQNC is more sensitive to the coupling mismatch than to the decay rate mismatch. In fact, the spectrum is appreciably modified already when $(G-g)/g = 10^{-5}$, and force noise increases significantly and in a broadband around resonance already when $(G-g)/g = 10^{-3}$. This modification is quite independent from the value of the decay rate mismatch, $(\Gamma-\gamma_m)/\gamma_m$, whose effect moreover is always concentrated in a narrow band around resonance and for larger values, $(\Gamma-\gamma_m)/\gamma_m = 0.5$. There is a weak dependence upon the sign of the two mismatches, which however is typically very small and can be negligible.

At large power, proportional to $ g^2 $, the force noise spectrum increases due to the uncancelled residual backaction noise, and the increase at large power is larger for larger mismatch parameters. At mechanical resonance, $ \omega=\omega_m $, both coupling and decay rate mismatches have an effect such that increase of force noise is larger when both mismatches are nonzero and opposite due to the effect of the negative mass yielding susceptibilities with opposite signs. Note that the effect of decay rate mismatch is instead hardly appreciable out of resonance, and noise increase is caused by the mismatch between the two couplings, regardless the value of the decay rate mismatch. 
The analysis allows us to conclude that CQNC is robust with respect to mismatch of the decay rates, up to $10\%$ mismatch, and especially off-resonance, where the advantage of backaction cancellation is more relevant. On the contrary, CQNC is very sensitive to the mismatch between the atomic and mechanical couplings with the cavity mode, which have to be controlled at $0.1 \%$ level or better. This means that in order to suppress backaction noise, the intensity of the cavity and atomic driving have to be carefully controlled in order to adjust the two couplings.

Finally, it is worth reminding that the scheme proposed in this section has been experimentally realized without squeezed light injection in a membrane-in-the-middle optomechanical cavity whose optical mode has been coupled to the collective spin mode of an atomic ensemble of Cs atoms which is at a 1-meter distance of the cavity \cite{cQNCNatureexp}. Moreover, this experimental system has been recently exploited for distant entanglement generation between the MO and atomic spin \cite{polzikDistantEntanglementOMS2020}.

\subsection{is signal amplification and noise suppression simultaneously possible in a standard OMS? \label{secAmplify}}
Up to now, we have shown that the QND and CQNC methods enable us to evade and cancel the backaction noise of measurement, respectively. But, there still exists an important question: is there any method to amplify the input signal in the output phase spectrum while keeping the backaction noise below the SQL? 
Here in this subsection, for the first time we clearly show that in the standard OMS the input signal cannot be amplified in the output cavity phase spectrum while the noise being reduced. Surprisingly, in the last section we will introduce the parametric sensing method for signal amplification and simultaneous noise suppression in a nonlinear hybrid OMS\cite{aliDCEforcesenning}. 

Consider the standard optomechanical Hamiltonian (\ref{E7}) in the absence of the atomic ensemble, and with $ G=0 $. The output cavity phase quadrature can be rewritten in the Fourier space as (see Eq.~(\ref{Pout}) with $ G=0 $)
\begin{eqnarray}
&& \hat P_a^{out}:=- r_a(\omega) \Big[  \hat {f} - \sqrt {\frac{\kappa }{{{\gamma _m}}}} \frac{1}{{g{\chi _m}}}\left( {(1 - \frac{1}{{{{\bar \chi}_a}\kappa }})\hat P_a^{in} - {\Delta _c}{\chi _a}\hat X_a^{in}} \right) \nonumber \\
&& \qquad \qquad  - g \sqrt {\frac{\kappa }{{{\gamma _m}}}} {\chi _a}\hat X_a^{in} \Big],
\end{eqnarray}
where $  r_a(\omega):= g{{\bar \chi}_a}{\chi _m}\sqrt {\kappa {\gamma _m}}$ plays the role of amplification amplitude of output phase spectrum and $ \bar \chi_a $ is simplified as
\begin{eqnarray} \label{effectivesusceptibility1}
&& \frac{1}{{{{\bar \chi}_a}}} = \frac{1}{{{\chi _a}}} - {\chi _a}{\Delta _c}\left( {{g^2}{\chi _m} - {\Delta _c}} \right). 
\end{eqnarray}
As is obviously seen, the output cavity phase quadrature includes the thermal mechanical noise, SN and backaction noise. Again, by defining the force noise as before, and with some algebraic manipulation, one can rewrite the output cavity spectrum for the zero detuning and in the limit of $ \omega/\kappa \ll 1 $ as 
\begin{eqnarray} \label{s_cavityOMS}
&& S_{p_a}^{\rm out}= R_c^{\rm st}(\omega) {S_F^{\rm st}}(\omega),
\end{eqnarray}
where $ {S_F^{\rm st}}(\omega) $ is the standard force noise spectrum (see Eq.~(\ref{sopt}), and the power of signal-response can be defined as
\begin{eqnarray}
&& R_c^{\rm st}(\omega):= r_c(\omega) r_c(-\omega)=\kappa \gamma_m \vert g \chi_m(\omega) \chi'_a(\omega) \vert^2. \label{signalpowerOMS}
\end{eqnarray}

It is clear that in the standard OMS, only at the mechanical resonance frequency $ \omega=\omega_m $, and at zero temperature ($ T=0 $) the noise can be suppressed and minimized, which occurs at the optimum cooperativity $ \mathcal{C}_{opt}=4g_{\rm opt}^2/\kappa \gamma_m= 1 $.
Now, the question is that if it is possible to amplify the signal at the output phase spectrum, i.e., $ R_c^{\rm st}(\omega) > 1$ when the force noise reaches to the SQL at $ g=g_{\rm opt} $.
It can be easily shown that in the standard OMS, the signal response can never be more than unity while the force noise is minimized at ($ \mathcal{C}_{\rm opt}=1 $) which means that no signal-amplification occurs, i.e., $ R_c^{\rm st}\vert_{\mathcal{C}_{\rm opt}=1}(\omega) \le 1 $. 
Note that as the optomechanical cooperativity increases ($ \mathcal{C}> 1 $), although the signal response increases but the backaction noise increases simultaneously above the SQL. In other words, in a standard OMS, it is impossible to have simultaneous signal amplification and force noise suppression.

In the next section, it will be shown that to have simultaneous signal amplification and noise suppression below the SQL in an \textit{ultra precision} quantum measurement, one needs to add another degree of freedom to a standard OMS, which can be possible by hybridizing the bare OMS with a BEC, together with executing parametric time-modulations on the atomic and mechanical modes.

\section{Parametric Single-quadrature force-sensing: Simultaneous signal amplification and backaction noise suppression using ultra-fast coherent time-modulation \label{secParametricsensing}}
In the previous sections we showed that for realizing an ultra precision quantum measurement below the SQL one should find ways to either evade or cancel the backaction noise of measurement.  Although in these methods the total noise of measurement is reduced, but the signal is not amplified at all. In this section, we present the so-called \textit{parametric} \textit{single-quadrature} force-sensing \cite{aliDCEforcesenning} beyond the SQL based on simultaneous signal amplification and backaction noise suppression via parametric amplification of the mechanical and Bogoliubov modes in a hybrid OMS. It is shown that using the parametric modulation, one can amplify the input signal while keeping the backaction noise below the SQL so that weak forces as low as $ 10^{-20}  \rm N/\sqrt{ \rm Hz}$ can be detectable \cite{aliDCEforcesenning}.

Based on the theory of linear quantum amplifiers \cite{clerkquantumnosie1}, in order to enhance the functionality of a linear amplifier it is necessary to add more degrees of freedom to the system so that the input signal is amplified more effectively. However, the price to pay for introducing extra degrees of freedom will be the manifestation of some added noises to the input signal. Here, we use the Bogoliubov mode of the BEC as an extra degree of freedom in a hybrid optomechanical cavity with a moving end-mirror in the red-detuned regime containing an interacting cigar-shaped BEC in the dispersive regime of atom-field interaction where the \textit{s}-wave scattering frequency of the BEC atoms as well as the spring coefficient of the cavity moving end-mirror (the MO) are parametrically modulated.

Note that, the most important advantage of using a BEC in an optomechanical cavity is that the Bogoliubov mode of the BEC behaves effectively as an MO (a moving mirror) with a controllable natural frequency while the ordinary mechanical oscillators (moving end mirrors or membranes) have fixed natural frequencies which cannot be changed after fabrication [for more details, see Sec.~(\ref{secBEC})]. Therefore, as was mentioned previously, to enhance the functionality of a quantum amplifier, we need to use more extra degrees of freedom and since the BEC has more controllability, thus, we have chosen it as another extra mode. 
It is shown that because of its large mechanical gain such a hybrid system with both the atomic and mechanical modulations is a much better amplifier in comparison to the (modulated) bare optomechanical system which can generate a stronger output signal while keeping the sensitivity nearly the same as that of the (modulated) bare one studied in Ref~\cite{optomechanicswithtwophonondriving}. Furthermore, the force measurement precision in the off-resonance region can be improved in such hybrid systems through the increase of the amplification bandwidth.

\subsection{hybrid OMS as a force sensor}
\begin{figure}
	\includegraphics[width=8.5cm]{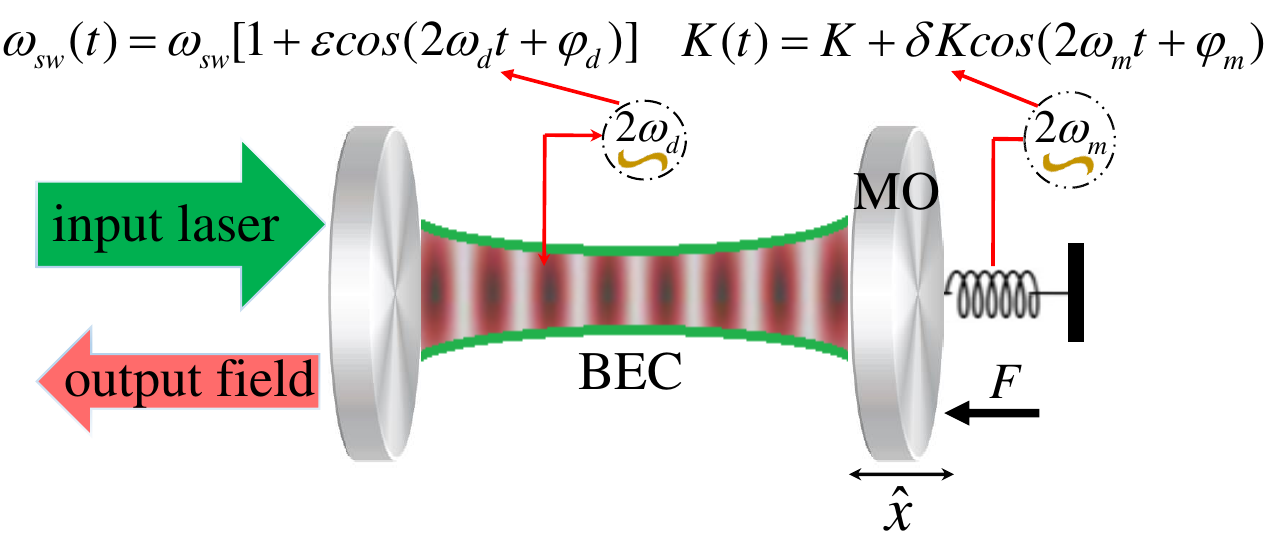}
	\caption{(Color online) Schematic of a hybrid optomechanical force sensor designed for the measurement of a weak force signal ($F$) exerting to the moving end-mirror of an optomechanical cavity which contains a BEC where the \textit{s}-wave scattering frequency of the condensate atoms as well as the spring coefficient of the cavity moving mirror are parametrically modulated, i.e., $ \omega_{sw}(t)=\omega_{sw}[1+\varepsilon \cos(2\omega_d t+\varphi_d)] $ and $ K(t)=K + \delta K \cos (2\omega_m t +\varphi_m) $. Also, the cavity mode is coherently driven by a classical laser field through the fixed mirror. Reprinted with permission from A. Motazedifard, A. Dalafi, F. Bemani, and M. H. Naderi, Phys. Rev. A 100, 023815 (2019). Copyright 2019 by the American Physical Society.}
	\label{fig11}
\end{figure}

As depicted in Fig.~(\ref{fig11}), the force sensor we are going to investigate is an optomechanical cavity with length $ L $ and damping rate $\kappa$ having a moving end-mirror with mass $m$, natural frequency $\omega_m$, and damping rate $ \gamma_m $ whose spring coefficient is parametrically modulated at twice its natural frequency. The cavity which is driven through the fixed mirror by a laser with frequency $\omega_{L}$ and wavenumber $k_{0}=\omega_{L}/c$ contains a BEC of $ N $ ultracold two-level atoms with mass $ m_a $ and transition frequency $ \omega_a $. Furthermore, we assume that the collision frequency of the condensate atoms is parametrically modulated through the modulation of the electromagnetic trap or the density of the BEC by changing the trap stiffness \cite{jaskulaBECSCE}. The total Hamiltonian of the system in the frame rotating at the driving laser frequency $ \omega_L $ can be written as

\begin{eqnarray} \label{H1total}
&& \hat H= \hbar \Delta_c \hat a^\dag \hat a + i\hbar E_L (\hat a^\dag - \hat a)+\hbar \omega_m \hat b^\dag \hat b-\hbar g_0 \hat a^\dag \hat a (\hat b + \hat b^\dag)\nonumber\\
&& +\frac{i \hbar}{2} (\lambda_m \hat b^{\dag 2}  e^{-2i\omega_m t}- \lambda_m^\ast \hat b^2 e^{2i\omega_m t}) + \hat H_F + \hat H_{BEC} .
\end{eqnarray}
The first three terms in the Hamiltonian describe, respectively, the free energy of the cavity mode, the coupling
between the cavity mode and the driving laser, and the free energy of the MO. Here, $ \Delta_c= \omega_c-\omega_L $ is the detuning of the optical mode from the driving laser frequency, $E_L$ is the pump rate of the external laser, and $ \hat a $ ($ \hat b $) is the annihilation operator of the cavity (MO) mode. The canonical position and momentum of the MO are $ \hat x= x_{\rm zpf}(\hat b + \hat b^\dag) $ and $ \hat p= \hbar (\hat b - \hat b^\dag)/(2ix_{\rm zpf} )$. 

The fourth term in Hamiltonian (\ref{H1total}) is the optomechanical interaction between the mechanical and optical modes with the single-photon optomechanical coupling $ g_0=x_{\rm zpf} \omega_c /L $. The fifth term describes the parametric driving of the MO spring coefficient at twice its natural frequency [$ K(t)=K+\delta K \cos (2\omega_m t + \varphi_m) $ with $ \varphi_m $ being the phase of external modulation] which is written in the RWA over time scales longer than $ \omega_m^{-1} $ where $ \lambda_m= \vert \lambda_m \vert e^{i\varphi_m} $ with $ \vert \lambda_m \vert= \delta K x_{\rm zpf}^2 / 2\hbar $ \cite{aliDCE3,optomechanicswithtwophonondriving}. Note that by fixing the phase of modulation $ \varphi_m $, it is always possible to take $ \lambda_m $ as a real number. It is worth to point out that this term can be considered as the mechanical phonon analog of the degenerate parametric amplification (DPA) which may lead to the DCE of mechanical phonons \cite{aliDCE3}.
The sixth term, $ \hat H_F $, accounts for the coupling of the MO to the input classical-force $ F $ to be measured which is given by 
\begin{eqnarray} \label{H_F}
&& \hat H_F=F(t) \hat x= x_{\rm zpf} F(t) (\hat b + \hat b^\dag).
\end{eqnarray}

The last term of Eq.~(\ref{H1total}) is the Hamiltonian of the dispersive atomic BEC inside the cavity which has been given by Eqs.(\ref{hBEC})-(\ref{h_sw}). In the presence of time modulation of the \textit{s}-wave scattering frequency of atomic collisions at twice the frequency of the Bogoliubov mode, the \textit{s}-wave scattering frequency becomes a time-depended function as  $ \omega_{sw}(t)= \omega_{sw}[1+\varepsilon \cos (2\omega_d t+\varphi_d)]$ where $ \varepsilon $ and $ \varphi_d $ are, respectively, the amplitude and the phase of modulation. Therefore, $ \hat H_{sw} $ in the RWA is given by \cite{aliDCEforcesenning,aliDCEsqueezing}
\begin{eqnarray} \label{H_sw_mod}
&& \hat H_{sw}(t) = \frac{i \hbar}{2} (\lambda_d \hat d^{\dag 2}  e^{-2i\omega_d t}- \lambda_d^\ast \hat d^2 e^{2i\omega_d t}),
\end{eqnarray}
where $ \lambda_{d}=-i\varepsilon \omega_{sw} e^{-i\varphi_d}/4 $ can be taken real by fixing the phase $ \varphi_d $.
It should be noted that the \textit{s}-wave scattering frequency, $ \omega_{sw} $, can be controlled experimentally by manipulating the transverse trapping frequency of the BEC through changing the waist radius of the optical mode $ w $ \cite{morsch}. Besides, the time modulation of the atomic collisions can be experimentally realized \cite{jaskulaBECSCE} by the time modulation of the scattering length via the modulation of the electromagnetic trap, or the modulation of the density of the BEC by changing the trap stiffness via the intensity modulation of the pump laser.

The Hamiltonian of Eq.~(\ref{H_sw_mod}) is a Bogoliubov-phonon analog of the DPA which can give rise to the generation of Bogoliubov-type Casimir phonons \cite{aliDCE3}. Here, it should be mentioned that in the BEC-Hamiltonian, we have ignored the cross-Kerr nonlinear coupling between the intracavity field and the Bogoliubov mode which is negligibly small in comparison to the radiation pressure interaction \cite{dalafi7}. Therefore, the total Hamiltonian of the system takes the form \cite{aliDCEforcesenning}
\begin{eqnarray}\label{Hamt}
&& \hat H_{tot}=\hbar \Delta_0 \hat a^\dag \hat a + \hbar \omega_m \hat b^\dag \hat b + \hbar \omega_d \hat d^\dag \hat d + i \hbar E_L(\hat a^\dag - \hat a ) \nonumber \\
&& \qquad   -\hbar g_0 \hat a^\dag \hat a (\hat b+\hat b^\dag) + \hbar G_{0} \hat a^\dag \hat a (\hat d +\hat d^\dag)+x_{zp} F(t) (\hat b + \hat b^\dag)  \nonumber \\
&& \qquad + i \frac{\hbar }{2} (\lambda_m\hat b^{\dag2} e^{-2i\omega_m t}- \lambda_m^\ast \hat b^2 e^{2i\omega_m t})\nonumber\\
&&\qquad+i\frac{ \hbar}{2} (\lambda_d \hat d^{\dag 2}  e^{-2i\omega_d t}- \lambda_d^\ast \hat d^2 e^{2i\omega_d t}), 
\end{eqnarray}
where $ \Delta_0=\Delta_c + NU_0/2 $ is the cavity Stark-shifted detuning.

\subsection{linearized quantum dynamics}
The linearized QLEs of the system can be obtained from the Hamiltonian of Eq.~(\ref{Hamt}). As has been shown~\cite{aliDCE3}, in the red-detuned regime and within the RWA where the two optomechanical and opto-atomic couplings are analogous to the beam-splitter interaction, the linearized QLEs describing the dynamics of the quantum fluctuations using the quadratures $ \delta \hat X_o=(\hat o+\hat o^{\dagger})/\sqrt{2} $ and $ \delta \hat P_o=(\hat o-\hat o^{\dagger})/\sqrt{2}i $ ($ o=a,b,d $) can be written as the following compact matrix form \cite{aliDCEforcesenning}
\begin{eqnarray} \label{udot}
&&\delta \dot {\hat u}(t)= A ~\delta \hat u(t) + \hat u_{in}(t),
\end{eqnarray}
where the vector of continuous-variable fluctuation operators and the corresponding vector of noises are, respectively, given by $ \delta\hat u=\Big(\delta \hat X_a,\delta \hat P_a,\delta \hat X_b, \delta \hat P_b,\delta \hat X_d,\delta \hat P_d\Big)^T $ and $ \hat  u_{in}(t)=\Big(\sqrt{\kappa}\hat X_a^{in},\sqrt{\kappa}\hat P_a^{in},\sqrt{\gamma_m}\hat X_b^{\prime in},\sqrt{\gamma_m}\hat P_b^{\prime in},\sqrt{\gamma_d} \hat X_d^{in},\sqrt{\gamma_d} \hat P_d^{in}\Big)^T $ in which  $ \hat X_o^{in}=(\hat o_{in}+\hat o_{in}^{\dagger})/\sqrt{2} $ and $ \hat P_o^{in}=(\hat o_{in}-\hat o_{in}^{\dagger})/\sqrt{2}i $ ($ o=a,b,d $). The time-independent drift matrix $ A $ is given by 
\begin{eqnarray} \label{A}
&&\!\!\!\!\!\!\!\!\!\!\!\!  A\!=\! \left( \begin{matrix}
{-\frac{\kappa}{2}} & {0} & {0} & {-g} & {0} & {G}  \\
{0} & {-\frac{\kappa}{2}} & {g} & {0} & {-G} & {0}  \\
{0} & {-g} & {\!\lambda_m\!-\!\frac{\gamma_m}{2}} & {0} & {0} & {0}  \\
{g} & {0} & {0} & {-(\lambda_m\!+\!\frac{\gamma_m}{2})} & {0} & {0}  \\
{0} & {G} & {0} & {0} & {\lambda_d -\frac{\gamma_d}{2}} & {0}  \\
{-G} & {0} & {0} & {0} & {0} & {-(\lambda_d\!+\!\frac{\gamma_d}{2} )}  \\
\end{matrix} \right),
\end{eqnarray}
where $ g=g_0 \bar a $ and $ G = G_0 \bar a $ are, respectively, the enhanced-optomechanical and opto-atomic coupling strengths in which $\bar a=E_L/\sqrt{\kappa^2/4+\bar \Delta_{0}^{2}} $ is the steady-state mean value of the optical mode. Here, $ \bar \Delta_{0}=\Delta_{0}-2g_0 \bar b+2G_0 \bar d $ is the effective cavity detuning where $ \bar b\approx g_0 \bar a^2/\omega_{m}  $ and $ \bar d\approx -G_0 \bar a^2/\omega_{d} $ are, respectively, the steady-state values of the mechanical and atomic mean fields in the RWA and in the high quality factors limit. Besides, $\gamma_{m}$ and $\gamma_{d}$ are the dissipation rates of the mechanical and the Bogoliubov modes, respectively.

Here, the red-detuned regime of cavity optomechanics is defined by the condition $ \bar\Delta_0 \approx \omega_m \approx \omega_d $. For this purpose, the frequency of the Bogoliubov mode of the BEC, i.e., $ \omega_d $, should be matched to the mechanical frequency ($ \omega_d \approx \omega_m $) which is possible through the manipulation of the \textit{s}-wave scattering frequency of the Bogoliubov mode via controlling the transverse frequency of the BEC trap \cite{morsch}. Besides, the effective detuning $\bar\Delta_0$ can be set in the red-detuning regime through the pump laser frequency.

Furthermore, the optical input vacuum noise $ \hat a_{in} $ as well as the Brownian noises $ \hat b_{in} $ and $ \hat d_{in} $ affecting, respectively, the MO and the Bogoliubov mode of the BEC, satisfy the Markovian correlation functions $ \langle \hat a_{in}(t) \hat a_{in}^\dag (t') \rangle= (1+\bar n_c^T) \delta (t-t') $, $ \langle \hat a_{in}^\dag (t) \hat a_{in}(t') \rangle= \bar n_c^T \delta (t-t') $; $ \langle \hat o_{in}(t) \hat o_{in}^\dag (t') \rangle= (1+\bar n_j^T) \delta (t-t') $ and $ \langle \hat o_{in}^\dag (t) \hat o_{in}(t') \rangle= \bar n_j^T \delta (t-t') $ with $\hat o= \hat b $  and $  \hat d $ where $ \bar n_j^T=[exp(\hbar \omega_j /k_B T)-1]^{-1} $ with $ j=c, m $ and $ d $ are the mean number of thermal excitations of the cavity, mechanical, and Bogoliubov modes at temperature $ T $. The quantum noise $\hat d_{in}$ originates from the other extra modes of the BEC as well as the fluctuations in the electromagnetic trap~\cite{dalafi4}. Besides,
\begin{eqnarray}
	&&\hat X_b^{\prime in}(t)=\hat X_b^{in}(t)+\sqrt{\frac{2}{\gamma_m}}\frac{x_{\rm zpf}}{\hbar}F(t) \sin\omega_m t,\label{md1}\\
	&&\hat P_b^{\prime in}(t)=\hat P_b^{in}(t)-\sqrt{\frac{2}{\gamma_m}}\frac{x_{\rm zpf}}{\hbar}F(t) \cos\omega_m t,\label{md2}
\end{eqnarray}
are the modified mechanical noises \cite{aliDCEforcesenning}.

Based on the Routh-Hurwitz criterion for the optomechanical stability condition\cite{aliDCE3}, the parameters $ \lambda_{m} $ and $ \lambda_{d} $ should satisfy the condition \cite{aliDCEsqueezing,aliDCEforcesenning,aliDCE3,optomechanicswithtwophonondriving}
\begin{eqnarray} \label{stability}
&&\!\!\!\!\!\!\! \lambda_{m(d)} \le \! \frac{\gamma_{m(d)}}{2} \left[1+ \mathcal{C}_{m(d)}\right]:= \lambda_{m(d)}^{max},
\end{eqnarray}
in which $ \mathcal{C}_m $($ \mathcal{C}_d $) is the collective optomechanical cooperativity associated with the mechanical mode (Bogoliubov mode) given by \cite{aliDCE3,aliDCEsqueezing,aliDCEforcesenning}
\begin{eqnarray}
&&  \mathcal{C}_{m(d)}=  \mathcal{C}_{0(1)} \frac{1+ \mathcal{C}_{1(0)} - \xi_{d(m)}^2}{(1+ \mathcal{C}_{1(0)}-\xi_{d(m)}^2)^2- \xi_{d(m)}^2  \mathcal{C}_{1(0)}^2} ~,
\end{eqnarray}
where $\mathcal{C}_0=4g^2/\kappa \gamma_m$ and $\mathcal{C}_1= 4G^2/\kappa \gamma_d $ are the optomechanical and opto-atomic cooperativities, respectively, and $ \xi_{d(m)}=2\lambda_{d(m)}/\gamma_{d(m)}$ plays the role of an effective dimensionless-amplitude of modulation \cite{aliDCE3,aliDCEforcesenning,aliDCEsqueezing}. The solution to the QLEs, i.e., Eq.~(\ref{udot}), in the Fourier space can be written as $ \delta\hat u(\omega)= \boldsymbol{  { \chi }}(\omega) \hat u_{in}(\omega) $ where $\boldsymbol{  { \chi }}(\omega)$ is the susceptibility matrix and the Fourier transforms of the modified mechanical noises, i.e., those of Eqs.(\ref{md1}) and (\ref{md2}) are as follows \cite{aliDCEforcesenning}
\begin{eqnarray}
	&&\hat X_b^{\prime in}(\omega)=\hat X_b^{in}(\omega)-\sqrt{\frac{2}{\gamma_m}}\frac{x_{\rm zpf}}{\hbar}\frac{i}{2}[F(\omega+\omega_m)-F(\omega-\omega_m)],\nonumber\\
	&&\hat P_b^{\prime in}(\omega)=\hat P_b^{in}(\omega)-\sqrt{\frac{2}{\gamma_m}}\frac{x_{\rm zpf}}{\hbar}\frac{1}{2}[F(\omega+\omega_m)+F(\omega-\omega_m)].\nonumber \\
\end{eqnarray}
Now, using the input-output theory for the field operators, the output $ P $-quadrature of the cavity field, i.e., $\delta \hat P_a^{out}(\omega)=-\sqrt{\kappa} \delta\hat P_{a}(\omega)+ \hat P_{in}^{a}(\omega)  $, is obtained as follows \cite{aliDCEforcesenning}
\begin{eqnarray} \label{paoutomega}
&& \!\!\!\! \!\!\! \!\!\!\!\! \delta \hat P_a^{out}(\omega)= \mathcal{A}(\omega) \hat P_a^{in}(\omega)+ \mathcal{B}(\omega) \hat X_b^{\prime in}(\omega) + \mathcal{D}(\omega) \hat X_d^{in}(\omega) ,
\end{eqnarray}
where $ \mathcal{A}(\omega)= 1-\kappa \chi_{22}(\omega) $, $ \mathcal{B}(\omega)=\sqrt{\kappa \gamma_{m}} \chi_{23}(\omega) $, and $ \mathcal{D}(\omega)=\sqrt{\kappa \gamma_{d}} \chi_{25}(\omega) $ and the relevant elements of the susceptibility matrix are given by
\begin{eqnarray}
&&\!\!\!\! \chi_{22}(\omega)=  \left[ \chi_0^{-1}(\omega)+g^2 \chi_{-m}+G^2 \chi_{-d}(\omega)  \right]^{-1} , \label{chi22} \nonumber \\
&&\!\!\!\! \chi_{23}(\omega)=g \left[ \chi_0^{-1}(\omega) \chi_{-m}^{-1}(\omega) \! +g^2 \! \! +G^2 \! \chi_{-d}(\omega)  \chi_{-m}^{-1}(\omega)   \right]^{-1} , \label{chi23} \nonumber \\
&&\!\!\!\! \chi_{25}(\omega)=-G \left[ \chi_0^{-1}(\omega) \chi_{-d}^{-1}(\omega) \! +G^2 \! \! +g^2 \! \chi_{-m}(\omega)  \chi_{-d}^{-1}(\omega)   \right]^{-1},  \label{chi25}\nonumber
\end{eqnarray}
with $ \chi_0^{-1}(\omega)= \kappa/2 - i\omega  $ and $ \chi_{- m(-d)}^{-1}(\omega)= \gamma_{m(d)}/2 - \lambda_{m(d)} - i \omega  $. It is clear that $ \chi_0(-\omega)= \chi_0(\omega)^\ast $ and $ \chi_{- m(-d)}(-\omega)= \chi_{- m(-d)}(\omega)^\ast $, so $ \chi_{ij}(-\omega)= \chi_{ij}(\omega)^\ast  $.

\subsection{single-quadrature force-sensing \label{sec.sensing}}
In the following, we will show how coherent modulations of both the atomic collisions frequency and the mechanical spring coefficient lead to the simultaneous signal amplification and backaction noise suppression which provides the best conditions for an ultra precision force sensing, i.e., the parametric force-sensing.

In the optomechanical force sensor demonstrated in Fig.~(\ref{fig11}), the imprint of the input mechanical signal is manifested in the  cavity output field through the optomechanical interaction. In other words, the MO position shift exerted by the external force leads to a change of the effective cavity length and therefore causes the variation of the optical cavity output phase. As a consequence, the signal corresponding to the exerted external force can be detected by measuring the spectrum of the optical output phase quadrature, $\hat P_a^{out}$, through methods like heterodyne, homodyne or \textit{synodyne} detections \cite{complexCQNC}. 
Very recently, it has been shown \cite{aliDCEsqueezing} that such a hybrid optomechanical system can be used as a quantum amplifier/squeezer through the coherent modulations. 
In the following, we will show how the present model of hybrid optomechanical system allows us for single-quadrature force-sensing with noise suppression and signal amplification which helps to surpass the SQL on force detection.

In order to measure and detect the input mechanical force, one should calculate the optical output phase quadrature spectrum. 
Since the signal has been coded in the input mechanical noise quadrature, for an efficient force-sensing, we should manipulate the system parameters such that the mechanical response to the input quadrature $ \hat X_b^{in} $ is amplified while the optical and atomic responses to the input noise quadratures $ \hat P_a^{in} $ and $ \hat X_d^{in} $ are attenuated. Since the mechanical response of the system is independent of the classical input signal force and depends only on the quantum properties of the system, in the following, we set aside the classical function $ F $ and calculate the optical output phase quadrature spectrum by considering just the input quantum noises. So the output optical power spectrum is obtained as \cite{aliDCEforcesenning}
\begin{eqnarray} \label{forcesensing1}
&& S_{P_{a}}^{out}(\omega)=R_m(\omega) \left[ (\bar n_m^T + \frac{1}{2})+n_{add}(\omega) \right],
\end{eqnarray}
where 
\begin{eqnarray}
&& R_m(\omega)=\vert \mathcal{B(\omega)} \vert^2=\kappa \gamma_m \vert \chi_{23}(\omega) \vert^2 \label{S_F}, \\
&& n_{add}(\omega)= (\bar n_c^T +\frac{1}{2}) \frac{\vert \mathcal{A}(\omega) \vert^2 }{\vert \mathcal{B}(\omega) \vert^2 }+ (\bar n_d^T + \frac{1}{2}) \frac{\vert \mathcal{D}(\omega) \vert^2}{\vert \mathcal{B}(\omega) \vert^2 } . \label{N_F}
\end{eqnarray}
Here, $ R_m(\omega) $ is the mechanical response to the input signal and $ n_{add}(\omega) $ is the added noise of measurement which originates from the contributions of the input optical and atomic vacuum noises to the phase quadrature of the output cavity field. As is seen from Eq.~(\ref{forcesensing1}), the added noise can be considered as an effective increase in the number of the thermal excitations of the mechanical reservoir due to the backaction of the optical and atomic modes. For a high precision force sensing and surpassing the SQL, one should amplify the mechanical response and suppress the added noise spectrum simultaneously. The SQL on force-sensing is defined as $ n_{add}^{\rm SQL}(\omega)=1/2 $ \cite{clerkquantumnosie1,forcedetection2} which has already been achieved experimentally \cite{schrepplerSQLexperiment}. In the following it is shown that through the mechanical and atomic modulations the SQL can be surpassed by suppressing the added backaction noise especially near the on-resonance frequency of the output $ P $-quadrature while the input force signal is amplified through the enhancement of the system mechanical response\cite{aliDCEforcesenning}.

The \textit{on-resonance} added noise and mechanical response are, respectively, given by \cite{aliDCEforcesenning} 
\begin{eqnarray} \label{signal and noise 1}
&& n_{add}(0)\!=\! \frac{(1-\xi_m)^2}{\mathcal{C}_0} \!\!  \left[\! \frac{\mathcal{G}_a}{(\sqrt{\mathcal{G}_a}\!-\!1)^2 } (\bar n_c^T\!+\!\frac{1}{2})\! +\! \frac{\mathcal{C}_1}{(1\!-\!\xi_d)^2} (\bar n_d^T\!+\!\frac{1}{2})\! \right], \label{N_F_0} \nonumber  \\
&& \\
&&R_m(0)=\mathcal{C}_0 \left(\frac{\sqrt{\mathcal{G}_a}-1}{1-\xi_m}\right)^2 , \label{S_F_0}
\end{eqnarray}
where the optical gain $ \mathcal{G}_a $, which is defined in the context of the linear amplifiers as the ratio of photons number in the output of the amplifier to that in the input \cite{forcedetection2}, is given by \cite{aliDCEsqueezing}
\begin{eqnarray}
&& \sqrt{\mathcal{G}_a}= \frac{\mathcal{C}_0-(1-\xi_m)+\mathcal{C}_1 \frac{1-\xi_m}{1-\xi_d}}{\mathcal{C}_0+(1-\xi_m)+\mathcal{C}_1 \frac{1-\xi_m}{1-\xi_d}} . \label{g_a}	
\end{eqnarray} 

As is seen from Eq.~(\ref{N_F_0}), the added noise is suppressed in the limit of $ \xi_m \to 1 $. However, as is seen from Eq.~(\ref{S_F_0}), in order to have signal amplification, the mechanical response should be increased simultaneously which is only possible when the optical gain is negligibly small or equal to zero, i.e., when $ \mathcal{G}_a= 0 $. To achieve zero gain, the \textit{impedance}-matching condition given by \cite{aliDCEforcesenning}
\begin{eqnarray} \label{impedance1}
&& \!\!\!\! \mathcal{C}_0+(\xi_m-1)[1-\mathcal{C}_1/(1-\xi_d)]=0, \quad    \mathcal{C}_0+ \mathcal{C}_1 \le 1 ,
\end{eqnarray}
should be satisfied. In other words, in order to have simultaneous noise suppression together with signal amplification, the numerical values of the cooperativities ($ \mathcal{C}_0 $ and $ \mathcal{C}_1 $) and also the atomic modulation $ \xi_{d} $ should be chosen so that the impedance-matching condition of Eq.~(\ref{impedance1}) is satisfied for any specified value of the mechanical modulation in the limit of $\xi_m\to 1 $.

In the case where there is neither mechanical nor atomic modulation (\textit{off-modulations}), i.e., $ \xi_{d}=\xi_m=0 $, we have \cite{aliDCEforcesenning}
\begin{eqnarray}\label{modoff}
&& \!\!\!  n_{add}^{\rm {off}}(0)= \! \frac{1}{\mathcal{C}_0} \left[\frac{(\mathcal{C}_0+\mathcal{C}_1-1)^2}{4} (\bar n_c^T+\frac{1}{2}) + \mathcal{C}_1 (\bar n_d^T + \frac{1}{2}) \right] \! ,\label{noise_no_modulation}  \\
&& \!\!\! R_m^{\rm {off}}(0)=\frac{4\mathcal{C}_0}{(1+\mathcal{C}_0+\mathcal{C}_1)^2} \quad. \label{signal_no_modulation}
\end{eqnarray}
As is evident, in this case the mechanical response is always smaller than unity under the impedance-matching condition ($ \mathcal{C}_0+\mathcal{C}_1=1 $) while the added noise is fairly large. This means that in the off-modulations case, the system is able to transduce the mechanical force but cannot amplify the signal and suppress the added noise. In other words, it cannot operate as a high precision measurement device. 

In the other special case where there is no atomic modulation ($ \xi_d=0 $) while the mechanical modulation is turned on, the impedance-matching condition reads $ \xi_m + \mathcal{C}_0/(1-\mathcal{C}_1)=1 $, and consequently
\begin{eqnarray} \label{signal and noise 2}
&&  n_{add}(0)=\frac{\mathcal{C}_1}{\mathcal{C}_0} (1-\xi_m)^2 (\bar n_d^T +\frac{1}{2}), \label{N_F_1}  \\
&&  R_m(0)=\frac{\mathcal{C}_0 }{(1-\xi_m)^2}. \label{S_F_1}
\end{eqnarray}
In this case, it is clear that in the limit of $ \xi_m \to 1 $ there is a large mechanical response to the input signal with no added optical noise while there is a small residual backaction noise due to the Bogoliubov mode of the BEC.

\begin{figure}
	\includegraphics[width=8.5cm]{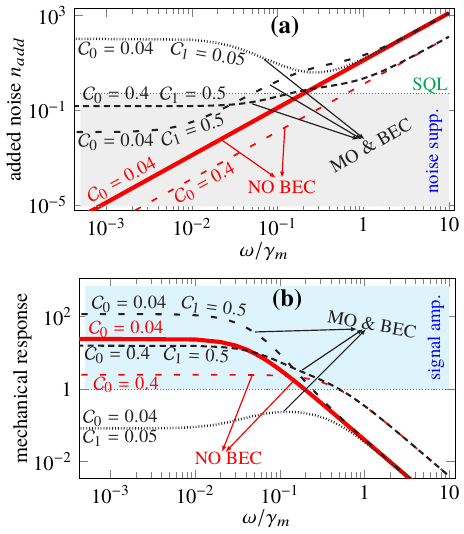}
	\caption{(Color online) (a) The added noise, $ n_{add}(\omega) $, and (b) the mechanical response to the input signal, $ R_m(\omega) $, vs dimensionless frequency $ \omega/\gamma_m $ for different ratios of cooperativities $ \mathcal{C}_{1}/\mathcal{C}_{0} $ under the impedance-matching condition of Eq.~(\ref{impedance1}). The red solid thick (indicated by $ \mathcal{C}_0=0.04 $) and the red loosely dashed (indicated by $ \mathcal{C}_0=0.4 $) curves correspond to the absence of BEC. The black solid thin (indicated by $ \mathcal{C}_0=0.04, \mathcal{C}_1=0.5 $), black densely dashed (indicated by $ \mathcal{C}_0=0.4, \mathcal{C}_1=0.5 $), and black dotted (indicated by $ \mathcal{C}_0=0.04, \mathcal{C}_1=0.05 $) curves correspond to the presence of the BEC when both atomic and mechanical modulations are turned on. The gray region (under the SQL line) in panel (a) and the cyan region in panel (b) correspond, respectively, to the situations where there are noise suppression and signal amplification. Here, we have set $ \gamma_m/\gamma_d=1 $. Reprinted with permission from A. Motazedifard, A. Dalafi, F. Bemani, and M. H. Naderi, Phys. Rev. A 100, 023815 (2019). Copyright 2019 by the American Physical Society.
	}
	\label{fig12}
\end{figure}

Figure (\ref{fig12}) shows the effect of cooperativities on the parametric force sensing for the different ratios $ \mathcal{C}_1/\mathcal{C}_0 $ under the impedance-matching condition. 
For each curve represented in Fig.~(\ref{fig12}), the effective amplitudes of modulations ($\xi_{d}$ and $ \xi_m $) can be obtained from the impedance-matching condition (\ref{impedance1}) for the specified values of cooperativities. 
Here, the red solid thick and red loosely dashed curves indicated, respectively, by $ \mathcal{C}_0=0.04 $ [with $ \xi_m=0.96 $] and $ \mathcal{C}_0=0.4 $ [with $ \xi_m=0.6 $] correspond to the absence of the BEC, i.e., $\mathcal{C}_{1}=0$. Besides, the black solid thin curve indicated by $ \mathcal{C}_0=0.04, \mathcal{C}_1=0.5 $ [with $ \xi_m=0.98, \xi_d=1.42  $], the black densely dashed curve indicated by $ \mathcal{C}_0=0.4, \mathcal{C}_1=0.5 $ [with $ \xi_m=0.84, \xi_d=1.32 $], and the black dotted curve indicated by $ \mathcal{C}_0=0.04, \mathcal{C}_1=0.05 $ [with $ \xi_m=0.30, \xi_d=0.94 $] correspond to the presence of the BEC when both atomic and mechanical modulations are turned on.

As is seen clearly in Fig.~(\ref{fig12}), in the absence of the BEC an acceptable amount of signal amplification is achievable near the on-resonance frequency through the mechanical modulation for small values of mechanical cooperativities while the added noise is nearly equal to zero which is due to the absence of an extra mode (the red solid thick curve indicated by $ \mathcal{C}_0=0.04 $). Nevertheless, the presence of the BEC with a large ratio of $\mathcal{C}_1/\mathcal{C}_0$ together with both atomic and mechanical modulations lead to much stronger signal amplification while the added noise does not increase very much and stays much below the SQL (see the black solid thin curve indicated by $ \mathcal{C}_0=0.04, \mathcal{C}_1=0.5 $).

However, the signal amplification is reduced substantially by decreasing the ratio of $\mathcal{C}_1/\mathcal{C}_0$ (the black densely dashed curve indicated by $ \mathcal{C}_0=0.4, \mathcal{C}_1=0.5 $). Especially, for lower values of $\mathcal{C}_1$ (the black dotted curve indicated by $ \mathcal{C}_0=0.04, \mathcal{C}_1=0.05 $) not only there is no signal amplification (the signal is attenuated) but also the added noise increases significantly. Therefore, equipping the system with an extra atomic mode of a BEC together with atomic modulation can enhance the ability of signal amplification substantially while the extra added noise can be kept much below the SQL in a specific parametric regime which is based on the so-called impedance-matching condition with a large ratio of $\mathcal{C}_1/\mathcal{C}_0$.

Here, we have not taken into account the effect of the classical fluctuation in the phase of the external laser that drives the cavity, i.e., the so-called laser phase noise (LPN) in our theoretical model. Although the effect of the LPN is very important especially for large values of the laser linewidth in an ultra precision measurement but as has been shown~\cite{cQNCMehmood2018}, as far as the laser linewidth is less than 1 kHz its effect becomes very negligible. In other words, for such low values of the laser line width, there is no necessity to take into account the effect of the LPN in a theoretical model.

It is worth comparing the presented method of force sensing which is based on \textit{parametric} modulations with those based on the backaction evasion \cite{clerkfeedback,clerkquantumnosie1} and CQNC techniques \cite{cQNCPRL,cQNCPRX,cQNCNatureexp,meystreCQNC2015,aliNJP} (see sections \ref{secBAnoise} and \ref{secCQNC}). The former is able to surpass the SQL by producing a large signal without suppressing the added noise while the latter can cancel the backaction noise completely without amplifying the signal. However, the force sensing scenario proposed in the present work, which is based on simultaneous signal amplification (mechanical response amplification) and noise suppression possesses the advantages of both the above-mentioned methods in that it provides a large signal to noise ratio. It should be pointed out that the improvement of force sensing in the present scheme\cite{aliDCEforcesenning} relies on small cooperativities with a large difference, i.e., $ \mathcal{C}_0+\mathcal{C}_1 < 1 $ and $ \mathcal{C}_0 \ll \mathcal{C}_1 $, which is achievable by taking a weak red-detuned driving together with the validity of the RWA [for more details see Ref\cite{aliDCEforcesenning}].

\subsection{sensitivity, signal-to-noise ratio and experimental discussion}\label{Sen and SNR}
Let us now calculate the system sensitivity to the external force as well as the signal-to-noise-ratio (SNR) so that the advantages of the present parametric force-sensing in comparison to the others are clarified more explicitly. 
Based on Eq.~(\ref{paoutomega}) and the explanations given in its previous paragraph, the output $ P $-quadrature of the cavity field can be rewritten as \cite{aliDCEforcesenning}
\begin{equation}\label{dPa}
\delta\hat P_{a}^{out}(\omega)=\delta\hat P_{a}^{out}(\omega)|_{F=0}-\frac{i\mathcal{B}(\omega)}{\sqrt{m\hbar\omega_{m}\gamma_{m}}}\tilde{F}(\omega),
\end{equation}
where $ \delta\hat P_{a}^{out}(\omega)|_{F=0}=\mathcal{A}(\omega) \hat P_a^{in}(\omega)+ \mathcal{B}(\omega) \hat X_b^{in}(\omega) + \mathcal{D}(\omega) \hat X_d^{in}(\omega) $ is the contribution of the quantum noises and the second term is the transduction force with $ \tilde{F}(\omega)=\frac{1}{2}[F(\omega+\omega_{m})-F(\omega-\omega_{m})] $ being the external force. In order to calculate the sensitivity of the device to the external force and the SNR we define the force operator as \cite{jacobtaylorsensivity}
\begin{equation}\label{force operator}
\delta \hat F(\omega)=\frac{\delta\hat P_{a}^{out}(\omega)}{\partial\delta\hat P_{a}^{out}(\omega)/\partial\tilde F(\omega)}=\delta\hat N(\omega)+\tilde{F}(\omega),
\end{equation}
where
\begin{equation}
\delta\hat N(\omega)=i\frac{\sqrt{m\hbar\omega_{m}\gamma_{m}}}{\mathcal{B}(\omega)}\delta\hat P_{a}^{out}(\omega)|_{F=0}
\end{equation}
is the \textit{noise force} operator. The power spectrum of the noise force, i.e., $ S_N(\omega)=\frac{1}{2}\langle{\delta\hat N(\omega),\delta\hat N(\omega)^{\dagger}}\rangle $ is simply obtained as \cite{aliDCEforcesenning}
\begin{equation} \label{forcenoise}
S_{N}(\omega)=m\hbar\omega_{m}\gamma_{m}\Big[(\bar n^T_{m}+\frac{1}{2})+n_{add}(\omega)\Big].
\end{equation}
Based on the standard definition of the SNR \cite{papoulisbookstochastic}, it is the ratio of the signal, i.e., the absolute value of $\tilde{F}(\omega)$, to the variance of the noises, i.e., the square root of $S_N(\omega)$,
\begin{equation}\label{SNR}
r(\omega)=\frac{|\tilde{F}(\omega)|}{\sqrt{S_{N}(\omega)}}=\frac{|\tilde{F}(\omega)|}{\sqrt{m\hbar\omega_{m}\gamma_{m}}\sqrt{(\bar n^T_{m}+\frac{1}{2})+n_{add}(\omega)}} .
\end{equation}

The sensitivity or the minimum detectable input of the device is the minimum magnitude of the input signal required to produce an output with $r(\omega)=1$ \cite{jacobtaylorsensivity,vitaliSNR}. Therefore, $ \mathcal{S}(\omega)=\sqrt{S_N(\omega)} $ is obtained as \cite{aliDCEforcesenning}
\begin{equation}\label{sensitivity}
\mathcal{S}(\omega)=\sqrt{m\hbar\omega_{m}\gamma_{m}}\Big[(\bar n^T_{m}+\frac{1}{2})+n_{add}(\omega)\Big]^{1/2}.
\end{equation}
As is seen from Eq.~(\ref{sensitivity}), the less the added noise the better the system sensitivity (especially for $n_{add}<1/2$ the SQL is surpassed).

The presented hybrid force-sensor\cite{aliDCEforcesenning} in this section can be realized in an experimental setup. For obtaining the numerical values of the sensitivity and the SNR, we use the experimentally feasible parameters~\cite{brennNatureBECOMS,ritterBECexp} as $ N=10^5 $ Rb atoms inside an optical cavity of length $ L=178 \mu$m with a damping rate of $ \kappa=2\pi\times 1.3 $MHz and the bare frequency $ \omega_{c}=2.41494\times 10^{15} $Hz corresponding to a wavelength of $ \lambda=780 $nm. The atomic $ D_{2} $ transition corresponding to the atomic transition frequency $ \omega_{a}=2.41419\times 10^{15} $Hz couples to the mentioned mode of the cavity. The atom-field coupling strength $ g_{a}=2\pi\times 14.1 $MHz and the recoil frequency of the atoms is $ \omega_{R}=23.7 $kHz. The movable end mirror can be assumed to have a mass of $ m=10^{-9}  $g and damping rate of $ \gamma_{m}=2\pi\times 100 $Hz which oscillates with frequency $ \omega_{m}=10^5 $Hz. In addition, the coherent modulation of the mechanical spring coefficient of the MO and also the time modulation of the \textit{s}-wave scattering frequency of atom-atom interaction of the BEC can be realized experimentally \cite{pontinmodulation,jaskulaBECSCE}.

In the following, we show how the red-detuned regime of cavity optomechanics, i.e., the condition $ \bar\Delta_{0}=\omega_d=\omega_m $ is compatible with the ranges of values of $ \mathcal{C}_0 $ and $ \mathcal{C}_1 $ based on the given experimental values \cite{aliDCEforcesenning}.
\textit{Firstly}, the condition $ \omega_d=\omega_m $ determines the \textit{s}-wave scattering frequency as $ \omega_{sw}=\omega_{m}-4\omega_{R} $ which for $ \omega_{R}=23.7 $ kHz leads to $ \omega_{sw}=0.22\omega_{R} $ and consequently $ \omega_{d}=4.22\omega_{R} $. As has been explained previously, the \textit{s}-wave scattering frequency is controllable experimentally through the transverse frequency of the electromagnetic trap of the BEC. 
\textit{Secondly}, for the specified values of $ \mathcal{C}_0 $ and $ \mathcal{C}_1 $ the value of the atom-laser detuning is determined by\cite{aliDCEforcesenning}
\begin{eqnarray}
&& \Delta_a=-\frac{g_{a}^{2}}{g_{0}}\sqrt{\frac{N \gamma_{m}}{8\gamma_{d}}\frac{\mathcal{C}_0}{\mathcal{C}_1}} .
\end{eqnarray}
For example, for the black solid curve of Fig.~(\ref{fig12}) (representing the presence of BEC with $ \mathcal{C}_0=0.04 $ and $ \mathcal{C}_1=0.5 $ together with the mechanical and atomic modulations) the atom-laser detuning is obtained as $ \Delta_a=-796.527 $ GHz which is an experimentally acceptable detuning to keep the system in the regime of dispersive atom-field interaction \cite{ritterBECexp}. In this way, the frequency of the external driving laser is determined by $ \omega_{L}=\omega_{a}-\Delta_a $ which for the black solid curve of Fig.~(\ref{fig12}) is obtained as $ \omega_{L}=2.41499\times 10^{15} $Hz.
\textit{Thirdly}, using the second part of the red-detuned condition, i.e., $ \bar\Delta_{0}=\omega_d $ the optical mean-field or the intracavity photon number is determined by the relation $ n_{cav}=\bar a^{2}=\frac{\omega_{m}\Delta_0-\omega_{m}^{2}}{2(g_{0}^2+G_{0}^2)} $ where $ \Delta_0=\omega_{c}-\omega_{L}-N\frac{g_{a}^2}{2\Delta_a} $. For the black solid curve of Fig.~(\ref{fig12}) the intracavity photon number is obtained as $ n_{cav}\approx 2155 $. Now, based on the relation $ n_{cav}=\frac{E_{L}^2}{\kappa^{2}/4+\omega_{m}^{2}} $ the pump rate of the external driving laser is determined which for the black solid curve of Fig.~(\ref{fig12}) is obtained as $ E_L=1.899\times 10^8 $Hz which is consistent with the experimental data given in Refs.~\cite{ritterBECexp,brennNatureBECOMS}. Similarly, for any other specified values of $ \mathcal{C}_0 $ and $ \mathcal{C}_1 $, one can calculate the frequency $\omega_{L}$ and the pump rate $E_L$ of the external driving laser which are necessary in an experimental setup to generate the theoretical results predicted in Fig.~(\ref{fig12}).

Let us now compare the present parametrically driven hybrid optomechanical force-sensor \cite{aliDCEforcesenning} with the parametrically driven bare one \cite{optomechanicswithtwophonondriving}. For this purpose, we compare the black solid curve of Fig.~(\ref{fig12}) with the red thick solid curve (representing the absence of BEC with $\mathcal{C}_0=0.04$ together with mechanical modulation) at zero temperature and at resonance frequency ($\omega\approx 0$). Based on Eq.~(\ref{sensitivity}) the sensitivity of the parametrically driven hybrid system (the black solid curve) is obtained as $\mathcal{S}|_{Hb}=5.82\times 10^{-20} \rm N/\sqrt{\rm Hz}$ while that of the parametrically driven bare system (red solid curve) is $\mathcal{S}|_{Br}=5.76\times 10^{-20} \rm N/\sqrt{\rm Hz}$. As is seen, both the (modulated) bare and the hybrid systems have nearly the same sensitivity. It should be noted that in order to have reliable results the SNR must be greater than a certain confident level, e.g., $r>3$ \cite{vitaliSNR}. To this end, either in the bare or in the hybrid system the input signal should be at least greater than $3\mathcal{S}$, i.e., $\tilde{F}>18\times 10^{-20} \rm N/\sqrt{\rm Hz}$ so that one can be assured that the signal has been detected in the system output\cite{aliDCEforcesenning}.

Although, the sensitivity and the SNR of the present (modulated) hybrid system are nearly the same as those of (modulated) bare one, the (modulated) hybrid system has a much greater mechanical gain $(R_{m|Hb}\approx 118)$ in comparison to the bare one $(R_{m|Br}\approx 25)$ (about 5 times larger) \cite{aliDCEforcesenning}. Here, it should be noted that in the absence of modulations the signal is not amplified because $ R_{m|\rm {off-mods}} < 1 $. In order to see the advantage of an amplifier with a larger mechanical gain let us look at Eq.~(\ref{dPa}). As is seen, the role of the mechanical gain is that it amplifies both the input signal, i.e., $\tilde{F}(\omega)$ and the mechanical thermal noise, i.e., $\delta\hat X_b(\omega) $ \cite{aliDCEforcesenning}. It is because of the fact that the signal is entered into the system through the channel of the input thermal noise which is a normal phenomenon in most amplifiers. Since the mechanical gain makes the input signal as well as the input thermal noise be amplified simultaneously, the increase of the mechanical gain does not help the enhancement of the sensitivity and the SNR \cite{aliDCEforcesenning}.

However, the important point is that if the signal is so weak that cannot be sensed by any instrument, one firstly needs a quantum instrument (a quantum detector/\textit{amplifier}) that can sense the weak signal and also amplify it so much that it can be sensible for a classical electronic device that receives the output of the quantum amplifier and gives us a photocurrent which is equivalent to the output cavity spectrum \cite{aliDCEforcesenning}. However, as has been already explained, the present amplifier (like others) amplifies both the signal and the thermal noise simultaneously. In this way, the SNR does not increase very much. Nevertheless, we will have an amplified signal (together with an amplified thermal noise) in the system output which is strong enough to be detected by an electronic device connected to the output of the amplifier \cite{aliDCEforcesenning}.

Therefore, although the thermal noise in the system output has been simultaneously amplified but the important points are \cite{aliDCEforcesenning}

i) The weak signal which was not previously detectable by the electronic device, has now been amplified so strongly that can be sensed by an electronic device which is connected to the output of our system.

ii) The output signal can be separated from the thermal noise (which is a white noise) by the well-known methods in electronics, especially if the signal frequency is known in advance. 
That is why the enhancement of the mechanical response is so important.

In short, in the context of the linearized regime of nonlinear hybrid OMSs, the nonlinear hybrid amplifier force-sensor \cite{aliDCEforcesenning} can act as a much better amplifier (because of its large mechanical gain) in comparison to the (modulated) bare one \cite{optomechanicswithtwophonondriving} which can amplify the input signal substantially while keeping the sensitivity nearly the same as that of the (modulated) bare one (for more details, see Refs.\cite{aliDCEforcesenning,optomechanicswithtwophonondriving}). However, it would be a future research to investigate the nonlinear amplifier in the nonlinear regime.

\section{Summary, conclusion and Perspectives \label{secOutlook}}

By considering the nonlinear hybrid OMSs as quantum sensors including atomic-BEC or an ensemble of ultracold atoms, we review and present three well-known methods for surpassing the SQL. As is well-known, in order to do an ultra precision quantum measurement below the SQL one needs to overcome the destructive effect of the backaction noise. In the first two methods, it is shown that one can evade or cancel the backaction noise to overcome the SQL while in the third method it has been shown how one can simultaneously suppress the backaction noise and amplify the input signal. It should be reminded that in the presented hybrid models, we use the atomic subsystem inside the cavity as an extra degree of freedom which plays the role of a mechanical mode interacting with the radiation pressure of the cavity. Although, one can consider real mechanical phononic modes in the OMS instead of atomic subsystem, but the controllability of the BEC is much more than the real mechanical oscillators. Nevertheless, experimental realization of the stable atomic-BEC or ultracold atomic-ensemble inside the optomechanical cavity is not very easy.

More surprisingly, it should be mentioned that very recently by exploiting the injection of squeezed vacuum states of light into the Virgo interferometer, the quantum radiation pressure backaction noise has been observed in Advanced Virgo gravitational wave detector \cite{virgoBAexperiment2020} for a macroscopic mirror with a mass of 42kg, as a quantum MO, at frequencies between 30-70 Hz \cite{virgoBAexperiment2020}. It is also expected that, in the future, the CQNC \cite{aliNJP,cQNCPRL,cQNCNatureexp,cQNCPRX} or parametric noise-suppression and signal-amplification \cite{aliDCEforcesenning,optomechanicswithtwophonondriving} which have been presented in two last sections (\ref{secCQNC}) and (\ref{secParametricsensing}) might be observed in macroscopic level in Advanced Virgo interferometer.

As a promising \textit{outlook} for future works, combination of the CQNC\cite{aliNJP}, optimized-quadrature force-sensing, complex-squeezing detection scheme, i.e., the so-called synodyne detection \cite{complexCQNC}, and single-quadrature parametric-sensing\cite{aliDCEforcesenning} might be explored as an effective method to perform ultra-precision quantum sensing in optomechanical sensors far below the SQL.

\begin{acknowledgements}
AMF would like to thank the Office of the Vice President for Research of the University of Isfahan and ICQTs. Also, the authors would like to express their gratitude to the referees whose valuable comments have improved the article substantially.
\end{acknowledgements}

\section*{Data Availability}
The data that supports the findings of this study are available within the article.

\end{document}